\documentclass[12pt,english]{article}
\usepackage{graphicx}
\usepackage{amstext}
\usepackage{caption}
\usepackage{etoolbox}
\usepackage{makeidx}
\include{epsf}
\usepackage{amsfonts}
\usepackage{authblk} 
\usepackage{sectsty}
\usepackage{amsmath,amssymb,epsfig}
\usepackage{amscd}
\usepackage{amsthm}
\usepackage{mathrsfs}
\usepackage{dsfont}
\usepackage[applemac]{inputenc}
\usepackage[english]{babel}
\usepackage{enumitem} 
\usepackage[]{latexsym}
\usepackage{braket}
\usepackage{caption}
\usepackage{hyperref}
\hypersetup{
pdftitle={},%
pdfauthor={},%
pdfsubject={},%
pdfkeywords={},%
colorlinks=true,%
linkcolor=blue,%
citecolor=red,%
linktocpage=true,%
hyperfootnotes=true,%
pageanchor=true
}

\newcommand{\be}{\begin{equation}}
\newcommand{\ee}{\end{equation}}
\def\beqa{\begin{eqnarray}}
\def\eeqa{\end{eqnarray}}
\def\bean{\begin{eqnarray*}}
\def\eean{\end{eqnarray*}}

\newcommand{\Tr}[1]{\:{\rm Tr}\,#1}

\renewenvironment{thebibliography}[1]
         {\section*{References}\frenchspacing\small
          \begin{list}{[\arabic{enumi}]}
         {\usecounter{enumi}\parsep=2pt\topsep 0pt
         \settowidth{\labelwidth}{[#1]}
         \leftmargin=\labelwidth\advance\leftmargin\labelsep
         \rightmargin=0pt\itemsep=1pt\sloppy}}{\end{list}}

 \numberwithin{equation}{section}

\input xy
\xyoption{all}

\newcommand{\grit}[1]{{\bfseries {\itshape {#1}}}}

\newcommand{\stsp}{\mathcal{S}}

\newcommand{\vsp}{\vspace{0.4cm}}
\newcommand{\gr}{\mathrm{g}}

\newtheorem{remark}{Remark}

\newtheorem{definition}{Definition}

\newtheorem{proposition}{Proposition}

\newtheorem*{proof*}{Proof}

\title{\textbf{A Pedagogical Intrinsic Approach to Relative Entropies as Potential Functions of Quantum Metrics: the $q$-$z$ Family}\vspace{0.5cm}}
\date{}

\author[1,2,3]{Florio M. Ciaglia}
\author[1,2,4]{Fabio Di Cosmo}
\author[1,2,5]{Marco Laudato}
\author[1,2,6]{Giuseppe Marmo}
\author[1,2,7]{Fabio M. Mele}
\author[1,2,8]{Franco Ventriglia}
\author[1,2,9]{Patrizia Vitale}

\affil[1]{\textit{\footnotesize Dipartimento di Fisica ``E. Pancini'', Universit\`a di Napoli Federico II, Complesso Universitario di Monte S. Angelo Edificio 6, via Cintia, 80126 Napoli, Italy.}}
\affil[2]{\textit{\footnotesize INFN-Sezione di Napoli, Complesso Universitario di Monte S. Angelo Edificio 6, via Cintia, 80126 Napoli, Italy.}}

\affil[3]{\footnotesize e-mail: \texttt{florio.m.ciaglia@gmail.com}}

\affil[4]{\footnotesize e-mail: \texttt{fabiodicosmo@gmail.com}}

\affil[5]{\footnotesize e-mail: \texttt{marcolaudato@hotmail.com}}

\affil[6]{\footnotesize e-mail: \texttt{marmo@na.infn.it}}

\affil[7]{\footnotesize e-mail: \texttt{mele.fabio91@gmail.com}}

\affil[8]{\footnotesize e-mail: \texttt{ventriglia@na.infn.it}}

\affil[9]{\footnotesize e-mail: \texttt{patrizia.vitale@na.infn.it}}

\begin{document}

\maketitle

\begin{abstract}
\small
The so-called $q$-z-\textit{R\'enyi Relative Entropies} provide a huge two-parameter family of relative entropies which includes almost all well-known examples of quantum relative entropies for suitable values of the parameters. 
In this paper we consider a log-regularized version of this family and use it as a family of potential functions to generate covariant $(0,2)$ symmetric tensors on the space of invertible quantum states in finite dimensions. 
The geometric formalism developed here allows us to obtain the explicit expressions of such tensor fields in terms of a basis of globally defined differential forms on a suitable unfolding space without the need to introduce a specific set of coordinates.
To make the reader acquainted with the intrinsic formalism introduced, we first perform the computation for the qubit case, and then, we extend the computation of the metric-like tensors to a generic $n$-level system. 
By suitably varying the parameters $q$ and $z$, we are able to recover well-known examples of quantum metric tensors that, in our treatment, appear written in terms of globally defined geometrical objects that do not depend on the coordinates system used. 
In particular, we obtain a coordinate-free expression for the von Neumann-Umegaki metric, for the Bures metric and for the Wigner-Yanase metric in the arbitrary $n$-level case.  
\end{abstract}

\tableofcontents

\section{Introduction}

  Information geometry is an approach to  classical information  by means of modern geometry \cite{amari, Amari-Nagaoka}. 
  In such a framework, families of classical probability distributions are endowed with the structure of a Riemannian manifold, which we provisionally call the statistical manifold\footnote{Strictly speaking what is called a  {\it Statistical Manifold} $P(\chi)$ is the infinite dimensional manifold of probability densities $p(x)$  on a measure space $\chi, x\in \chi$, while a   {\it Statistical Model}  is the manifold of  families of probability densities $p(x;\xi)$ parametrized by a set of n variables $\xi=[\xi_1, ..., \xi_n]$, so that the map $\xi \rightarrow p(x;\xi)$ is injective.},  in order to address natural questions such as the definition of  distance between two probability distributions, or the notion of parallel transport, necessary to take derivatives on the statistical manifold, if one wants to perform any calculation on it.  
In classical information geometry a metric tensor and a dual pair of affine connections on the space of probability distributions may be obtained from a divergence function (say a potential function), which is a two-points function defined on two copies of the statistical manifold\footnote{A dynamical framework where two copies of the statistical manifold are connected with the tangent bundle by means of the Hamilton-Jacobi theory can be found in \cite{CDCFMMP17}.} and, roughly speaking, may be interpreted as a measure of the separation between probabilities. 
The metric is essentially associated with the Hessian matrix of the divergence function and the connections with triple derivatives of the divergence.  
Since the divergence need not be symmetric with respect to the exchange of its arguments, the triple derivatives give rise to two dually paired connections, and these connections need not be metric.
It is possible to describe these connection by means of a tensor field known as skewness tensor \cite{a.a.v.v.-differential_geometry_in_statistical_inference}.

In this paper we switch the focus to the quantum world, where  probability distributions are replaced by quantum states (non-commutative analogue of probabilities). Paraphrasing what we said in the beginning, Quantum Information Geometry is an approach to quantum information by means of geometry. 
However, the set of quantum states is not a differential manifold, it is  a disjoint union of differential manifolds of different dimensions\footnote{The only exception is the two-dimensional case, in which the space of quantum states is a smooth manifold with boundary known as the Bloch ball \cite{grab, orbite}} \cite{grab, orbite}. 
Therefore, the issue of developing a coordinate-free differential calculus, which would supply  an intrinsic definition of the metric and skewness tensor together with operative tools to perform calculations,  is not obvious.
Here, we will focus on the space of invertible quantum states (density matrices with maximal rank), which is a differential manifold in the standard sense \cite{Marsden}. 
Following the ideas of  some of the authors \cite{vitale}, we decided to unfold the space of invertible quantum states to a ``bigger manifold'', namely, the product of the unitary group times the open interior of the simplex.
In this manner, being the unfolding manifold parallelizable, the differential calculus on the subalgebra of covariant tensor fields pulled-back from the space of invertible quantum states benefits of the existence of left-invariant and right-invariant vector fields and forms. 
In particular, we exploit these tools to perform the calculations needed to extract the metric tensors from the family of quantum potential functions known as q-z-relative entropies \cite{DATTA} without introducing explicit coordinates. 
As in \cite{vitale}, metric tensors for states that are not of maximal rank are obtained via a limiting procedure.

The paper is organized as follows. 
The first part of the article contains a pedagogical and detailed formulation of the canonical formalism of Information Geometry, both classical and quantum.
In this part of the paper we aim at giving a reformulation of some well-known and some new results in Information Geometry within a fully coordinate-free picture. 
Specifically, in Section \ref{gfromS} we will give a careful definition of the notion of potential functions for geometric tensors, and explore their symmetry properties further developing the work done in \cite{vitale}.
In Subsection \ref{monodpi} we will focus on the quantum picture, and we will recall the notion of quantum stochastic maps, the notion of monotonicity property for a family of metric tensors on the space of invertible quantum states, and the notion of Data Processing Inequality (DPI) for a family  of quantum divergence functions on the space of invertible quantum states.
The coordinate-free picture just introduced will allow us to prove that the DPI for a family of quantum divergence functions implies the monotonicity property for the associated family of quantum metric tensors.

The second part is dedicated to the application of the formalism developed in the first part to an explicit example, i.e., the family of quantum relative  entropies known as q-z-relative entropies \cite{DATTA}. 
In Section \ref{qzrelentropy} we review some examples of well known quantum relative entropies, together with their relevant properties, and we show how to retrieve all of them from the family $S_{q,z}$ of $q-z$-relative entropies   \cite{DATTA}  by means of suitable choices on the particular values of the parameters $(q,z)$. 
Our goal is to compute the metric tensors associated with th family of $q-z$-relative entropies for a generic n-level system. 
In order to do this without introducing explicit coordinates, in Section \ref{unfolding} we unfold the space of invertible quantum states $\mathcal S_n$ of an n-level quantum system into the manifold $\mathcal{M}_{n}=SU(n)\times\Delta_{n}^{0}$ given by the Cartesian product of the Lie group $SU(n)$ and the open interior of a $n$-dimensional simplex. 
Indeed, any $\rho\in\mathcal S_n$ can be parametrized in terms of a diagonal matrix $\rho_{0}$ and unitary transformation $U\in SU(n)$ by means of $\rho=U\rho_0 U^{-1}$.
Clearly, the unitary matrix $U$ is determined only up to unitary transformations by elements in the commutant of $\rho_0$. 
The diagonal matrices associated with states, form a simplex, therefore, if we limit the analysis to faithful states, we can consider the space to be parametrized by some homogeneous space of $SU(n)$, i.e., $SU(n)$ quotiented by the stability group of the state times the open ``part'' of the simplex to which the diagonal part of the state belongs. 
As the homogeneous spaces of $SU(n)$ are not parallelizable , we shall consider the differential calculus we are going to use as carried on the group $SU(n)$ times the open part of the simplex. 
In Section \ref{2level} we  tackle the problem of computing a metric out of  the potential function $S_{q,z}$ adopting the techniques developed in Sec. \ref{gfromS},  for qubits, and we retrieve some notable limits such as the Bures metric tensor and the Wigner-Yanase metric tensor.
In Sec \ref{Nlev} we extend the result to n-level systems and, after non-trivial calculations fully explained in the  \ref{subsec: explicit computations}, we obtain the explicit formula of equation \eqref{gNlevel} for the (pullback of the) metric tensor.   
Although the actual computations are non-trivial, the final expression is in our opinion remarkably simple. 
Many interesting examples of quantum metric tensors which are already available in the literature can be easily obtained as special cases of our formula. 
This represents the main result of our work. 
In Sec. \ref{limandex} we study the expression of the metric tensor when we take some special limits for the parameters $q$ and $z$. 
For example, we are able to obtain explicit expression for the Bures and the  Wigner-Yanase metric tensor for generic n-level systems. 
Section \ref{sec: Conclusions and Outlook} presents some concluding remarks.
 
\section{Quantum metrics from potential functions}\label{gfromS}

In this section we will provide an intrinsic definition of the coordinates-based formulae used in information geometry to derive a metric tensor and a skewness tensor  from a divergence function \cite{amari, Amari-Nagaoka, Matumoto}.
We will recast most of the well-known material on divergence functions and their symmetry properties using the intrinsic language of differential geometry.
This will turn out to be very useful when dealing with quantum information geometry, where we have to take in consideration nonlinear manifolds like the space of pure quantum states.

From a purely abstract point of view, the aim of this section is to answer the following question: given a finite-dimensional, real smooth manifold $M$ and a smooth two point-function $S\colon M\times M$, is it possible to give a coordinate-free algorithm by means of which we can extract symmetric covariant tensor fields of order $2$ and $3$?
Note that we are not assuming $M$ to be a statistical model, that is, $M$ is a generic abstract manifold.
From this, it follows that we are considering a generalization of what is done in classical information geometry, where the starting point is a smooth manifold the points of which are in one-to-one correspondence with some family of probability distributions on some measurable space.
Furthermore, we are not imposing any requirement on the two-point function $S$, unlike it is often done in classical information geometry where $S$ is assumed to be a divergence function (i.e., a non-negative function vanishing on the diagonal of $M\times M$).

We will give a positive answer to the above-mentioned question exhibiting an explicit extraction algorithm that does not depend on the coordinates used and works for every finite-dimensional, real smooth manifold.
It turns out that the class of functions for which this algorithm is well-defined is broader than the class of divergence functions of classical information geometry.
We call a generic function for which the extraction algorithm applies, a \grit{potential function}.

A similar attitude toward the definition of an extraction algorithm was pursued in \cite{vitale}.
However, the results presented there are not as general as those presented here.
Indeed, the principal ingredient of \cite{vitale} is a smooth two-point function $F$ which happens to be a divergence function, that is, it is non-negative and vanishes on the diagonal, while here we start placing no restriction of our two-point function, and work our way back to the minimum set of properties that it must satisfy in order for the extraction algorithm to work.
What is more, it follows from proposition $1$ in \cite{DATTA} that not all elements in the family of $q$-$z$-\textit{R\'enyi Relative Entropies} we will be using are divergence functions.

The coordinate-free point of view developed in this section allows to set up an abstract framework in which computations may be performed without the need to introduce coordinates, so that the geometrical properties are not hided by the explicit system of coordinates adopted, and thus are always clearly exposed.
Furthermore, in this framework it is possible to give a qualitative analysis of the symmetry properties of the tensor fields in terms of the symmetry properties of the potential functions from which they are extracted.
In the context of quantum information theory, this will lead us to easily show that the so-called data processing inequality for quantum divergence functions implies the so-called monotonicity property for the associated quantum metric tensors.
Although this is result is not new, we believe that the coordinate-free geometrical framework introduced here permits a clearer analysis of the subject and is more in line with the modern attitude toward the geometrization of physics and information theory.

\vsp

Roughly speaking, in the following we will introduce the notions of left and right lift of a vector field, along with the diagonal immersion of a manifold into its double.
Then, after some important properties connecting the diagonal immersion with the left and right lifts are proved, we introduce a coordinate-free algorithm to extract covariant tensors of order $(0,2)$ and $(0,3)$ from a two-point function.
This will lead us to define the class of \grit{potential functions}, i.e., the class of two-point functions generalizing the concept of divergence functions of classical information geometry.
Finally, we will analyze how potential functions behave with respect to smooth maps between differential manifolds.

\vsp

Let $M$ be a differential manifold, $TM$ its tangent bundle, and $\tau\colon TM\rightarrow M$ the canonical projection.
A point in the tangent bundle $TM$ is a couple $(m\,,v_{m})$, where $m\in M$, and $v_{m}\in T_{m}M$ is a tangent vector at $m$.
Note that, in general, $TM$ is not a cartesian product, hence, the notation $(m\,,v_{m})$ should be treated with care because the second factor $v_{m}$ is not independent from the first one.
A vector field $X\in\mathfrak{X}(M)$ may be thought of as a derivation of the associative algebra $\mathcal{F}(M)$ of smooth functions on $M$, or as a section of the tangent bundle $TM$, that is, a map $X\colon M\rightarrow TM$ such that $\tau\circ X=id_{M}$.
In the latter case, we may write the evaluation of a  vector field on $m\in M$ as $X(m)=(m\,,v^{X}_{m})$.

Let $M\times M$ denote the so-called double manifold of $M$.
We have two canonical projections $pr_{l}\colon M\times M\rightarrow M$ and $pr_{l}\colon M\times M\rightarrow M$ acting as:
\be
\begin{split}
pr_{l}(m_{1}\,,m_{2})&:=m_{1}\\
pr_{r}(m_{1}\,,m_{2})&:=m_{2}\,.
\end{split}
\ee
Given $f\in\mathcal{F}(M)$, we may define the following functions on $M\times M$ by means of $pr_{l}$ and $pr_{r}$:
\be
\begin{split}
f_{l}\colon M\times M\rightarrow \mathbb{R}\,,&\;\;\;\;\; f_{l}:=pr_{l}^{*}f\\
f_{r}\colon M\times M\rightarrow \mathbb{R}\,,&\;\;\;\;\; f_{r}:=pr_{r}^{*}f\,.
\end{split}
\ee
This means that on $M\times M$ we have identified two different subalgebras of $\mathcal{F}(M\times M)$, the left and the right subalgebras:
\be
\begin{split}
\mathcal{F}_{l}(M\times M)&:=\left\{f_{l}\in\mathcal{F}(M\times M)\colon\;\exists f\in\mathcal{F}(M) \mbox{ such that }\;f_{l}=pr_{l}^{*} f\right\}\\
&\\
\mathcal{F}_{r}(M\times M)&:=\left\{f_{r}\in\mathcal{F}(M\times M)\colon\;\exists f\in\mathcal{F}(M) \mbox{ such that }\;f_{r}=pr_{r}^{*} f\right\}\,.
\end{split}
\ee
The tangent space $T_{(m_{1},m_{2})} M\times M$ at $(m_{1}\,,m_{2})\in M\times M$ may be split into the direct sum $T_{m_{1}}M\oplus T_{m_{2}}M$.
Accordingly, we may write the evaluation of a vector field $\mathbb{X}\in\mathfrak{X}(M\times M)$ at $(m_{1}\,,m_{2})$ as:
\be
\mathbb{X}(m_{1}\,,m_{2})=(m_{1}\,,v^{\mathbb{X}}_{m_{1}}\,;m_{2}\,,v^{\mathbb{X}}_{m_{2}})\,.
\ee
This motivates the following:

\begin{definition}\label{def: left and right lift of vector fields}
Let $X\in\mathfrak{X}(M)$ be a smooth vector field.
We defined the left and right lift of $X$ to be, respectively, the vector fields $\mathbb{X}_{l},\mathbb{X}_{r}\in\mathfrak{X}(M\times M)$ defined as:
\be
\mathbb{X}_{l}(m_{1}\,,m_{2})=(m_{1}\,,v^{X}_{m_{1}}\,;m_{2}\,,0)\,,
\ee
\be
\mathbb{X}_{r}(m_{1}\,,m_{2})=(m_{1}\,,0\,;m_{2}\,,v^{X}_{m_{1}})\,.
\ee
\end{definition}

By direct computation, it is possible to prove the following:

\begin{proposition}\label{prop: important properties of left and right lift}
Let $X,Y\in\mathfrak{X}(M)$,  and $f\in\mathcal{F}(M)$, and denote with $L$ the Lie-derivative.
The following equalities hold:
\be
[\mathbb{X}_{l}\,,\mathbb{Y}_{l}]=\left([X\,,Y]\right)_{l}\,,\;\;\;\;\;[\mathbb{X}_{r}\,,\mathbb{Y}_{r}]=\left([X\,,Y]\right)_{r}\,,\;\;\;\;\;[\mathbb{X}_{l}\,,\mathbb{Y}_{r}]=0\,,
\ee
\be
(fX)_{l}=f_{l}\mathbb{X}_{l}\,,\;\;\;\;\;(fX)_{r}=f_{r}\mathbb{X}_{r}\,,\;\;\;\;\;L_{\mathbb{X}_{l}}f_{r}=L_{\mathbb{X}_{r}}f_{l}=0\,.
\ee
\end{proposition}

There is a natural immersion $i_{d}$ of $M$ into its double $M\times M$ given by:
\be
M\ni m\mapsto i_{d}(m)=(m\,,m)\in M\times M\,.
\ee
The map $i_{d}$ allows us to immerse $M$ in the diagonal of its double, and, by means of the pullback operation, gives an intrinsic and coordinates-free definition of the procedure of ``restricting to the diagonal'' used in information geometry.
Indeed, the pullback of a function to a submanifold can be identified with the restriction of the function to the submanifold.
Note that the same is not true for covariant tensors of higher order for which a "restriction" in the sense of evaluation at specific points is always possible, however this does not coincide with the value that the pulled-back covariant tensor will take at the same point as an element of the submanifold.

By using the tangent functor it is possible to associate vector fields on $M$ with vector fields on $M\times M$ along the immersion $i_{d}$ of $M$ into $M\times M$.
We have the following proposition:

\begin{proposition}
Let $X\in\mathfrak{X}(M)$, then $X$ is $i_{d}$-related to $\mathbb{X}_{l} + \mathbb{X}_{r}$, that is \cite{Marsden}:
\be
Ti_{d}\circ X= \mathbb{X}_{lr}\circ i_{d},,
\ee
where $\mathbb{X}_{lr}\equiv(\mathbb{X}_{l} + \mathbb{X}_{r})$, and $Ti_{d}$ denotes the tangent map of $i_{d}$.
\begin{proof*}
By direct computation, we have:
\be
Ti_{d}\circ X(m)=Ti_{d}(m\,,v^{X}_{m})=(m\,,v^{X}_{m}\,;m\,,v^{X}_{m})\,.
\ee
and:
\be
\mathbb{X}_{lr}\circ i_{d}(m)=\mathbb{X}_{lr}(m\,,m)=(m\,,v^{X}_{m}\,;m\,,v^{X}_{m})\,, 
\ee
and the proposition follows.
\end{proof*}
\end{proposition}

Now, we are ready to introduce the coordinate-free algorithm to extract covariant $(0,2)$ tensor from a two-point function.
In order to do so, we define the following maps:

\begin{definition}\label{def: covariant 2-tensors from lie derivative}
Let $S\in\mathcal{F}(M\times M)$.
We define the following bilinear, $\mathbb{R}$-linear maps from $\mathfrak{X}(M)\times\mathfrak{X}(M)$ to $\mathcal{F}(M)$:
\be
g_{ll}\left(X\,,Y\right):=i_{d}^{*}\left(L_{\mathbb{X}_{l}}L_{\mathbb{Y}_{l}} S\right)\,,\;\;\;\;\;g_{rr}\left(X\,,Y\right):=i_{d}^{*}\left(L_{\mathbb{X}_{r}}L_{\mathbb{Y}_{r}} S\right)\,,
\ee
\be
g_{lr}\left(X\,,Y\right):=i_{d}^{*}\left(L_{\mathbb{X}_{l}}L_{\mathbb{Y}_{r}} S\right)\,,\;\;\;\;\;g_{rl}\left(X\,,Y\right):=i_{d}^{*}\left(L_{\mathbb{X}_{r}}L_{\mathbb{Y}_{l}} S\right)\,.
\ee
\end{definition}

Notice that, at the moment, these maps  do not have definite symmetry properties.
To prove that these maps give a coordinate-free version of the formulae for metric-like tensors used in information geometry, we start with the following proposition:

\begin{proposition}\label{prop: nec and suf conditions for building 2 tensors from S}
Consider the maps in definition \ref{def: covariant 2-tensors from lie derivative}.
Then:
\begin{enumerate}
\item\label{cond: 1} $g_{lr},g_{rl}$ are covariant $(0\,,2)$ tensors on $M$, and $g_{lr}(X\,,Y)=g_{rl}(Y\,,X)$;
\item\label{cond: 2} $g_{ll}$ is a symmetric covariant $(0\,,2)$ tensor on $M$ if and only if:
\be\label{eqn: necessary and sufficient conditions for gll 2}
i_{d}^{*}\left(L_{\mathbb{X}_{l}}S\right)=0\;\;\forall X\in\mathfrak{X}(M)\,;
\ee
\item\label{cond: 3} $g_{rr}$ is a symmetric covariant $(0\,,2)$ tensor on $M$ if and only if:
\be\label{eqn: necessary and sufficient conditions for grr 2}
i_{d}^{*}\left(L_{\mathbb{X}_{r}}S\right)=0\;\;\forall X\in\mathfrak{X}(M)\,.
\ee
\end{enumerate}

\begin{proof*}
To show \ref{cond: 1} we have to show that $g_{lr}$ and $g_{rl}$ are bilinear with respect to vector fields, and $\mathcal{F}(M)$-linear.
We start with $g_{lr}$.
According to proposition \ref{prop: important properties of left and right lift}, we have:
\be
g_{lr}(fX + hY\,,Z)=i_{d}^{*}\left(L_{f_{l}\mathbb{X}_{l} + h_{l}\mathbb{Y}_{l}}L_{\mathbb{Z}_{r}} S\right)\,.
\ee
The linearity of the pullback, together with the properties of the Lie derivative, imply:
\be
i_{d}^{*}\left(L_{f_{l}\mathbb{X}_{l} + h_{l}\mathbb{Y}_{l}}L_{\mathbb{Z}_{r}} S\right)=i_{d}^{*}\left(f_{l}L_{\mathbb{X}_{l}}L_{\mathbb{Z}_{r}} S\right) + i_{d}^{*}\left(h_{l}L_{\mathbb{Y}_{l}}L_{\mathbb{Z}_{r}} S\right)\,.
\ee
Since $L_{\mathbb{X}_{l}}L_{\mathbb{Z}_{r}} S$ and $L_{\mathbb{Y}_{l}}L_{\mathbb{Z}_{r}} S$ are smooth functions, we have that 
$$
i_{d}^{*}\left(f_{l}\,L_{\mathbb{X}_{l}}L_{\mathbb{Z}_{r}} S\right)=i_{d}^{*}f_{l} \, i_{d}^{*}\left(L_{\mathbb{X}_{l}}L_{\mathbb{Z}_{r}} S\right)\,,
$$
$$
i_{d}^{*}\left(f_{l}\,L_{\mathbb{Y}_{l}}L_{\mathbb{Z}_{r}} S\right)=i_{d}^{*}f_{l} \, i_{d}^{*}\left(L_{\mathbb{Y}_{l}}L_{\mathbb{Z}_{r}} S\right)\,,
$$
and thus:
\be
g_{lr}(fX + hY\,,Z)=f\,i_{d}^{*}\left(L_{\mathbb{X}_{l}}L_{\mathbb{Z}_{r}} S\right) + h\,i_{d}^{*}\left(L_{\mathbb{Y}_{l}}L_{\mathbb{Z}_{r}} S\right)=f\,g_{lr}(X\,,Z) + h\,g_{lr}(Y\,,Z)\,.
\ee
According to last equality of proposition \ref{prop: important properties of left and right lift}, we have $L_{\mathbb{X}_{l}}f_{r}=L_{\mathbb{X}_{r}}f_{l}=0$ for all $X$ and $f$.
Taking this equality into account, we may proceed as above, and show that:
\be
g_{lr}(Z\,,fX + hY)=f\,g_{lr}(Z\,,X) + h\,g_{lr}(Z\,,Y)\,.
\ee
This proves that $g_{lr}$ is a covariant $(0\,,2)$ tensor field on $M$.
With exactly the same procedure, we can prove that $g_{rl}$ is a covariant $(0\,,2)$ tensor field on $M$.
Finally, the equality $g_{lr}(X\,,Y)=g_{rl}(Y\,,X)$ follows from direct computation.

To show \ref{cond: 2}, again, we have to show that $g_{ll}$ is bilinear with respect to vector fields, and $\mathcal{F}(M)$-linear.
The linearity and $\mathcal{F}(M)$-linearity on the first factor are proved analogously to the previous case.
Concerning the second factor, we start with the following chain of equalities:
$$
g_{ll}(Z\,,fX+hY)=i_{d}^{*}\left(L_{\mathbb{Z}_{l}}L_{f_{l}\mathbb{X}_{l}+h_{l}\mathbb{Y}_{l}}S\right)=i_{d}^{*}\left(L_{\mathbb{Z}_{l}}\left(f_{l}L_{\mathbb{X}_{l}}S\right)\right) + i_{d}^{*}\left(L_{\mathbb{Z}_{l}}\left(h_{l}L_{\mathbb{Y}_{l}}S\right)\right)=
$$
$$
=i_{d}^{*}\left(L_{\mathbb{Z}_{l}}f_{l}\,L_{\mathbb{X}_{l}}S\right) + i_{d}^{*}\left(f_{l}L_{\mathbb{Z}_{l}}L_{\mathbb{X}_{l}}S\right) + i_{d}^{*}\left(L_{\mathbb{Z}_{l}}h_{l}\,L_{\mathbb{Y}_{l}}S\right) + i_{d}^{*}\left(h_{l}L_{\mathbb{Z}_{l}}L_{\mathbb{Y}_{l}}S\right)=
$$
\be
=i_{d}^{*}\left(L_{\mathbb{Z}_{l}}f_{l}\right)\,i_{d}^{*}\left(L_{\mathbb{X}_{l}}S\right) + f\,g_{ll}(Z\,,X) + i_{d}^{*}\left(L_{\mathbb{Z}_{l}}h_{l}\right)\,i_{d}^{*}\left(L_{\mathbb{Y}_{l}}S\right) + h\,g_{ll}(Z\,,Y)\,.
\ee
It is then clear that:
\be
g_{ll}(Z\,,fX+hY)= f\,g_{ll}(Z\,,X) + h\,g_{ll}(Z\,,Y)
\ee
is equivalent to:
\be\label{eqn: necessary and sufficient conditions for gll 1}
i_{d}^{*}\left(L_{\mathbb{Z}_{l}}f_{l}\right)\,i_{d}^{*}\left(L_{\mathbb{X}_{l}}S\right)  + i_{d}^{*}\left(L_{\mathbb{Z}_{l}}h_{l}\right)\,i_{d}^{*}\left(L_{\mathbb{Y}_{l}}S\right)=0\,.
\ee
Being $f$ and $h$ arbitrary functions, equation \eqref{eqn: necessary and sufficient conditions for gll 1} is satisfied if and only if:
\be
i_{d}^{*}\left(L_{\mathbb{X}_{l}}S\right)=0\;\;\forall X\in\mathfrak{X}(M)
\ee
as claimed.
Now, we prove that $g_{ll}$ is a symmetric tensor:
\be
g_{ll}\left(X\,,Y\right)=i_{d}^{*}\left(L_{\mathbb{X}}L_{\mathbb{Y}}S\right)=i_{d}^{*}\left(L_{\mathbb{Y}}L_{\mathbb{X}}S\right) + i_{d}^{*}\left(L_{[\mathbb{X}\,,\mathbb{Y}]}S\right)=i_{d}^{*}\left(L_{\widetilde{Y}}L_{\widetilde{X}}S\right)=g_{ll}\left(Y\,,X\right)\,,
\ee 
where, in the last passage, we have used  the first equality of proposition \ref{prop: important properties of left and right lift}.
With exactly the same procedure we can prove  \ref{cond: 3}.
\end{proof*}
\end{proposition}

Interestingly, when $S$ satisfies condition \eqref{eqn: necessary and sufficient conditions for gll 2} and condition \eqref{eqn: necessary and sufficient conditions for grr 2}, the covariant tensor fields are all related to one another.
In order to clearly see this, we recall the following proposition (see \cite{Marsden} page 239):

\begin{proposition}\label{prop: marsden is everywhere}
Let $\phi\colon N\rightarrow M$ be a smooth map between smooth manifolds.
Let $X\in\mathfrak{X}(N)$ and $Y\in\mathfrak{X}(M)$ be $\phi$-related, that is $T\phi\circ X=Y\circ \phi$, then:
\be
L_{X}\,\phi^{*}(f)=\phi^{*}\left(L_{Y}\,f\right)\;\;\;\;\forall f\in\mathcal{F}(M)\,.
\ee
\end{proposition}

This means that $X$ and $Y$ agree along the image of $N$ into $M$.
In particular, since $X\in\mathfrak{X}(M)$ is $i_{d}$-related to $\mathbb{X}_{l} + \mathbb{X}_{r}$, we have that:
\be\label{eqn: lie derivative and id-relatedness}
L_{X}\,i_{d}^{*}(f)=i_{d}^{*}\left(L_{\mathbb{X}_{l} + \mathbb{X}_{r}}\,f\right)\;\;\;\;\forall f\in\mathcal{F}(M\times M)\,.
\ee
Now, we are ready to prove

\begin{proposition}\label{prop: all the covariant tensor fields are related when S behaves well}
Let $S$ be a smooth function on $M\times M$ satisfying condition \eqref{eqn: necessary and sufficient conditions for gll 2} and condition \eqref{eqn: necessary and sufficient conditions for grr 2}.
Then:
\be
g_{ll}=g_{rr}=-g_{lr}=-g_{rl}\,.
\ee
In particular, all these tensors are symmetric.
\begin{proof*}
According to definition \ref{def: covariant 2-tensors from lie derivative}, we have:
$$
g_{ll}\left(X\,,Y\right) + g_{lr}\left(X\,, Y\right)=i_{d}^{*}\left(L_{\mathbb{X}_{l}}L_{\mathbb{Y}_{l} + \mathbb{Y}_{r}}\,S\right)= i_{d}^{*}\left(L_{\mathbb{Y}_{l} + \mathbb{Y}_{r}}L_{\mathbb{X}_{l}}\,S\right) + i_{d}^{*}\left(L_{\left[\mathbb{X}_{l}\,,\mathbb{Y}_{l} +  \mathbb{Y}_{r}\right]}\,S\right)\,.
$$
Now, recalling that $\left[\mathbb{X}_{l}\,,  \mathbb{Y}_{r}\right]=0$ because of the third equality in proposition \ref{prop: important properties of left and right lift}, that $S$ satisfies condition \eqref{eqn: necessary and sufficient conditions for gll 2}, and recalling equation \eqref{eqn: lie derivative and id-relatedness}, we have

\be
g_{ll}\left(X\,,Y\right) + g_{lr}\left(X\,, Y\right)=i_{d}^{*}\left(L_{\mathbb{Y}_{l} + \mathbb{Y}_{r}}L_{\mathbb{X}_{l}}\,S\right)=L_{Y}\,i_{d}^{*}\left(L_{\mathbb{X}_{l}}\,S\right)=0\,.
\ee
This proves that $g_{ll}=-g_{lr}$.
Proceeding analogously, we obtain $g_{rr}=-g_{rl}$.
Then, being $g_{ll}$ and $g_{rr}$ symmetric (see proposition \ref{prop: nec and suf conditions for building 2 tensors from S}), and being $g_{lr}\left(X\,, Y\right)=g_{rl}\left( Y \,,X\right)$ (see proposition \ref{prop: nec and suf conditions for building 2 tensors from S}), we obtain $g_{lr}=g_{rl}$, and thus $g_{ll}=g_{rr}$ which completes the proof.
\end{proof*}
\end{proposition}

\begin{remark}\label{rem: vanishing of pullback of the differential on the diagonal is not good enough}
Note that conditions \eqref{eqn: necessary and sufficient conditions for gll 2} and \eqref{eqn: necessary and sufficient conditions for grr 2} are are not equivalent to $i^{*}_{d}(\mathrm{d}S)=0$.
For instance, take $M=\mathbb{R}$, and 
$$
S=\frac{x^{2} - y^{2}}{2}\,.
$$
We have 
$$
\mathrm{d}S=x\mathrm{d}x - y\mathrm{d}y\,,
$$
and thus $i_{d}^{*}\mathrm{d}S=0$, while, an easy calculation shows that
$$
i_{d}^{*}(L_{\mathbb{X}_{l}}S)=x\neq0\;\;\mbox{ if } \mathbb{X}_{l}=\frac{\partial}{\partial x}\,,
$$ 
and 
$$
i_{d}^{*}(L_{\mathbb{X}_{r}}S)=-y\neq0\;\;\mbox{ if } \mathbb{X}_{r}=\frac{\partial}{\partial y}\,.
$$
It can be checked that the maps $g_{ll}$ and $g_{rr}$ associated with $S$ do not define tensor fields because they are not $\mathcal{F}(M)$-linear in the second factor.
\end{remark}

Motivated by proposition \ref{prop: all the covariant tensor fields are related when S behaves well}, we give the following definition:

\begin{definition}[{\bf Potential function}]\label{def: potential function}
Let $S$ be a smooth function on $M\times M$.
We call $S$ a {\bf potential function} if it satisfies condition \eqref{eqn: necessary and sufficient conditions for gll 2} and condition \eqref{eqn: necessary and sufficient conditions for grr 2}, that is:
\be
i_{d}^{*}\left(L_{\mathbb{X}_{l}}S\right)=0\;\;\forall X\in\mathfrak{X}(M)\,,
\ee
\be
i_{d}^{*}\left(L_{\mathbb{X}_{r}}S\right)=0\;\;\forall X\in\mathfrak{X}(M)\,.
\ee
We denote with $g$ the symmetric covariant $(0,2)$ tensor field associated with $S$ (see proposition \ref{prop: all the covariant tensor fields are related when S behaves well}).
\end{definition}

We stress that proposition \ref{prop: all the covariant tensor fields are related when S behaves well} gives necessary and sufficient conditions for $S$ to give rise to a (unique) symmetric covariant $(0,2)$ tensor field on $M$.
This gives a formal and intrinsic characterization of potential functions.

To make contact with the coordinate-based formulae of information geometry, we introduce  coordinate chart $\{x^{j}\}$ on $M$, and a coordinate chart $\{q^{j}\,,Q^{j}\}$ on $M\times M$.
Then, we have:
\be
X=X^{j}(\mathbf{x})\,\frac{\partial}{\partial x^{j}}\,,\;\;\;\;\;\mathbb{X}_{l}=X^{j}(\mathbf{q})\,\frac{\partial}{\partial q^{j}}\,,\;\;\;\;\;\mathbb{X}_{r}=X^{j}(\mathbf{Q})\,\frac{\partial}{\partial Q^{j}}\,.
\ee
Consequently, it is easy to see that:
{\footnotesize \be\label{eqn: coordinate expression of g}
g=\left.\left(\frac{\partial^{2}\,S}{\partial q^{j}\partial q^{k}}\right)\right|_{d}\,\mathrm{d}x^{j}\otimes_{s}\mathrm{d}x^{k}=\left.\left(\frac{\partial^{2}\,S}{\partial Q^{j}\partial Q^{k}}\right)\right|_{d}\,\mathrm{d}x^{j}\otimes_{s}\mathrm{d}x^{k}=-\left.\left(\frac{\partial^{2}\,S}{\partial q^{j}\partial Q^{k}}\right)\right|_{d}\,\mathrm{d}x^{j}\otimes_{s}\mathrm{d}x^{k}\,,
\ee}
and these expressions are in complete accordance with the ones used in information geometry \cite{amari, Amari-Nagaoka, Matumoto}.

\begin{remark}
If $S$ is not a potential function, we cannot define the tensor $g_{ll}$, or the tensor $g_{rr}$, or both.
However, we can always define the tensors $g_{lr}$ and $g_{rl}$.
These tensors will not be symmetric, and we can decompose them into symmetric and anti-symmetric part.
For example, let $M=\mathbb{R}^{2}$, let $\{x^{j}\}_{j=0,1}$ be a global Cartesian coordinates system on $M$, and let $\{q^{j}\,,Q^{j}\}_{j=0,1}$ be a global Cartesian coordinates on $M\times M$.
Consider the function:
\be
S(q^{0}\,,q^{1}\,;Q^{0}\,,Q^{1})=-\frac{1}{2}\left((q^{0} - Q^{0})^{2} + (q^{1} - Q^{1})^{2} +q^{0}Q^{1} - q^{1}Q^{0}\right)\,.
\ee
An explicit calculation shows that:
\be
g_{lr}= \mathrm{d}x^{0}\otimes\mathrm{d}x^{0} + \mathrm{d}x^{1}\otimes\mathrm{d}x^{1} + \mathrm{d}x^{0}\wedge\mathrm{d}x^{1}\,.
\ee
\end{remark}

The coordinate expressions in equation \eqref{eqn: coordinate expression of g} allow us to give a ``local'' characterization of potential functions:

\begin{proposition}\label{prop: potential function iff diagonal is a critical "surface"}
A function  $S\in\mathcal{F}(M\times M)$ is a potential function according to definition \ref{def: potential function} if and only if every point $(m\,,m)$ on the diagonal of $M\times M$ is  a critical point for $S$.

\begin{proof*}
The proof follows upon comparing the local expression for $(m\,,m)$ to be a critical point for $S$ with the coordinate expressions of condition \eqref{eqn: necessary and sufficient conditions for gll 2} and condition \eqref{eqn: necessary and sufficient conditions for grr 2} in the coordinate system $\{q^{j}\,,Q^{j}\}$ introduced before.
\end{proof*}
\end{proposition}

This characterization of potential functions allows us to better understand what kind of tensor field is $g$.
Specifically, resorting to the theory of multivariable calculus it is possible to prove the following proposition:

\begin{proposition}\label{prop: nec and suf conditions for g to be positive or negative semidefinite}
Let $S$ be a potential function on $M\times M$.
Then:
\begin{enumerate}
\item $g$ is positive-semidefinite if and only if every point on the diagonal is a local minimum for $S$.
In particular, $g$ is a metric if and only if every point of the diagonal is a nondegenerate local minimum for $S$;
\item $g$ is negative-semidefinite if and only if every point on the diagonal is a local maximum for $S$.
\end{enumerate}
\end{proposition}

It is now easy to see the relation between the class of potential functions introduced here and the class of divergence functions of classical information geometry:

\begin{definition}[{\bf Divergence function}]\label{def: divergence function}
A smooth function $S$ on $M\times M$ such that:
\be
S(m_{1}\,,m_{2})\geq0\,,\;\;\;\;S(m_{1}\,,m_{2})=0\;\Longleftrightarrow m_{1}=m_{2}\,,
\ee
is called a {\bf divergence function}.
\end{definition}
According to proposition \ref{prop: potential function iff diagonal is a critical "surface"} $S$ is a potential function (see definition \ref{def: potential function}) and thus it gives rise to a symmetric covariant $(0\,,2)$ tensor field $g$ on $M$.
According to proposition \ref{prop: nec and suf conditions for g to be positive or negative semidefinite}, the tensor field $g$ is positive-semidefinite.

In information geometry, divergence (contrast) functions give rise to metric tensors by means of the second derivatives, and to symmetric covariant $(0,3)$ tensors by means of third derivatives.
These tensors are referred to as skewness tensors \cite{Amari-Nagaoka, Matumoto}. 
We will now give an intrinsic definition for these skewness tensors using again Lie derivatives.

Let $S$ be a potential function on $M\times M$.
For $j=1,...,8$, define the following maps $T_{j}:\mathfrak{X}(M)\times\mathfrak{X}(M)\times\mathfrak{X}(M)\rightarrow\mathcal{F}(M)$:
\be
T_{1}(X,Y,Z):=i_{d}^{*}\left(L_{\mathbb{X}_{l}}L_{\mathbb{Y}_{l}}L_{\mathbb{Z}_{l}}\,S\right)\,, \;\;\;\;\;
T_{2}(X,Y,Z):=i_{d}^{*}\left(L_{\mathbb{X}_{r}}L_{\mathbb{Y}_{r}}L_{\mathbb{Z}_{r}}\,S\right)\,,\ee
\be
T_{3}(X,Y,Z):=i_{d}^{*}\left(L_{\mathbb{X}_{l}}L_{\mathbb{Y}_{l}}L_{\mathbb{Z}_{r}}\,S\right)\,, \;\;\;\;\;
T_{4}(X,Y,Z):=i_{d}^{*}\left(L_{\mathbb{X}_{r}}L_{\mathbb{Y}_{r}}L_{\mathbb{Z}_{l}}\,S\right)\,,
\ee
\be
T_{5}(X,Y,Z):=i_{d}^{*}\left(L_{\mathbb{X}_{l}}L_{\mathbb{Y}_{r}}L_{\mathbb{Z}_{r}}\,S\right)\,,  \;\;\;\;\;
T_{6}(X,Y,Z):=i_{d}^{*}\left(L_{\mathbb{X}_{r}}L_{\mathbb{Y}_{l}}L_{\mathbb{Z}_{l}}\,S\right)\,,
\ee
\be
T_{7}(X,Y,Z):=i_{d}^{*}\left(L_{\mathbb{X}_{l}}L_{\mathbb{Y}_{r}}L_{\mathbb{Z}_{l}}\,S\right)\,,  \;\;\;\;\;
T_{8}(X,Y,Z):=i_{d}^{*}\left(L_{\mathbb{X}_{r}}L_{\mathbb{Y}_{l}}L_{\mathbb{Z}_{r}}\,S\right)\,,
\ee
Following the line of reasoning developed above, patient but simple calculations show that:
\be\label{eqn: 3 tensor 1}
T_{12}(X\,,Y\,,Z):= T_{1}(X\,,Y\,,Z) - T_{2}(X\,,Y\,,Z)\,,
\ee
\be\label{eqn: 3 tensor 2}
T_{34}(X\,,Y\,,Z):= T_{3}(X\,,Y\,,Z) - T_{4}(X\,,Y\,,Z)\,,
\ee
\be\label{eqn: 3 tensor 3}
T_{56}(X\,,Y\,,Z):= T_{5}(X\,,Y\,,Z) - T_{6}(X\,,Y\,,Z)\,,
\ee
\be\label{eqn: 3 tensor 4}
T_{78}(X\,,Y\,,Z):= T_{7}(X\,,Y\,,Z) - T_{8}(X\,,Y\,,Z)
\ee
are actually tensors fields on $M$.
Recalling that $X$ is $i_{d}$ related to $\mathbb{X}_{l} + \mathbb{X}_{r}$, and applying equation \eqref{eqn: lie derivative and id-relatedness}, we have:
\be
T_{1}(X\,,Y\,,Z) + T_{6}(X\,,Y\,,Z)=i_{d}^{*}\left(L_{\mathbb{X}_{l}+\mathbb{X}_{r}}L_{\mathbb{Y}_{l}}L_{\mathbb{Z}_{l}}\,S\right)=L_{X}g(Y\,,Z) \,,
\ee
\be
T_{2}(X\,,Y\,,Z) + T_{5}(X\,,Y\,,Z)=i_{d}^{*}\left(L_{\mathbb{X}_{l} + \mathbb{X}_{r}}L_{\mathbb{Y}_{r}}L_{\mathbb{Z}_{r}}\,S\right)=L_{X}g(Y\,,Z) \,,
\ee
and thus $T_{12}=T_{56}$.
Similarly, it can be shown that $T_{34}=T_{78}$.
Furthermore:
\be
T_{3}(X\,,Y\,,Z) + T_{5}(X\,,Y\,,Z)=i_{d}^{*}\left(L_{\mathbb{X}_{l}}L_{\mathbb{Y}_{l}+\mathbb{Y}_{r}}L_{\mathbb{Z}_{r}}\,S\right)=L_{Y}\,g(X\,,Z) + g\left([X\,,Y]\,,Z\right)\,,
\ee
\be
T_{4}(X\,,Y\,,Z) + T_{6}(X\,,Y\,,Z)=i_{d}^{*}\left(L_{\mathbb{X}_{r}}L_{\mathbb{Y}_{l}+\mathbb{Y}_{r}}L_{\mathbb{Z}_{r}}\,S\right)=L_{Y}\,g(X\,,Z) + g\left([X\,,Y]\,,Z\right)\,,
\ee
and thus $T_{34}=-T_{56}$, from which it follows that
\be
T_{12}=T_{56}=-T_{34}=-T_{78}\;.
\ee
This means that we can define a single symmetric tensor field $T$ of order $3$ on $M$ starting with a potential function $S$.
For instance, we set:
\be\label{eqn: 3 tensor total}
T(X\,,Y\,,Z):=i_{d}^{*}\left(L_{\mathbb{X}_{l}}L_{\mathbb{Y}_{l}}L_{\mathbb{Z}_{r}}\,S - L_{\mathbb{X}_{r}}L_{\mathbb{Y}_{r}}L_{\mathbb{Z}_{l}}\,S\right)\,.
\ee
We have thus proved the following proposition:

\begin{proposition}
Let $S$ be a potential function on $M\times M$.
Then, all the maps defined in equations \eqref{eqn: 3 tensor 1}, \eqref{eqn: 3 tensor 2}, \eqref{eqn: 3 tensor 3}, and \eqref{eqn: 3 tensor 4} define the same symmetric covariant $(0\,,3)$ tensor field $T$ on $M$.
\end{proposition}

For the sake of simplicity, we write $T$ as in equation \eqref{eqn: 3 tensor total}.
In the coordinate charts $\{x^{j}\}$ and   $\{q^{j}\,,Q^{j}\}$ introduced above, we have:
\be
T=\left.\left(\frac{\partial^{3}S}{\partial q^{j}\partial q^{k}\partial Q^{l}} - \frac{\partial^{3}S}{\partial Q^{j}\partial Q^{k}\partial q^{l}}\right)\right|_{d}\mathrm{d}x^{j}\otimes_{s}\mathrm{d}x^{k}\otimes_{s}\mathrm{d}x^{l}\,,
\ee
and this expression is the one conventionally used in information geometry for the skewness tensor in information geometry \cite{Amari-Nagaoka, Matumoto}.

\subsection{Potential functions and smooth mappings}

Now that we have a formal intrinsic characterization for potential functions, we may ask what happens to a potential function through the pullback operation.
This will be of capital importance when we analyze the monotonicity properties of metric tensors on the space of invertible quantum states.

Suppose $\phi\colon N\rightarrow M$ is a smooth map between differential manifolds.
Let $i_{N}$ and $i_{M}$ denote, respectively, the diagonal immersions of $N$ and $M$ into their doubles $N\times N$ and $M\times M$.
Let $\Phi\colon N\times N\rightarrow M\times M$ be the map defined by:
\be
(n\,,n)\mapsto \Phi(n\,,n):=(\phi(n)\,,\phi(n))\,.
\ee
A direct calculation shows that:
\be\label{eqn: potential functions and differentiable mappings, commuting diagrams}
\Phi\circ i_{N}=i_{M}\circ \phi\,.
\ee
Furthermore:

\begin{proposition}\label{prop: potential functions and differentiable mapping, relatedness of vector fields}
Let $X\in\mathfrak{X}(N)$ be $\phi$-related to $Z\in\mathfrak{X}(M)$, that is, $T\phi\circ X=Z\circ \phi$.
Then $\mathbb{X}_{l}$ is $\Phi$-related to $\mathbb{Z}_{l}$, that is, $T\Phi\circ\mathbb{X}_{l}=\mathbb{Z}_{l}\circ \Phi$, and $\mathbb{X}_{r}$ is $\Phi$-related to $\mathbb{Z}_{r}$, that is, $T\Phi\circ\mathbb{X}_{r}=\mathbb{Z}_{r}\circ \Phi$

\begin{proof*}
By hypothesis, it is $T\phi\circ X=Z\circ \phi$.
We want to cast this equality in a more useful form.
We start noting that:
\be
T\phi\circ X(n)=T\phi(n\,,v^{X}_{n})=\left(\phi(n)\,,T_{n}\phi(v^{x}_{n})\right)\,,
\ee
\be
Z\circ\phi(n)=Z(\phi(n))=\left(\phi(n)\,,v_{\phi(n)}^{Z}\right)\,,
\ee
from which it follows that $T\phi\circ X=Z\circ \phi$ implies:
\be\label{eqn: potential functions and differentiable mapping, relatedness of vector fields 1}
T_{n}\phi(v^{X}_{n})=v_{\phi(n)}^{Z}\,.
\ee
Now, we have
\be
T\Phi(n_{1}\,,v_{n_{1}}\,;n_{2}\,,v_{n_{2}})=\left(\phi(n_{1})\,,T_{n_{1}}\phi(v_{n_{1}})\,;\phi(n_{2})\,,T_{n_{2}}\phi(v_{n_{2}})\right)\,,
\ee
and thus:
\be\label{eqn: potential functions and differentiable mapping, relatedness of vector fields 2}
T\Phi\circ\mathbb{X}_{l}(n_{1}\,,n_{2})=T\Phi(n_{1}\,,v^{X}_{n_{1}}\,,n_{2}\,,0)=\left(\phi(n_{1})\,,T_{n_{1}}\phi(v^{X}_{n_{1}})\,;\phi(n_{2})\,,0\right)\,.
\ee
On the other hand:
\be\label{eqn: potential functions and differentiable mapping, relatedness of vector fields 3}
\mathbb{Z}_{l}\circ\Phi(n_{1}\,,n_{2})=\mathbb{Z}_{l}(\phi(n_{1})\,,\phi(n_{2}))=\left(\phi(n_{1})\,,v_{\phi(n_{1})}^{Z}\,;\phi(n_{2})\,,0\right)\,.
\ee
Plugging equation \eqref{eqn: potential functions and differentiable mapping, relatedness of vector fields 1} into equation \eqref{eqn: potential functions and differentiable mapping, relatedness of vector fields 3}, and then comparing equation \eqref{eqn: potential functions and differentiable mapping, relatedness of vector fields 3} with equation \eqref{eqn: potential functions and differentiable mapping, relatedness of vector fields 2} we obtain:
\be
T\Phi\circ\mathbb{X}_{l}=\mathbb{Z}_{l}\circ\Phi
\ee
as claimed.
Proceeding analogously, we prove that $\mathbb{X}_{r}$ is $\Phi$-related to $\mathbb{Z}_{r}$.
This completes the proof.
\end{proof*}
\end{proposition}

With the help of proposition \ref{prop: potential functions and differentiable mapping, relatedness of vector fields} we are able to analyze the behavior of potential functions with respect to smooth maps.
Specifically, we have the following:

\begin{proposition}\label{prop: potential functions, metrics and differentiable mappings}
Let $\phi\colon N\rightarrow M$ be a smooth map between smooth manifold, and let $\Phi\colon N\times N\rightarrow M\times M$ be defined as $\Phi(n_{1}\,,n_{2}):=(\phi(n_{1})\,,\phi(n_{2}))$.
Let $S$ be a potential function on $M\times M$ then $\Phi^{*}S$ is a potential  function on $N\times N$, and the symmetric covariant tensor extracted from $\Phi^{*}S$ is equal to the pullback by means of $\phi$ of the symmetric covariant tensor extracted from $S$.
 
\begin{proof*}
Suppose that $S$ is a potential function on $M\times M$.
Take a generic $X\in\mathfrak{X}(N)$, and consider a vector field $Z\in\mathfrak{X}(M)$ which is $\phi$-related to $X$.
Then:

\be
\begin{split}
i_{N}^{*}\left(L_{\mathbb{X}_{l}}\Phi^{*}S\right)&=i_{N}^{*}\Phi^{*}\left(L_{\mathbb{Z}_{l}}S\right)=\left(\Phi\circ i_{N}\right)^{*}\left(L_{\mathbb{Z}_{l}}S\right)=\\
&=\left(i_{M}\circ\phi\right)^{*}\left(L_{\mathbb{Z}_{l}}S\right)=\phi^{*}i_{M}^{*}\left(L_{\mathbb{Z}_{l}}S\right)=0\,,
\end{split}
\ee
where we used equation \eqref{eqn: lie derivative and id-relatedness}, proposition \ref{prop: potential functions and differentiable mapping, relatedness of vector fields},  equation \eqref{eqn: potential functions and differentiable mappings, commuting diagrams}, and condition \eqref{eqn: necessary and sufficient conditions for gll 2}.
In a similar way, it is possible to show that $i_{N}^{*}\left(L_{\mathbb{X}_{r}}\Phi^{*}S\right)=0$, and this means that $\Phi^{*}S$ is a potential function on $N\times N$.

Denote with $g_{N}$ the symmetric covariant tensor field on $N$ generated by the potential function $\Phi^{*}S$, and with $Z$ ($W$) the vector field on $M$ which is $\phi$-related to $X$ ($Y$).
Recalling equation \eqref{eqn: lie derivative and id-relatedness}, proposition \ref{prop: potential functions and differentiable mapping, relatedness of vector fields}, and equation \eqref{eqn: potential functions and differentiable mappings, commuting diagrams}, we have:

$$
g_{N}(X\,,Y)=i_{N}^{*}\left(L_{\mathbb{X}_{l}}L_{\mathbb{Y}_{l}}\Phi^{*}S\right)=i_{N}^{*}\Phi^{*}\left(L_{\mathbb{Z}_{l}}L_{\mathbb{W}_{l}}S\right)=
$$
$$
=\left(\Phi\circ i_{N}\right)^{*}\left(L_{\mathbb{Z}_{l}}L_{\mathbb{W}_{l}}S\right)=\left(i_{M}\circ \phi\right)^{*}\left(L_{\mathbb{Z}_{l}}L_{\mathbb{W}_{l}}S\right)=
$$
$$
=\phi^{*}i_{M}^{*}\left(L_{\mathbb{Z}_{l}}L_{\mathbb{W}_{l}}S\right)=\phi^{*}\left(g_{M}(Z\,,W)\right)\,.
$$
Being $g_{N}(X\,,Y)$ a function, we may evaluate it at $n$:

$$
\left(g_{N}(X\,,Y)\right)(n)=\left(\phi^{*}\left(g_{M}(Z\,,W)\right)\right)(n)=
$$
\be
=\left(g_{M}(Z\,,W)\right)(\phi(n))=\left. g_{M} \right|_{\phi(n)}\left(\left. Z \right|_{\phi(n)}\,,\left. W \right|_{\phi(n)}\right)
\ee
Now, by the very definition of the pullback $\phi^{*}g_{M}$ we have:

\be
\left((\phi^{*}g_{M})(X\,,Y)\right)(n)=g_{M}|_{\phi(n)}(\phi_{*}X|_{\phi(n)}\,,\phi_{*}Y|_{\phi(n)})\,.
\ee
Being $Z$ and $W$ $\phi$-related to, respectively $X$ and $Y$, we have:

\be
\left. Z \right|_{\phi(n)}=\phi_{*}X|_{\phi(n)} \,,\;\;\;\;\;\left. W \right|_{\phi(n)}=\phi_{*}Y|_{\phi(n)}\,,
\ee
and thus the symmetric covariant tensor we can extract from $\Phi^{*}S$ coincides with the pullback $\phi^{*}g_{M}$ we can extract from $S$.

\end{proof*}
\end{proposition}

We are now in the position to say something about the relation between the symmetry properties of $g$ and the symmetry properties of the potential function $S$ with which it is associated.
At this purpose, let $G$ be a Lie group acting on $M$ by means of diffeomorphisms $\phi_{\gr}$ with $\gr\in G$.
Then $G$ acts on $M\times M$ by means of the maps $\Phi_{\gr}(m_{1}\,,m_{2}):=(\phi_{\gr}(m_{1})\,,\phi_{\gr}(m_{2}))$.
Let $S$ be a potential function on $M\times M$.
It then follows from proposition \ref{prop: potential functions, metrics and differentiable mappings} that:
\be
\left(\phi_{\gr}^{*}g - g\right)(X\,,Y)=-i_{d}^{*}\left(L_{\mathbb{X}_{r}}L_{\mathbb{Y}_{r}}\left(S - \Phi_{\gr}^{*}S\right)\right)=0\,.
\ee
From this equation we conclude that if $S$ is invariant under the action of $G$ on $M\times M$  associated with the action of $G$ on $M$, then 
\be
\left(\phi_{\gr}^{*}g - g\right)(X\,,Y)=0\;\;\;\;\forall X,Y\in\mathfrak{X}(M)\,,
\ee
and thus $G$ is a symmetry group for the metric-like  tensor $g$ associated with $S$, that is:
\be
\phi_{\gr}^{*}g=g\;\;\;\;\forall \gr\in G\,.
\ee

\subsection{Quantum divergence functions and monotonicity}\label{monodpi}

We will now use the geometric tools developed in the previous section in order to define the monotonicity property for quantum metric tensors, to define the data processing inequality (DPI) for quantum divergence functions, and to prove that quantum divergence functions satisifying the data processing inequality give rise to quantum metric tensors satisfying the monotonicity property.
Essentially, the monotonicity property is a quantum version of the so-called invariance criterion of classical information geometry \cite{amari,cencov}, where classical stochastic mappings are replaced with quantum stochastic mappings.
Consequently, we will introduce the notion of quantum stochastic mapping according to \cite{petz1}.
This class of maps plays a prominent role not only in the definition of the monotonicity property for quantum metric tensors, but also for the definition of the data processing inequality for quantum divergence functions.

Denote with $\mathbb{N}_{2}$ the set of natural number without $\{0\}$ and $\{1\}$.
Let $j\in\mathbb{N}_{2}$, and let  $\stsp_{j}\subset\mathcal{B}(\mathcal{H}_{j})$ be the manifold of invertible quantum states associated  with a system with Hilbert space $\mathcal{H}_{j}$ where $\mathrm{dim}(\mathcal{H}_{j})=j$.
The notion of quantum stochastic map is then formulated in terms of completely-positive trace preserving  (CPTP) maps.
Specifically, we say that a  CPTP map $\phi$ from $\mathcal{B}(\mathcal{H}_{j})$ to $\mathcal{B}(\mathcal{H}_{k})$ is stochastic if $\phi(\stsp_{j})\subseteq\stsp_{k}$ \cite{petz1}.
Note that the family of quantum stochastic map form a category precisely as the family of classical stochastic map  \cite{cencov}.

In Holevo's books \cite{holevo-probabilistic_and_statistical_aspects_of_quantum_theory} and \cite{holevo-statistical_structure_of_quantum_theory} there is an interesting discussion on the theoretical and operational relevance of the class of quantum stochastic maps.
Once we have fixed this class of maps between invertible density matrices, we are ready to give a definition of the monotonicity property for quantum Riemannian metric tensors \cite{cencov_morozowa-markov_invariant_geometry_on_state_manifolds, petz1, petz_sudar-geometries_of_quantum_states}.
Clearly, since the family of quantum stochastic maps may connect systems with different dimensions, we must not consider a single tensor field defined on the manifold of invertible density matrices of a single quantum system, but, rather, a family of  tensor fields.

\begin{definition}\label{def: monotonicity of quantum metrics}
Let $\{S_{j}\}_{j\in\mathbb{N}_{2}}$ be a family of functions such that $S_{j}$ is a divergence function on $\stsp_{j}\times\stsp_{j}$ for all $j\in\mathbb{N}_{2}$.
Assume that each $S_{j}$ generates a metric tensor $g_{j}$ on $\stsp_{j}$ for each $j\in\mathbb{N}_{2}$.
We say that the family $\{g_{j}\}_{j\in\mathbb{N}_{2}}$ of metric tensors has the monotonicity property if:
\be
g_{j}(X\,,X)\geq\left(\phi^{*}g_{k}\right)(X\,,X)\,,
\ee
for all $X\in\mathfrak{X}(\stsp_{j})$ and for all stochastic maps $\phi$.
By the very definition of the pullback operation, the  monotonicity property is equivalent to:
\be
\left. g_{j}\right|_{\rho}\left(\left. X\right|_{\rho}\,,\left. X\right|_{\rho}\right)\geq \left. g_{k}\right|_{\phi(\rho)}\left(\left. \phi_{*}X\right|_{\phi(\rho)}\,,\left. \phi_{*}X\right|_{\phi(\rho)}\right)\,,
\ee
where $\rho\in\stsp_{j}$.
\end{definition}

Roughly speaking, the monotonicity property for a family of quantum Riemannian metric tensors  ensures that the notion of geodesical distance between invertible density matrices, as encoded in the family of quantum Riemannian metric tensors, does not increase under quantum stochastic maps.
We will now rephrase the monotonicity property of the family $\{g_{j}\}_{j\in\mathbb{N}_{2}}$ in terms of the behavior of the family $\{S_{j}\}_{j\in\mathbb{N}_{2}}$ of divergence functions with respect to stochastic maps.
We have the following proposition:

\begin{proposition}\label{prop: monotonicity on quantum divergence functions}
Let $\{g_{j}\}_{j\in\mathbb{N}_{2}}$ be a family of monotone metrics generated by the family of divergence functions $\{S_{j}\}_{j\in\mathbb{N}_{2}}$ according to definition \ref{def: monotonicity of quantum metrics}.
Let $\phi\colon\stsp_{j}\rightarrow\stsp_{k}$ be a stochastic map, and let $\Phi\colon\stsp_{j}\times\stsp_{j}\rightarrow\stsp_{k}\times\stsp_{k}$ be defined as
\be
\Phi(\rho_{1}\,,\rho_{2}):=(\phi(\rho_{1})\,,\phi(\rho_{2}))\,.
\ee
Then, setting $S_{jk}^{\Phi}=\left(S_{j} - \Phi^{*}S_{k}\right)$, the monotonicity property of $\{g_{j}\}_{j\in\mathbb{N}_{2}}$ is equivalent to:
\be
g_{jk}^{\phi}(X\,,X):=-i_{j}^{*}\left(L_{\mathbb{X}_{l}}L_{\mathbb{X}_{r}} S_{jk}^{\Phi}\right)\geq 0
\ee
for all $X\in\mathfrak{X}(\stsp_{j})$ and for all stochastic maps $\phi$.

\begin{proof*}
According to proposition \ref{prop: potential functions, metrics and differentiable mappings}, we know that $\phi^{*}g_{k}$ is the metric-like tensor generated by the divergence function $\Phi^{*}S_{k}$. 
This means that we may write:
\be
\left(\phi^{*}g_{k}\right)(X\,,Y)=-i_{j}^{*}\left(L_{\mathbb{X}_{l}}L_{\mathbb{Y}_{r}} \Phi^{*}S_{k}\right)
\ee
where we used definition \ref{def: covariant 2-tensors from lie derivative}.
Again using definition \ref{def: covariant 2-tensors from lie derivative}, we write:
\be
g_{j}(X\,,Y)=-i_{j}^{*}\left(L_{\mathbb{X}_{l}}L_{\mathbb{Y}_{r}} S_{j}\right)\,.
\ee
Comparing these two equations, it then follows that
\be
g_{j}(X\,,X)\geq\left(\phi^{*}g_{k}\right)(X\,,X)\,
\ee
is equivalent to:
\be
g_{jk}^{\phi}(X\,,X):=-i_{j}^{*}\left(L_{\mathbb{X}_{l}}L_{\mathbb{X}_{r}} S_{jk}^{\Phi}\right)\equiv -i_{j}^{*}\left(L_{\mathbb{X}_{l}}L_{\mathbb{X}_{r}} \left(S_{j} - \Phi^{*}S_{k}\right)\right)\geq 0
\ee
as claimed.
\end{proof*}
\end{proposition}

As anticipated before, there is a very interesting connection between this result and the so-called data processing inequality (DPI) for quantum divergences:

\begin{definition}
We say that $\{S_{j}\}_{j\in\mathbb{N}_{2}}$ satisfies the data processing inequality (DPI) if:
\be
S_{j}(\rho_{1}\,,\rho_{2})\geq S_{k}(\phi(\rho_{1})\,,\phi(\rho_{2}))
\ee
for all $\rho_{1},\rho_{2}$ and for all stochastic maps $\phi$.
\end{definition}

The operational meaning of this inequality is to ensure that the information-theoretical content encoded in the family of quantum two-point functions does not increase under quantum stochastic maps.
Then, the following proposition shows that {\itshape the DPI ``implies'' the monotonicity property}:

\begin{proposition}\label{prop: DPI implies monotonicity}
If the family $\{S_{j}\}_{j\in\mathbb{N}_{2}}$ satisfies the DPI, then it generates a family $\{g_{j}\}_{j\in\mathbb{N}_{2}}$ of metric tensors satisfying the monotonicity property.

\begin{proof*}
The function $\Phi^{*}S_{k}$ is a potential function because $S_{k}$ is a potential function (see proposition \ref{prop: potential functions, metrics and differentiable mappings}).
According to the DPI, we have:
$$
S_{jk}^{\Phi}(\rho_{1}\,,\rho_{2}):=S_{j}(\rho_{1}\,,\rho_{2}) - \Phi^{*}S_{k}(\rho_{1}\,,\rho_{2})=
$$
\be
=S_{j}(\rho_{1}\,,\rho_{2})-
 S_{k}(\phi(\rho_{1})\,,\phi(\rho_{2}))\geq 0\,. 
\ee
From this, we conclude that $S_{jk}^{\Phi}$ is a non-negative potential function vanishing on the diagonal of $\stsp_{j}\times\stsp_{j}$.
This means that every point on the diagonal of $M\times M$ is a local minimum for $S^{\Phi}_{jk}$.
Then, according to proposition \ref{prop: nec and suf conditions for g to be positive or negative semidefinite} the metric-like tensor $g_{jk}^{\phi }$ it generates is positive-semidefinite.
In particular it is:
\be
g_{jk}^{\phi}(X\,,X)\geq 0\,.
\ee
According to proposition \ref{prop: monotonicity on quantum divergence functions} this is equivalent to the monotonicity property for the family $\{g_{j}\}_{j\in\mathbb{N}_{2}}$, and the proposition is proved.
\end{proof*}
\end{proposition}

This result may be seen as a sort of generalization to the quantum case of the invariance criterion of classical information geometry \cite{amari, cencov}.
Furthermore, the abstract coordinate-free framework in which proposition \ref{prop: DPI implies monotonicity} is contextualized may prove to be useful for a generalization to the  infinite dimensional case.

\section{$q$-$z$-Relative entropies}\label{qzrelentropy}

In the previous section we have seen how to use potential functions in order to define covariant tensors on smooth manifolds in a coordinate-free fashion.
In the following sections, we will exploit this idea in the context of finite-dimensional quantum systems.
In particular, we will select a family of quantum relative entropies and compute the family of metric-like tensors on the space of invertible quantum states which is associated to it.

\vsp

Relative entropies (entropic functionals) play a  central role in the context of quantum information theory \cite{ohya_petz-quantum_entropy_and_its_use, petz-quantum_information_theory_and_quantum_statistics, tomamichel-quantum_information_theory_and_quantum_statistics}.
On of the first examples of quantum relative entropy is given by the so-called \textit{von Neumann-Umegaki relative entropy (conditional expectation)} \cite{araki-relative_entropy_of_states_of_von_neumann_algebras, umegaki-conditional_expectation_in_an_operator_algebra_IV}:
\begin{equation}
D(\rho|\varrho)=\text{Tr}\,\rho(\log\rho-\log\varrho).
\end{equation}
In the asymptotic memoryless setting, it yields fundamental limits on the performance of information-processing tasks connected with quantum hypothesis testing \cite{datta2}, and thus it may be seen as the quantum generalization of the Kullback-Leibler divergence function used in classical information geometry. 

Another important family of quantum relative entropies is the $q$\textit{-R\'enyi relative entropies} ($q$-RRE)\footnote{Differently from the  notation used in \cite{DATTA, datta}, here we follow \cite{vitale} and use the notation $q$ instead of $\alpha$. This is just a relabeling of the parameter (i.e., $\alpha=q$) which, as  will be clear in the following, helps to compare our results with those in   \cite{vitale} and should not be confused with the parameter $\alpha$ in \cite{amari}, related to  $\alpha$-divergences, where instead $\alpha=2q-1$.}:
\begin{equation}
\widetilde{D}_{q}(\rho|\varrho)=\frac{1}{q-1}\log\text{Tr}\,\bigl(\rho^q\varrho^{1-q}\bigr),
\end{equation}
where $q\in(0,1)\cup(1,\infty)$. 
These quantum relative entropies are able to describe the cut-off rates in quantum binary state discrimination \cite{datta6}. 

From the point of view of metric tensors on the space of invertible quantum states, it is well known that the Bures metric \cite{bengtsson} may be obtained from the potential function:
\begin{equation}
D^{2}_{B}(\rho|\varrho)=4\bigl[1-\text{Tr}\,\bigl(\rho\varrho\bigr)^{\frac{1}{2}}\bigr]\,.
\end{equation}
As explained in section \ref{bures}, this potential function is is basically related to the so-called \textit{root fidelity} $\sqrt{F}(\rho,\varrho)=\text{Tr}\,\left(\sqrt{\sqrt{\rho}\varrho\sqrt{\rho}}\right)$.

The Wigner-Yanase metric tensor, on the other hand, may be extracted from the Wigner-Yanase skew information function \cite{gibilisco, hasegawa_petz-noncommutative_extension_of_information_geometry_II}:
\begin{equation}
\mathcal{I}(\rho|\varrho)=4\bigl[1-\text{Tr}\,\bigl(\rho^{\frac{1}{2}}\varrho^{\frac{1}{2}}\bigr)\bigr].
\end{equation}

Several efforts were made in order to find a sort of universal family of quantum relative entropies  containing this plethora of different divergence functions. 
A first (partial) result was achieved by the $q$-\textit{quantum R\'enyi divergence}\footnote{Note that the $q$-QRD reduces to the $q$-RRE when $[\rho,\,\varrho]=0$, that is, $D_{q}$ may be seen as a noncommutative generalization of $\widetilde{D}_{q}$ \cite{datta}.} ($q$-QRD):
\begin{equation}\label{Sqrd}
D_{q}(\rho|\varrho)=\frac{1}{q-1}\log\text{Tr}\,\bigl(\varrho^{\frac{1-q}{2q}}\rho\varrho^{\frac{1-q}{2q}}\bigr)^{q},
\end{equation}
where again $q\in(0,1)\cup(1,\infty)$. 
However, the data-processing inequality (DPI)
\begin{equation}
D_{q}(\Phi(\rho)|\Phi(\varrho))\leq D_{q}(\rho|\varrho)\,,
\end{equation}
where $\Phi$ is a completely positive trace preserving map (CPTP) acting on a pair of semidefinite Hermitian operators $\rho$ and $\varrho$, is not satisfied for $q\in(0,1/2)$\footnote{See for instance \cite{dpiqrd} where it is shown that the Riemannian metric derived from the sandwiched R\'enyi $q$-divergence \eqref{Sqrd} is monotone if and only if $q\in(-\infty,-1]\cup[\frac{1}{2},\infty)$.}.

Recently, a new family of two-point functions which includes all the previous examples was defined \cite{DATTA, jaksic_ogata_pautrat_pillet-entropic_fluctuations_in_quantum_mechanics_an_introduction}. 
It is the so-called $q$-$z$-\textit{R\'enyi Relative Entropy} ($q$-$z$-RRE)
\begin{equation}
D_{q,z}(\rho|\varrho)=\frac{1}{q-1}\log\text{Tr}\,\bigl(\rho^{\frac{q}{2z}}\varrho^{\frac{1-q}{z}}\rho^{\frac{q}{2z}}\bigr)^z\,,
\end{equation}
which we may rewrite as:
\begin{equation}
D_{q,z}(\rho|\varrho)=\frac{1}{q-1}\log\text{Tr}\,\bigl(\rho^{q/z}\varrho^{(1-q)/z}\bigr)^z.
\end{equation}

\begin{remark}
In general, the product of two Hermitian matrices is not a Hermitian matrix. However, the product matrix $\rho^{q/z}\varrho^{(1-q)/z}$ has real, non-negative eigenvalues, even though it is not in general a hermitian matrix. It means that the trace functional
\begin{equation}
\label{trf}
f_{q,z}(\rho|\varrho)=\text{Tr}\,\bigl(\rho^{q/z}\varrho^{(1-q)/z}\bigr)^z
\end{equation}
is well defined as the sum of the the $z$-th power of the eigenvalues of the product matrix \cite{DATTA} and it can be developed in Taylor series.
\end{remark}

As is shown in \cite{DATTA}, by suitably varying the parameters $q$ and $z$ it is possible to recover the $q$-RRE family
\begin{equation}
D_{q,1}(\rho|\varrho):=\lim_{z\to1} D_{q,z}(\rho|\varrho)\equiv \widetilde{D}_{q}(\rho|\varrho)=\frac{1}{q-1}\log\text{Tr}\,\bigl(\rho^q\varrho^{1-q}\bigr),
\end{equation}
the $q$-QRD family
\begin{equation}
D_{q,q}(\rho|\varrho):=\lim_{z\to q} D_{q,q}(\rho|\varrho)\equiv D_{q}(\rho|\varrho)=\frac{1}{q-1}\log\text{Tr}\,\bigl(\varrho^{\frac{1-q}{2q}}\rho\varrho^{\frac{1-q}{2q}}\bigr),
\end{equation}
and the von Neumann-Umegaki relative entropy:
\begin{equation}
D_{1,1}(\rho|\varrho):=\lim_{z=q\to1} D_{q,z}(\rho|\varrho)\equiv D(\rho|\varrho)=\text{Tr}\,\rho(\log\rho-\log\varrho).
\end{equation}

\begin{figure}[h!]
\begin{center}
\includegraphics[scale=3]{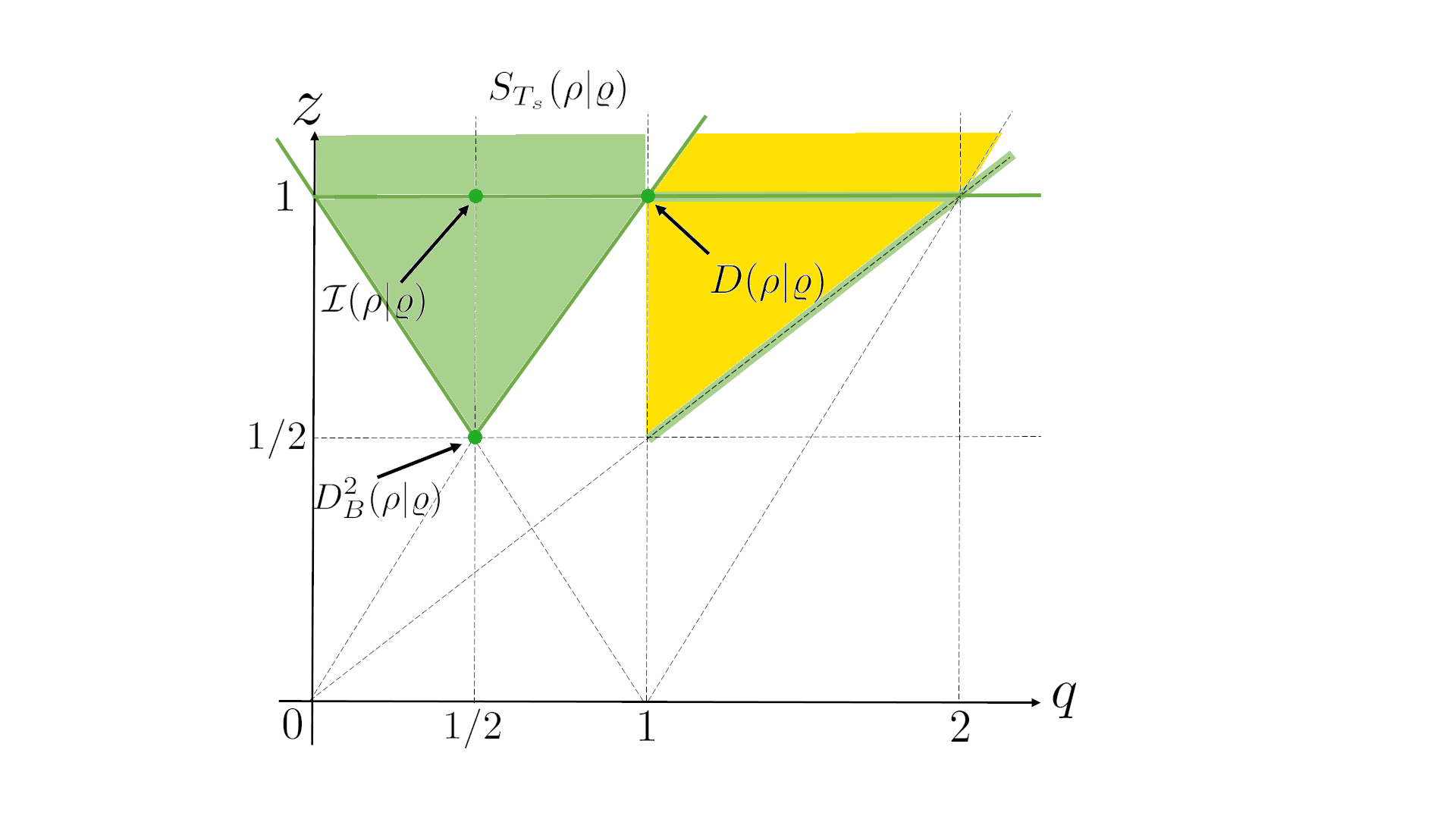}
\caption{\textit{In this figure, a schematic overview of the various relative entropies unified by the $q-z$-relative entropy is shown. The green region indicates the range of the parameters in which the DPI was proven, while the yellow region indicates where it is just conjectured.}}
\label{dattafig}
\end{center}
\end{figure}

The data processing inequality for the $q$-$z$-RRE was studied in \cite{Beigi17, carlen_frank_lieb-some_operator_and_trace_function_convexity_theorems, Epstein31, FL16, Hiai22} and it is not established yet in full generality. To prove it, one has to show that the trace functional \eqref{trf} is jointly concave when $q\leq 1$, or jointly convex when $q\geq1$. The results of these analysis are well summarized in \cite{DATTA, carlen_frank_lieb-some_operator_and_trace_function_convexity_theorems} and it results that the DPI holds only for certain range of the parameters as sketched in fig. \ref{dattafig}.

Since we are interested in computing the metric tensors from the two-parameter family of $D_{q,z}$, we will manipulate the explicit expression of the $q$-$z$-\textit{R\'enyi Relative Entropies} in order to get a functional form that is better suited to geometrical computations.
At this purpose, we start considering the following regularization of the logarithm, the so-called q-logarithm:
\begin{equation}\label{qlog}
\log_q\rho=\frac{1}{1-q}(\rho^{1-q}-1)\qquad\text{with}\qquad\lim_{q\to1}\log_q\rho=\log\rho.
\end{equation}
Then, inspired by Petz \cite{petz1}, we will consider a rescaling by a factor $1/q$. 
In this way, the resulting two-parameter family of two-point functions will be symmetric under the exchange of $q\rightarrow(1-q)$. 
We refer to the resulting family of quantum relative entropies as ``regularized $q$-$z$-RRE'', and write it as:
\begin{equation}
\label{sqz}
S_{q,z}(\rho|\varrho)=\frac{1}{q(1-q)}\biggl[1-\text{Tr}\,\biggl(\rho^{\frac{q}{z}}\varrho^{\frac{1-q}{z}}\biggr)^z\,\biggr].
\end{equation}

Since the analysis of the DPI involves only the trace functional, we are ensured that the DPI holds for the same range of parameters of the $q$-$z$-RRE. 
Moreover, in the limit $z\to1$, it is possible to recover the expression for the Tsallis relative entropy in \cite{vitale}
\begin{equation}\label{tsallis}
S_{q,1}(\rho|\varrho):=\lim_{z\to1}S_{q,z}(\rho|\varrho)\equiv S_{Ts}(\rho|\varrho)=\frac{1}{q(1-q)}\biggl[1-\text{Tr}\,\biggl(\rho^{q}\varrho^{1-q}\biggr)\biggr],
\end{equation}
in the limit $z=q\to1$, we recover the von Neumann-Umegaki relative entropy
\begin{equation}
S_{1,1}(\rho|\varrho):=\lim_{z=q\to1}S_{q,z}(\rho|\varrho)\equiv D(\rho|\varrho)=\text{Tr}\,\rho(\log\rho-\log\varrho),
\end{equation}
in the limit $z=q=1/2$, we recover the divergence function of the Bures metric tensor
\begin{equation}
S_{\frac{1}{2},\frac{1}{2}}(\rho|\varrho):=\lim_{z=q\to\frac{1}{2}}S_{q,z}(\rho|\varrho)\equiv D^{2}_{B}(\rho|\varrho)=4\bigl[1-\text{Tr}\,\bigl(\rho\varrho\bigr)^{\frac{1}{2}}\bigr],
\end{equation}
and finally, in the limit $z=1$, $q\to1/2$, we recover the divergence function of the Wigner-Yanase metric tensor:
\begin{equation}
S_{\frac{1}{2},1}(\rho|\varrho):=\lim_{z=1,q\to\frac{1}{2}}S_{q,z}(\rho|\varrho)\equiv \mathcal{I}(\rho|\varrho)=4\bigl[1-\text{Tr}\,\bigl(\rho^{\frac{1}{2}}\varrho^{\frac{1}{2}}\bigr)\bigr]\;.
\end{equation}
All these special cases belong to the range of parameters for which  $S_{q,z}$ is actually a quantum divergence function satisfying the DPI.
Consequently, in accordance with the results of subsection \ref{monodpi}, the family of quantum metric tensors we can extract from $S_{q,z}$ satisfies the monotonicity property.

\begin{remark}
It is well known that the family of (quantum) f-Divergences introduced by Petz  satisfies the data processing inequality \cite{petz-from_f-divergence_to_quantum_quasi-entropies_and_their_use}, and thus the family of metric tensors that can be extracted from it will satisfy the monotonicity property.
Taking into account that there are elements of the family of q-z-RREs that also belong to the family of f-Divergences, e.g., the von Neumann-Umegaki relative entropy, it is natural to ask if there is a relation between the f-Divergences and those q-z-RREs satisfying the data processing inequality and thus leading to families of metric tensors satisfying the monotonicity property.
Unfortunately, according to the results of section \ref{gfromS},  a metric tensor may be extracted from different potential functions, and thus, in principle,  an f-Divergence and a q-z-RRE need not be related even if they give rise to the same metric tensors.

\end{remark}

\section{Unfolding of the space of invertible quantum states}\label{unfolding}

We want to perform calculations without referring to explicit coordinate systems, therefore, we will unfold the manifold $\stsp_{n}$ of invertible density matrices to the more gentle manifold $\mathcal{M}_{n}=SU(n)\times\Delta^{0}_{n}$, where $\Delta^{0}_{n}$ is the open interior of the n-dimensional simplex $\Delta_{n}$, that is:
\be
\Delta^{0}_{n}:=\left\{\vec{p}\in\mathbb{R}^{n}\colon\,p^{j}>0\,,\;\,\sum_{j=1}^{n}\,p^{j}=1\right\}\,.
\ee
This manifold is parallelizable since it is the Cartesian product of parallelizable manifolds, and thus, we have a global basis of vector fields and a global basis of differential one-forms at our disposal.
We will use these bases to perform coordinate-free computations in any dimension.
However, before entering the description of these basis, we want to explain why $\mathcal{M}_{n}$ may be thought of as an unfolding manifold for $\stsp_{n}$.
To do this, let us start selecting an orthonormal basis $\{|j\rangle\}_{j=0,...,n-1}$ in $\mathcal{H}$.
Associated with it there is an orthonormal basis $\{\mathbf{E}_{jk}\}_{j,k=0,...,(n-1)}$ in $\mathcal{B}(\mathcal{H})$ defined setting $\mathbf{E}_{jk}:=|j\rangle\langle k|$.
Now, consider an invertible density matrix $\rho\in\stsp_{n}$.
It is well known that $\rho$ can be diagonalized, and that its eigenvalues are strictly positive and sum up to one.
This means that, denoting with $\vec{p}\in\Delta^{0}_{n}$ a vector the components of which coincide with the eigenvalues of $\rho$, we can find a $\mathbf{U}\in SU(n)$ such that:

\be
\rho=\mathbf{U}\,\rho_{0}\,\mathbf{U}^{\dagger}\,,
\ee
where $\rho_{0}$ is a diagonal matrix in the sense that its only nonzero components with respect to the canonical basis $\{\mathbf{E}_{jk}\}_{j,k=0,...,(n-1)}$ of $\mathcal{B}(\mathcal{H})$ are those relative to $\{\mathbf{E}_{jj}\}_{j=0,...,(n-1)}$.
It is clear that every $\rho_{0}$ can be identified with a point $\vec{p}$ in $\Delta^{0}_{n}$ and vice versa.
This one-to-one correspondence is given by the map $\rho_{0}=p^{j}\mathbf{E}_{jj}$.

\begin{remark}
It is important to point out that the correspondence between $\rho_{0}$ and $\vec{p}$ explicitly depends on the choice of the basis $\{\mathbf{E}_{jk}\}_{j=0,...,(n-1)}$ as a reference basis.
For instance, if we consider a multipartite system for which:
\be\label{eqn: decomposition of algebra of composite system}
\mathcal{B}(\mathcal{H}_{N})=\bigotimes_{j=1}^{r}\,\mathcal{B}(\mathcal{H}_{n_{j}})\,,
\ee
where $N=n_{1}\,n_{2}\cdots n_{r}$, and we select an orthonormal basis in $\mathcal{H}_{N}$ which is made up of separable vectors with respect to the decomposition \eqref{eqn: decomposition of algebra of composite system},
  the orthonormal basis in $\mathcal{B}(\mathcal{H}_{N})$ turns out to be composed of separable elements with respect to the decomposition \eqref{eqn: decomposition of algebra of composite system}.
Consequently, the reference density matrix $\rho_{0}$ associated with the probability vector $\vec{p}$ is  separable, and this clearly has consequences with respect to the entanglement properties of the system.
Specifically, when we unfold the quantum state $\rho$ into the couple $(\mathbf{U}\,,\vec{p})$, all the information regarding the entanglement properties of $\rho$ will be encoded in $\mathbf{U}$ because $\vec{p}$ is associated with the separable state $\rho_{0}$.
\end{remark}

The diagonalization procedure for $\rho\in \stsp_{n}$ provides us with a map:

\be
\begin{split}
\pi_{n}\colon SU(n)\times & \Delta^{0}_{n}\rightarrow\stsp_{n}\\
(\mathbf{U}\,,\vec{p})\mapsto\,\pi_{n}(\mathbf{U}\,,\vec{p})=\mathbf{U}\,\rho_{0}&\,\mathbf{U}^{\dagger}\,\mbox{ with } \rho_{0}=p^{j}\,\mathbb{E}_{jj}\,.
\end{split}
\ee
Obviously, the map $\pi_{n}$ is a surjection because for a given $\rho\in\stsp_{n}$ there is an infinite number of elements $(\mathbf{U}\,,\vec{p})\in\mathcal{M}_{n}$ such that $\pi_{n}(\mathbf{U}\,,\vec{p})=\rho$.
It is in this sense that we think of $\mathcal{M}_{n}$ as an unfolding manifold for $\stsp_{n}$.
Now that we have the map $\pi_{n}$, we proceed to prove the following:

\begin{proposition}\label{prop: the unfolding map of the invertible density matrices is a surjective differentiable map}
The map $\pi_{n}\colon\mathcal{M}_{n}\rightarrow\stsp_{n}$ is a surjective differentiable map, and the kernel of its differential at $(\mathbf{U}\,,\vec{p})\in\mathcal{M}_{n}$ is given by $(\imath\mathbf{H}\,,\vec{0})$, where $\mathbf{H}$ is a self-adjoint matrix such that $[\mathbf{H}\,,\rho_{0}]=\mathbf{0}$.

\begin{proof*} 
The surjectivity of the map $\pi_{n}$ follows from the spectral decomposition for every density matrix $\rho$.
Then, consider the following curve $\gamma_{t}$ on $\mathcal{M}_{n}$:

\be
\gamma_{t}(\mathbf{U}\,,\vec{p})=(\mathbf{U}\exp(\imath t\mathbf{H}),\vec{p}_{t})\,,
\ee
where $\mathbf{H}$ is self-adjoint and traceless, and $\vec{p}_{t}$ is any curve in the interior of the n-simplex $\Delta^{0}_{n}$ starting at $\vec{p}_{0}=\vec{p}$ and such that $\left.\frac{\mathrm{d} \vec{p}_{t}}{\mathrm{d}t}\right|_{t=0}=\vec{a}$ with $\sum_{j}a^{j}=0$.
The differential:

$$
\left(T\pi_{n}\right)_{(\mathbf{U},\,\vec{p})}\colon T_{(\mathbf{U},\,\vec{p})}\mathcal{M}_{n}\rightarrow T_{\rho}\stsp_{n}
$$
of $\pi$ at $(\mathbf{U}\,,\vec{p})$ is:

\be
\begin{split}
\left(T\pi_{n}\right)_{(\mathbf{U},\vec{p})}\left(\left.\frac{\mathrm{d}\gamma(\mathbf{U},\vec{p})}{\mathrm{d} t}\right|_{t=0}\right)&=\frac{\mathrm{d}}{\mathrm{d}t}\left(\mathbf{U}\exp(\imath t\mathbf{H})\,\rho_{0}(t),\exp(-\imath t\mathbf{H})\mathbf{U}^{\dagger}\right)_{t=0}=\\
&=\mathbf{U}\,\left(\imath\,[\mathbf{H}\,,\rho_{0}] + a^{j}\mathbf{E}_{jj}\right)\,\mathbf{U}^{\dagger}\,,
\end{split}
\ee
and it then follows that the tangent vector $(\imath\mathbf{H},\vec{a})$ at $(\mathbf{U},\vec{p})$ is sent to the null vector $\mathbf{0}$ at $\rho=\mathbf{U}\,\rho_{0}\,\mathbf{U}^{\dagger}$ if and only if $\vec{a}=\vec{0}$ and $[\mathbf{H}\,,\rho_{0}]=\mathbf{0}$.
\end{proof*}

\end{proposition}

The global differential calculus on $\mathcal{M}_{n}$ is easily defined considering the projection maps:

\be
\begin{split}
pr_{SU(n)}\colon & \mathcal{M}_{n}=SU(n)\times \Delta^{0}_{n}\rightarrow SU(n)\,,\;(\mathbf{U}\,,\vec{p})\mapsto pr_{SU(n)}(\mathbf{U}\,,\vec{p})=\mathbf{U}\,,\\
pr_{\Delta^{0}_{n}}\colon & \mathcal{M}_{n}=SU(n)\times \Delta^{0}_{n}\rightarrow \Delta^{0}_{n}\,,\;(\mathbf{U}\,,\vec{p})\mapsto pr_{\Delta^{0}_{n}}(\mathbf{U}\,,\vec{p})=\vec{p}\,.
\end{split}
\ee
Then, since $SU(n)$ is a Lie group, we have, for instance, a basis of globally defined left-invariant differential one-forms $\{\theta^{j}\}_{j=1,...,n^{2}-1}$ and a basis of globally defined left-invariant vector fields $\{X_{j}\}_{j=1,...,n^{2}-1}$ which is dual to $\{\theta^{j}\}_{j=1,...,n^{2}-1}$.
Consequently, we can take the pullback of every $\theta^{j}$ by means of $pr_{SU(n)}$ and obtain a set of globally defined differential one-forms on $\mathcal{M}_{n}$.
With an evident abuse of notation, we will keep writing $\{\theta^{j}\}_{j=1,...,n^{2}-1}$ for this set of one-forms.
Regarding $\Delta^{0}_{n}$, we will construct an ``overcomplete'' basis of differential one-forms as follows.
First of all, we define $n$ functions $\mathrm{P}^{j}\colon\Delta^{0}_{n}\rightarrow\mathbb{R}$:
\be
\vec{p}\mapsto\,\mathrm{P}^{j}(\vec{p})=p^{j}\,.
\ee
These functions are globally defined and smooth, and thus their differential $\mathrm{d}\mathrm{P}^{j}=\mathrm{d}p^{j}$ are globally defined differential one-forms.
Clearly, we have $n$ of them, and since $\mathrm{dim}(\Delta^{0}_{n})=n-1$, these one-forms are not functionally independent.
Indeed, it holds:

\be
\sum_{j=1}^{n}\,\mathrm{P}^{j}(\vec{p})=\sum_{j=1}^{n}\,p^{j}=1\,,
\ee
and thus:
\be
\sum_{j=1}^{n}\,\mathrm{d}\mathrm{P}^{j}=\sum_{j=1}^{n}\,\mathrm{d}p^{j}=0\,.
\ee
Now, the set $\{\mathrm{d}p^{j}\}_{j=1,...,n}$ of globally defined differential one-forms on $\Delta^{0}_{n}$ is a basis of the module of one-forms on $\Delta^{0}_{n}$, that is, for every differential one-form $\omega$ on $\Delta^{0}_{n}$, it always exists a decomposition:

\be
\omega=\omega_{j}\,\mathrm{d}p^{j}\,,
\ee
where $\omega_{j}\in\mathcal{F}(\Delta^{0}_{n})$.
This is the sense in which $\{\mathrm{d}p^{j}\}_{j=1,...,n}$ is an overcomplete basis for the space of differential one-forms on $\Delta^{0}_{n}$.
Now, similarly to what we have done for $SU(n)$, we consider the pullback of $\mathrm{d}p^{j}$ by means of $pr_{\Delta^{0}_{n}}$, and we obtain a set of globally defined differential one-forms on $\mathcal{M}_{n}$.
Again with an abuse of notation, we will keep writing this set as $\{\mathrm{d}p^{j}\}_{j=1,...,n}$.
Eventually, the set $\left(\{\theta^{j}\}_{j=1,...,n^{2}-1}\,,\{\mathrm{d}p^{j}\}_{j=1,...,n}\right)$ is a basis of the module of differential one-forms on $\mathcal{M}_{n}$.

\section{The qubit case}
\label{2level}

Now that we have the global basis for a differential calculus on $\mathcal{M}_{n}$ (for all n), we may proceed with the explicit computations.
First of all, we consider the regularized $(q-z)$-R\'{e}nyi relative entropy introduced in equation \eqref{sqz}:

\be
S_{q,z}(\rho\,,\varrho)=\frac{1}{q(1-q)}\left[1-\Tr\left(\rho^{\frac{q}{z}}\,\varrho^{\frac{1-q}{z}}\right)^z\,\right]\,,
\ee
and take its pullback to $\mathcal{M}_{n}\times\mathcal{M}_{n}$ by means of the map:

\be
\begin{split}
&\Pi_{n}\colon\mathcal{M}_{n}\times\mathcal{M}_{n}\rightarrow\bar{\stsp}_{n}\times\bar{\stsp}_{n}\\
&\left(\mathbf{U}\,,\vec{p}_{1}\,;\mathbf{V}\,,\vec{p}_{2}\right)\mapsto \left(\pi_{n}(\mathbf{U}\,,\vec{p}_{1})\,;\pi_{n}(\mathbf{V}\,,\vec{p}_{2})\right)\,.
\end{split}
\ee
The result is the following function on $\mathcal{M}_{n}\times\mathcal{M}_{n}$:

\be
\mathbb{D}_{q,z}\left(\mathbf{U}\,,\vec{p}_{1}\,;\mathbf{V}\,,\vec{p}_{2}\right)=\frac{1}{q(1-q)}\left(1 - \Tr\left[\left(
(\mathbf{U}\, \rho_{0}\,\mathbf{U}^{\dagger})^{\frac{q}{z}} 
(\mathbf{V}\,\varrho_{0}\,\mathbf{V}^{\dagger})^{\frac{1-q}{z}}\right)^{z}\right]\right)\,.
\ee
At the moment, we do not know if  $\mathbb{D}_{q,z}$ is a potential function, but  we can always extract a covariant tensor field from it by computing (see proposition \ref{prop: nec and suf conditions for building 2 tensors from S}):

\be
g_{q,z}(X\,,Y):=-i_{d}^{*}\left(L_{\mathbb{X}_{l}}\,L_{\mathbb{Y}_{r}}\,\mathbb{D}_{q,z}\right)\,.
\ee
Here we will consider the particular case of the qubit ($n=2$).
Without entering into the details of the calculations, for which we refer to \ref{subsec: explicit computations}, we simply state that we have:
$$
g_{q,z}(X\,,Y)=\frac{z}{q(1-q)}i_{d}^{*}\left[\text{Tr}\left((AB)^{z-1}\,\left(i_{\mathbb{X}_{l}}\mathrm{d}A\right)\,\left(i_{\mathbb{Y}_{r}}\mathrm{d}B\right)\right)\right] +
$$
{\footnotesize
\be\label{ABmetric}
+\frac{1}{q(1-q)}i_{d}^{*}\left[\sum_{m=0}^{\infty}m\,c_{m}(z)\sum_{a=0}^{m-2}\text{Tr}\left((AB-\mathds1)^{a}\,\left(i_{\mathbb{X}_{l}}\mathrm{d}A\right)\,B\,(AB-\mathds1)^{m-a-2}\,A\,\left(i_{\mathbb{Y}_{r}}\mathrm{d}B\right)\right)\right]\,,
\ee}
where we have introduced the notation:
\begin{equation}
A=\rho^{\frac{q}{z}}=U\,\rho_0^{\frac{q}{z}}\,U^{-1}\,,\qquad\qquad B=\varrho^{\frac{1-q}{z}}=V\,\varrho_0^{\frac{1-q}{z}}\,V^{-1}\,.
\end{equation}
Performing the pull-back along the map $i_{d}$, which essentially amounts to put $U=V, U^{-1}=V^{-1}$ and $\rho_0=\varrho_0$, the first term in \eqref{ABmetric} gives
{\footnotesize
\be\label{1part}
\begin{split}
&\frac{z}{q(1-q)}i_{d}^{*}\left[\text{Tr}\left((AB)^{z-1}\,\left(i_{\mathbb{X}_{l}}\mathrm{d}A\right)\,\left(i_{\mathbb{Y}_{r}}\mathrm{d}B\right)\right)\right]=\\
&=\frac{z}{q(1-q)}\biggl\{\text{Tr}\left(\rho_0^{\frac{z-1}{z}}\mathrm{d}\rho_0^{\frac{q}{z}}(X) \mathrm{d}\rho_0^{\frac{1-q}{z}}(Y)\right)+ \\ 
& + \text{Tr}\left(\rho_0^{\frac{z-1}{z}}\bigl[U^{-1}\mathrm{d}U(X),\rho_0^{\frac{q}{z}}\bigr]\,\bigl[U^{-1}\mathrm{d}U(Y),\rho_0^{\frac{1-q}{z}}\bigr]\right)\biggr\}\;,
\end{split}
\ee}
where we have used the relation
\be
\mathrm{d}\rho^\alpha=\mathrm{d}(U\rho_0^{\alpha} U^{-1})=U\mathrm{d}\rho_0^{\alpha} U^{-1}+U\bigl[U^{-1}\mathrm{d}U,\rho_0^{\alpha} \bigr]U^{-1}\;,
\ee
and the fact that the mixed terms vanish (see \ref{subsec: explicit computations}). 
Now, for a two-level system, we have a basis in $\mathcal{B}(\mathcal{H})$ made up of the $(2\times 2)$ identity matrix $\sigma_0$ and the Pauli matrices $\sigma_{1},\sigma_{2},\sigma_{3}$.
We select an orthonormal basis $\{|j\rangle\}_{j=0,1}$ in $\mathcal{H}$ made up of eigenvector of $\sigma_{3}$, so that, in this basis, $\sigma_{3}$ is the diagonal matrix $\sigma_{3}=\mathbf{E}_{00} - \mathbf{E}_{11}$.
According to section \ref{unfolding}, $\rho_{0}$ is a diagonal density matrix, and, in the qubit case, it is characterized in terms of a single real parameter $-1\leq w\leq 1$ so that it can be written as
\be
\rho_0=\begin{pmatrix}\frac{1+w}{2} & 0\\
0 & \frac{1-w}{2}
\end{pmatrix}=\frac{1}{2}(\sigma_0+w\sigma_3)\,.
\ee
Therefore, for any power $\rho_0^{\alpha}$ of $\rho_0$, we have
\be
\rho_0^{\alpha}=\begin{pmatrix}\left(\frac{1+w}{2}\right)^{\alpha} & 0\\
0 & \left(\frac{1-w}{2}\right)^{\alpha}
\end{pmatrix}=\frac{1}{2}(a_{\alpha}+b_{\alpha})\sigma_0+\frac{1}{2}(a_{\alpha}-b_{\alpha})\sigma_3\;,
\ee
with
\be
a_{\alpha}=\left(\frac{1+w}{2}\right)^{\alpha}\,,\qquad\qquad b_{\alpha}=\left(\frac{1-w}{2}\right)^{\alpha}\;.
\ee
The first term in \eqref{1part} then yields
\be\label{g1transv}
\begin{split}
\frac{z}{q(1-q)} \text{Tr}\left(\rho_0^{\frac{z-1}{z}}\mathrm{d}\rho_0^{\frac{q}{z}}(X) \mathrm{d}\rho_0^{\frac{1-q}{z}}(Y)\right)&=\frac{1}{z}\text{Tr}\left(\rho_0^{-1}\mathrm{d}\rho_0(X)\,\mathrm{d}\rho_0(Y)\right)=\\
&=\frac{1}{z(1-w^2)}\,\mathrm{d}w(X)\,\mathrm{d}w(Y)\;.
\end{split}
\ee
Moreover, being $[\sigma_j,\sigma_k]=2i\varepsilon_{jk}^\ell\sigma_\ell$ and $U^{-1}\mathrm{d}U=i\sigma_k\theta^k$, with $\{\theta^k\}_{k=0,\dots,3}$ a basis of left-invariant 1-forms on $U(2)$, the left-invariant Maurer-Cartan 1-form, the commutators in the second term of \eqref{1part} are given by
\be
\bigl[U^{-1}\mathrm{d}U(X),\rho_0^{\alpha}\bigr]=(a_{\alpha}-b_{\alpha})\varepsilon_{3k}^\ell\theta^k(X)\,\sigma_\ell\;,
\ee
and, remembering that $\sigma_j\sigma_k=\delta_{jk}+i\varepsilon_{jk}^{\ell}\sigma_\ell$, we get
\be\label{g1tang}
\begin{split}
&\text{Tr}\left(\rho_0^{\frac{z-1}{z}}\bigl[U^{-1}\mathrm{d}U(X),\rho_0^{\frac{q}{z}}\bigr]\,\bigl[U^{-1}\mathrm{d}U(Y),\rho_0^{\frac{1-q}{z}}\bigr]\right)= \\
&=\frac{z}{q(1-q)}(a_{\frac{q}{z}}-b_{\frac{q}{z}})(a_{\frac{1-q}{z}}-b_{\frac{1-q}{z}})(a_{\frac{z-1}{z}}+b_{\frac{z-1}{z}})\delta_{jk}\theta^j(X)\,\theta^k(Y)+\\
&+i\,\frac{z}{q(1-q)}(a_{\frac{q}{z}}-b_{\frac{q}{z}})(a_{\frac{1-q}{z}}-b_{\frac{1-q}{z}})(a_{\frac{z-1}{z}}-b_{\frac{z-1}{z}})\varepsilon_{3jk}\theta^j(X)\,\theta^k(Y)\;,
\end{split}
\ee
where there is a sum over repeated indices $j$ and $k$ with $j,k\neq3$.
 
Regarding the second term in \eqref{ABmetric}, after the pull-back it yields
{\footnotesize \be\label{2part}
\begin{split}
&\frac{1}{q(1-q)}i_{d}^{*}\left[\sum_{m=0}^{\infty}m\,c_{m}(z)\sum_{a=0}^{m-2}\text{Tr}\left((AB-\mathds1)^{a}\,\left(i_{\mathbb{X}_{l}}\mathrm{d}A\right)\,B\,(AB-\mathds1)^{m-a-2}\,A\,\left(i_{\mathbb{Y}_{r}}\mathrm{d}B\right)\right)\right]=\\
&=\frac{1}{q(1-q)}\,\sum_{n=0}^{\infty}n\,c_n(z)\sum_{m=0}^{n-2}\,\left(\text{Tr}\Bigl((\rho_0^{\frac{1}{z}}-\mathds1)^m\,\mathrm{d}\rho_0^{\frac{q}{z}}(X)\rho_0^{\frac{1-q}{z}}(\rho_0^{\frac{1}{z}}-\mathds1)^{n-m-2}\,\rho_0^{\frac{q}{z}}\,\mathrm{d}\rho_0^{\frac{1-q}{z}}(Y)\Bigr)+\right.\\
&+\left.\text{Tr}\Bigl((\rho_0^{\frac{1}{z}}-\mathds1)^m\bigl[U^{-1}\mathrm{d}U(X),\rho_0^{\frac{q}{z}}\bigr]\rho_0^{\frac{1-q}{z}}(\rho_0^{\frac{1}{z}}-\mathds1)^{n-m-2}\,\rho_0^{\frac{q}{z}}\bigl[U^{-1}\mathrm{d}U(Y),\rho_0^{\frac{1-q}{z}}\bigr]\Bigr)\right)\;.
\end{split}
\ee}
In the first term of \eqref{2part}, the matrices involved in the trace are all diagonal and hence they commute with each other. 
Therefore, using the relation
{\small \be
\sum_{n=0}^{\infty}n\,c_n(z)\sum_{m=0}^{n-2}(\rho_0^{\frac{1}{z}}-\mathds1)^m(\rho_0^{\frac{1}{z}}-\mathds1)^{n-m-2}=\sum_{n=0}^{\infty}n(n-1)\,c_n(z)(\rho_0^{\frac{1}{z}}-\mathds1)^{n-2}=z(z-1)\rho_0^{\frac{z-2}{z}}\;,
\ee}
we get:
{\small \be\label{g2transv}
\begin{split}
&\frac{1}{q(1-q)}\sum_{n=0}^{\infty}n\,c_n(z)\sum_{m=0}^{n-2}\text{Tr}\Bigl((\rho_0^{\frac{1}{z}}-\mathds1)^m\,\mathrm{d}\rho_0^{\frac{q}{z}}(X)\rho_0^{\frac{1-q}{z}}(\rho_0^{\frac{1}{z}}-\mathds1)^{n-m-2}\,\rho_0^{\frac{q}{z}}\,\mathrm{d}\rho_0^{\frac{1-q}{z}}(Y)\Bigr)=\\
&=\frac{z-1}{z}\text{Tr}\left(\rho_0^{-1}\mathrm{d}\rho_0(X)\,\mathrm{d}\rho_0(Y)\right)=\frac{z-1}{z}\,\frac{1}{1-w^2}\,\mathrm{d}w(X)\, \mathrm{d}w(Y)\;.
\end{split}
\ee}
The evaluation of the second term in \eqref{2part} is a little bit trickier. 
Indeed, as discussed in  \ref{subsec: explicit computations}, we have to use binomial expansions, partial sums of geometric series, and a lot of algebra, in order to obtain:
{\footnotesize\be\label{g2tang}
\begin{split}
&\quad\frac{1}{q(1-q)}\sum_{n=0}^{\infty}n\,c_n(z)\sum_{m=0}^{n-2}\text{Tr}\Bigl((\rho_0^{\frac{1}{z}}-\mathds1)^m\bigl[U^{-1}\mathrm{d}U(X),\rho_0^{\frac{q}{z}}\bigr]\rho_0^{\frac{1-q}{z}}(\rho_0^{\frac{1}{z}}-\mathds1)^{n-m-2}\,\rho_0^{\frac{q}{z}}\bigl[U^{-1}\mathrm{d}U(Y),\rho_0^{\frac{1-q}{z}}\bigr]\Bigr)= \\
&=\frac{z}{q(1-q)}(a_{\frac{q}{z}}-b_{\frac{q}{z}})(a_{\frac{1-q}{z}}-b_{\frac{1-q}{z}})(a_{\frac{z-1}{z}}-b_{\frac{z-1}{z}})\,\frac{(a_{\frac{1}{z}}+b_{\frac{1}{z}})}{(a_{\frac{1}{z}}-b_{\frac{1}{z}})}\,\delta_{jk}\,\theta^j(X)\,\theta^k(Y)+\\
&-i\frac{z}{q(1-q)}(a_{\frac{q}{z}}-b_{\frac{q}{z}})(a_{\frac{1-q}{z}}-b_{\frac{1-q}{z}})(a_{\frac{z-1}{z}}-b_{\frac{z-1}{z}})\,\varepsilon_{3jk}\,\theta^j(X)\,\theta^k(Y)\;,
\end{split}
\ee}
where, again, there is a sum over repeated indices $j$ and $k$ with $j,k\neq3$.

Finally, summing the four terms \eqref{g1transv}, \eqref{g1tang}, \eqref{g2transv} and \eqref{g2tang}, the covariant metric-like tensor on the unfolding space of the space of invertible quantum states for a two-level system is given by:
\be\label{qubitmetric}
\begin{split}
g_{q,z}&=\frac{1}{1-w^2}\,dw\otimes dw+\frac{2wz}{q(1-q)}\frac{(a_{\frac{q}{z}}-b_{\frac{q}{z}})(a_{\frac{1-q}{z}}-b_{\frac{1-q}{z}})}{(a_{\frac{1}{z}}-b_{\frac{1}{z}})}\,(\theta^1\otimes\theta^1+\theta^2\otimes\theta^2)\\
&=g_{q,z}^{transv}+g_{q,z}^{tang}\;,
\end{split}
\ee
i.e., the imaginary terms in Eqs. \eqref{g1tang} and \eqref{g2tang} erase each other and the metric splits into a component transversal to the orbits of the unitary group, which is the usual Fisher-Rao metric, and a component tangential to the orbits of the unitary group.

\subsection{Weak radial limit to pure states}\label{radiallimit}

For a two-level system the full density matrix $\rho$ can be written in terms of the Pauli matrices as
\be
\rho=U\rho_0U^{-1}=\frac{1}{2}(\sigma_0+w\vec x\cdot\vec\sigma)\qquad,\qquad-1\leq w \leq1\;,\;U\in SU(2)
\ee
with the parameters $x_j$, $j=1,2,3$, functions of the unitary matrix elements through the relation
\be
U\sigma_3U^{-1}=\vec x\cdot\vec\sigma
\ee
Taking the square on both sides, it follows that
\be
\sum_{j=1}^3x_j^2=1\;,
\ee
i.e., the manifold of parameters is the three dimensional ball $B^2$ of unit radius, usually known as Bloch ball. This is a stratified manifold, each stratum being an orbit of $SL(2,\mathbb C)$ \cite{grab, orbite}. There are two strata. One is provided by rank two-states, the other one by rank one-pure states, characterized by $w^2=1$. The latter actually coincides with the topological boundary $S^2$ of the Bloch ball.

However, the metric defined in equation \eqref{qubitmetric} only holds for invertible density states, that is, inside the bulk of the Bloch ball. 
It is possible to extend the metric to pure states, that is, to the boundary of the Bloch ball for a two-level system, by performing the so-called \textit{weak radial limit} \cite{wrl1, wrl2}. This procedure can be summarized by means of the following steps:
\begin{enumerate}
\item we first consider an invertible quantum state $\bar\rho$ strictly inside the Bloch sphere with  $a>b$, where $a=\frac{1+w}{2}$ and $b=\frac{1-w}{2}$ denote the elements of the probability vector associated to the corresponding diagonal density matrix $\bar\rho_0$;
\item we then evaluate the scalar product $g_{q,z}(X_1,X_2)\bigl|_{\bar\rho}$ of two tangential vectors, say $X_1,X_2$, at the point $\bar\rho$, so that only the tangential part of the metric contributes;
\item we finally perform the radial limit $w\rightarrow1$ along the radius passing through $\bar\rho$, up to the pure state $\bar\rho_P$ with eigenvalues $a=1, b=0$.
\end{enumerate}
This limiting procedure therefore yields the following expression of the metric $g^0_{q,z}$ for pure states living on the boundary of the Bloch ball
\be\label{wrlpure}
g^0_{q,z}=\frac{2z}{q(1-q)}(\theta^1\otimes\theta^1+\theta^2\otimes\theta^2)\;,
\ee
which, again for $z\rightarrow1$, reduces to the metric $g^0_q$ for pure states obtained in \cite{vitale} starting from the Tsallis relative entropy, and it is singular for $q\rightarrow1,0$. Performing the limit $q\rightarrow1$\footnote{The case $q=0$ can be treated in the same way.} in equation \eqref{qubitmetric}, the expression for rank two density states yields
\be\label{q=1metric}
g\equiv\lim_{q\rightarrow1,0}g_{q,z}=\frac{1}{1-w^2}dw\otimes dw+2w\log{\left(\frac{1+w}{1-w}\right)}(\theta^1\otimes\theta^1+\theta^2\otimes\theta^2)\;,
\ee
where we have used the definition \eqref{qlog} of the q-logarithm function. Notice that equation \eqref{q=1metric} does not depend on the parameter $z$ and it actually coincides with the result of \cite{vitale}. Moreover, both in equation \eqref{qubitmetric} and in equation \eqref{q=1metric}, we recognize the transversal contribution and the round metric on the 2-sphere, with different coefficients. Now, in agreement with equation \eqref{wrlpure}, we see that for $q=0,1$ the coefficient of the tangential component diverges when we take the limit $w\rightarrow\pm1$. This essentially means that it is not possible to perform the radial limit procedure to recover the metric for pure states or in other words, the radial limit and the limit $q\rightarrow1,0$ commute and give a negative result.

This is coherent with what we know from Petz classification theorem \cite{petz1,petz2}. Indeed, the metric \eqref{qubitmetric} can be recast in the Petz form
\be\label{gpetz}
g^{Petz}=\frac{1}{1-w^2}dw\otimes dw+\frac{w^2}{(1+w)f\left(\frac{1-w}{1+w}\right)}(\theta^1\otimes\theta^1+\theta^2\otimes\theta^2)\;,
\ee
with operator monotone function $f: [0,\infty[\rightarrow\mathbb R$, such that $f(t)=tf(1/t)$, given by\footnote{As proved in Sec. \ref{monodpi}, quantum divergence functions satisfying the DPI give rise to quantum metrics possessing the monotonicity property. Therefore, since the DPI is satisfied in the range of parameters we are considering in this work (see Sec. \ref{qzrelentropy}) and according to Petz classification theorem, we are ensured that the function \eqref{omf} is operator monotone for $z\in\mathbb R^+$ and $q\in[0,1]$.}
\be\label{omf}
f(t)=\frac{q(1-q)}{4z}\frac{(t-1)(t^{\frac{1}{z}}-1)}{(t^{\frac{q}{z}}-1)(t^{\frac{1-q}{z}}-1)}\qquad,\qquad t=\frac{1-w}{1+w}\;.
\ee
For $q=1,0$, instead we have
\be\label{omfvn}
f(t)=\frac{t-1}{4\ln{t}}
\ee
and, according to \cite{wrl1,wrl2}, the radial limit is well defined if and only if $f(0)\neq0$ which is actually the case for $q\neq1,0$ but it is not verified for $q=1,0$.

Note that, up to a normalization factor, in the limit $z\rightarrow1$ the function \eqref{omf} reduces to the operator monotone function reproducing the Petz metric obtained from the Tsallis relative entropy \cite{andai}
\be
f(t)=\frac{q(1-q)}{4}\frac{(t-1)^2}{(t^q-1)(t^{1-q}-1)}\;,
\ee
and accordingly \eqref{omfvn} is the operator monotone function reproducing the metric \eqref{q=1metric} obtained from the von Neumann relative entropy. 

\subsection{The $z\rightarrow1, q\rightarrow\frac{1}{2}$ limit: Wigner-Yanase metric}\label{WY2}

In the limit $z\rightarrow1$, we recover the result of \cite{vitale}. Indeed, as already discussed in equation \eqref{tsallis}, for such value of the parameter $z$ the $q-z$-relative entropy \eqref{sqz} reduces to the Tsallis $q$-relative entropy and coherently the metric \eqref{qubitmetric} becomes
\be\label{2levtsallis}
g_q\equiv g_{q,1}=\frac{1}{1-w^2}\,dw\otimes dw+\frac{2}{q(1-q)}(a_q-b_q)(a_{1-q}-b_{1-q})\,(\theta^1\otimes\theta^1+\theta^2\otimes\theta^2)\;.
\ee
Thus, performing the limit $q\rightarrow\frac{1}{2}$, we get
\be\label{qubitWY}
g_{\frac{1}{2},1}=\frac{1}{1-w^2}\,dw\otimes dw+8(1-\sqrt{1-w^2})\,(\theta^1\otimes\theta^1+\theta^2\otimes\theta^2)\equiv g_{WY}\;,
\ee
which is the (pullback of the) so-called Wigner-Yanase information metric \cite{gibilisco_isola-a_characterization_of_wigner-yanase_skew_information_among_statistically_monotone_metrics, gibilisco, hasegawa_petz-noncommutative_extension_of_information_geometry_II}. Indeed, as already discussed in Sec. \ref{qzrelentropy}, for such values of the parameters, the $q-z$-relative entropy reduces to
\be
S_{\frac{1}{2},1}(\rho|\varrho)=\mathcal{I}(\rho|\varrho)=4\bigl[1-\text{Tr}\,\bigl(\rho^{\frac{1}{2}}\varrho^{\frac{1}{2}}\bigr)\bigr]\;,
\ee
which, as discussed in \cite{gibilisco_isola-a_characterization_of_wigner-yanase_skew_information_among_statistically_monotone_metrics, gibilisco, hasegawa_petz-noncommutative_extension_of_information_geometry_II}, is the divergence function for the Wigner-Yanase metric. Moreover, for such values of the parameters ($z=1$ and $q=\frac{1}{2}$), the operator monotone function \eqref{omf} gives
\be\label{WYomf}
f(t)=\frac{1}{16}\frac{(t-1)^2}{(\sqrt{t}-1)^2}=\frac{1}{16}(\sqrt{t}+1)^2\;,
\ee
which, up to a normalization factor, is the operator monotone function associated with the Wigner-Yanase metric \cite{gibilisco_isola-a_characterization_of_wigner-yanase_skew_information_among_statistically_monotone_metrics, gibilisco, hasegawa_petz-noncommutative_extension_of_information_geometry_II}. This can be easily checked by substituting the function \eqref{WYomf} in the expression \eqref{gpetz}, where we set $t=\frac{1-w}{1+w}$, and thus obtaining the metric \eqref{qubitWY}.  

\subsection{The $z\rightarrow\frac{1}{2}, q\rightarrow\frac{1}{2}$ limit: Bures metric}\label{bures}

As anticipated in Sec. \ref{qzrelentropy}, in the limit $z\rightarrow\frac{1}{2}, q\rightarrow\frac{1}{2}$, the $q-z$-relative entropy \eqref{sqz} reduces to
\be
S_{\frac{1}{2}\frac{1}{2}}(\rho|\varrho)=4\bigl[1-\text{Tr}(\rho\varrho)^{1/2}\bigr]\;, 
\ee 
which coincides with the divergence function of the Bures metric
\be
D_{B}^{2}(\rho|\varrho)=4\bigl[1-\sqrt{F}(\rho,\varrho)\bigr]\;, 
\ee
where $\sqrt{F}(\rho,\varrho)$ is the so-called \textit{root fidelity}. Indeed, according to Uhlmann's fidelity theorem \cite{bengtsson}, the root fidelity is given by
\be
\begin{split}
\sqrt{F}(\rho,\varrho)&=\text{Tr}\left|\sqrt{\varrho}\sqrt{\rho}\right|\\
&=\text{Tr}\sqrt{\sqrt{\varrho}\sqrt{\rho}\sqrt{\rho}\sqrt{\varrho}}\\
&=\text{Tr}\sqrt{\sqrt{\varrho}\rho\sqrt{\varrho}}\\
&=\text{Tr}\sqrt{\rho\varrho}\;,
\end{split}
\ee
where in the last equality we used the property that, for any pair of square matrices $A$ and $B$, the eigenvalues of $AB$ and $BA$ are the same from which it follows that the matrix $\rho\tilde\rho$ has real, non-negative eigenvalues (even though it is not in general self-adjoint), and the trace functional $\text{Tr}(\cdot)^{1/2}$ in this expression is well-defined as the sum of the square roots of these eigenvalues, which are the same as those of $\varrho^{\frac{1}{2}}\rho\varrho^{\frac{1}{2}}$ \cite{DATTA}.

Accordingly, for $z=q=\frac{1}{2}$, the $q-z$-metric \eqref{qubitmetric} then reduces to (pullback of) the Bures metric
\be\label{2levbures}
g_{\frac{1}{2}\frac{1}{2}}=\frac{1}{1-w^2}\,dw\otimes dw+4w^2\,(\theta^1\otimes\theta^1+\theta^2\otimes\theta^2)\equiv g_{Bures}\;,
\ee 
as can be easily checked by direct computation from equation \eqref{qubitmetric} or alternatively noticing that, for such values of the parameters, the operator monotone function \eqref{omf} yields
\be
f(t)=\frac{1+t}{8}\qquad,\qquad t=\frac{1-w}{1+w}\;,
\ee
which, up to a normalization factor, is actually the operator monotone function associated to the Bures metric \cite{andai, kubo} as can be verified by substituting it in equation \eqref{gpetz} thus obtaining the metric \eqref{2levbures}.

\section{The n-level case}
\label{Nlev}

Here, we will discuss the result of the computations and simply refer to \ref{subsec: explicit computations} for all the details.
The final result is:

\be\label{gNlevel}
g_{q,z}=g^{\perp}_{q,z} + g^{\parallel}_{q,z}=\sum_{\alpha=1}^{n}\,p_\alpha \mathrm{d}\ln p_\alpha\otimes \mathrm{d}\ln p_\alpha+\frac{z}{q(1-q)}\sideset{}{'}\sum_{j,k=1}^{n^2-1}\mathcal C_{jk}\,\theta^j\otimes\theta^k\,,
\ee
where $\{p_{\alpha}\}_{\alpha=1,..n}$ denote the eigenvalues of $\rho$, the coefficients $C_{jk}$ are given by:

\be\label{gNlevelcoeff}
\mathcal C_{jk}=\sideset{}{'}\sum_{\alpha,\beta=1}^{n}\mathcal E_{\alpha\beta}\,\Re{\bigl[M_j^{\alpha\beta}M_k^{\beta\alpha}\bigr]}\;,
\ee
with $M_{j}^{\alpha\beta}$ being numerical coefficients depending on the choice of a basis in the Lie algebra of $SU(n)$, and with:

\be
\mathcal E_{\alpha\beta}:=\frac{(p_\alpha-p_\beta)(p_\alpha^{\frac{q}{z}}-p_\beta^{\frac{q}{z}})(p_\alpha^{\frac{1-q}{z}}-p_\beta^{\frac{1-q}{z}})}{(p_\alpha^{\frac{1}{z}}-p_\beta^{\frac{1}{z}})}\,.
\ee 
Note that, as it is shown in \ref{subsec: explicit computations}, when $\alpha=\beta$, as well as for all those points such that $p_{\alpha}=p_{\beta}$, the terms in the sum vanish.
We decided to highlight this fact using the notation $\sideset{}{'}\sum$ in \eqref{gNlevel} and \eqref{gNlevelcoeff}.

It turns out that the coefficients $C_{jk}$ are symmetric in $j$ and $k$, and thus $g_{q,z}$ is a symmetric tensor.
Whenever the parameters $q$ and $z$ are such that the regularized $q,z$-RRE $S_{q,z}$ of equation \eqref{sqz} is a divergence function in the sense of definition \ref{def: divergence function}, we have that $\mathbb{D}_{q,z}$ is a non-negative potential function, and that $g_{q,z}$ is a positive-semidefinite symmetric covariant tensor field which is the pullback of the positive-semidefinite symmetric covariant tensor field on $\stsp_{n}$ extracted from $S_{q,z}$ (see proposition \ref{prop: potential functions, metrics and differentiable mappings} and proposition \ref{prop: nec and suf conditions for g to be positive or negative semidefinite}).
The invertibility of the covariant tensor on $\stsp_{n}$ must be checked with a case by case analysis, and this may be done using the results of proposition \ref{prop: the unfolding map of the invertible density matrices is a surjective differentiable map} regarding the kernel of the unfolding map $\pi_{n}$.

If the values of $q$ and $z$ for which the $q$-$z$-\textit{regularized R\'enyi Relative Entropy} is a divergence function in the sense of definition \ref{def: divergence function} are such that the data processing inequality is satisfied, then  the family of symmetric covariant tensors we may extract will necessarily satisfy the monotonicity property. 
In particular, according to the formulae in the introduction of this chapter, the family of metric tensors associated with the von Neumann-Umegaki relative entropies, with the Tsallis relative entropies,  with the Wigner-Yanase skew informations, and with the Bures divergences, all satisfies the monotonicity property.

Equation \eqref{gNlevel} points out another interesting fact.
The first term in the expression of $g_{q,z}$ is precisely the Fisher-Rao metric tensor related to the component of the ``classical'' probability vector $\vec p=(p_1,\dots,p_n)$ identified with the diagonal elements of the invertible density matrix.
Consequently, since the monotonicity property is connected to the data processing inequality, since the data processing inequality depends on the explicit values of $q$ and $z$, and since the Fisher-Rao contribution to $g_{q,z}$ does not depend on $(q,z)$, the ``obstruction'' to the monotonicity property is completely encoded in the unitary contribution to $g_{q,z}$.

\section{Special limits and explicit examples}\label{limandex}

We collect here some special limits and explicit examples arising from the general form given in equation \eqref{gNlevel} (or \eqref{gNlevelapp}).
Specifically, we start recovering the results for the qubit given in section \ref{2level} from the general formalism of section \ref{Nlev}.
This allows us to take confidence with the computation of all the numerical coefficients needed to explicitly use equation \eqref{gNlevel} (or \eqref{gNlevelapp}) in higher-dimensional cases.
Then, in analogy to what we have done in the qubit case, we consider the various limits for the parameters $q$ and $z$ thus providing the explicit expressions of both the Wigner-Yanase and the Bures metric on the space of (invertible) quantum states of a generic n-level system. 
Finally, as a less trivial example, we give a detailed analysis of the computations in the case of a three-level system.

\subsection{The qubit case}\label{subsec: the qubit case}

A quantum state $\rho$ of a two-level system ($n=2$) may be written as:
\be
\rho_0=\begin{pmatrix}\frac{1+w}{2} & 0\\
0 & \frac{1-w}{2}\end{pmatrix}=\left(\frac{1+w}{2}\right)\tau_{11}+\left(\frac{1-w}{2}\right)\tau_{22}\;,
\ee
with:
\be\label{eigenval}
p_1=\frac{1+w}{2}\equiv a\qquad,\qquad p_2=\frac{1-w}{2}\equiv b\;.
\ee
Referring to section \ref{2level} and \ref{subsec: explicit computations} for the notation, we write the Pauli matrices  in terms of the $\mathbf{E}_{\alpha\beta}$ matrices as
\be\label{2levsigmatau}
\begin{split}
&\sigma_0=\begin{pmatrix}1 & 0\\
0 & 1\end{pmatrix}=\mathbf{E}_{11}+\mathbf{E}_{22}\qquad\quad,\;\;\;\qquad\sigma_1=\begin{pmatrix}0 & 1\\
1 & 0\end{pmatrix}=\mathbf{E}_{12}+\mathbf{E}_{21}\;,\\
&\sigma_2=\begin{pmatrix}0 & -i\\
i & 0\end{pmatrix}=i(\mathbf{E}_{21}-\mathbf{E}_{12})\qquad,\qquad\sigma_3=\begin{pmatrix}1 & 0\\
0 & -1\end{pmatrix}=\mathbf{E}_{11}-\mathbf{E}_{22}\;.
\end{split}
\ee
Then, according to Eqs. \eqref{sigmatau} and \eqref{2levsigmatau}, in the qubit case we have:
\begin{eqnarray}
&M_0^{11}=M_0^{22}=1\,,\qquad\quad M_0^{12}=M_0^{21}=0\\
&M_1^{11}=M_1^{22}=0\,,\qquad\quad M_1^{12}=M_1^{21}=1\\
&\;\;M_2^{11}=M_2^{22}=0\,,\qquad\quad M_2^{21}=-M_2^{12}=i\\
&M_3^{11}=-M_3^{22}=1\,,\qquad\quad M_3^{12}=M_3^{21}=0\,.
\end{eqnarray}
Now, taking into account equation \eqref{eigenval}, the explicit expression of the transversal part $g^{\perp}_{q,z}$ of the metric given by equation \eqref{gNlevel} reads:
\be
\begin{split}
g_{q,z}^{\perp}&=\frac{1}{p_1}dp_1\otimes dp_1+\frac{1}{p_2}dp_2\otimes dp_2\\
&=\frac{1}{2(1+w)}dw\otimes dw+\frac{1}{2(1-w)}dw\otimes dw\\
&=\frac{1}{1-w^2}dw\otimes dw\;.
\end{split}
\ee
On the other hand, equations \eqref{gtangtheta} and \eqref{jkcoeff} give the following expression for the tangential part $g_{q,z}^{\parallel}$:
\be
\begin{split}
g_{q,z}^{\parallel}&=\frac{z}{q(1-q)}\sideset{}{'}\sum_{j,k=1}^{3}\left(\mathcal E_{12}\,\Re{\bigl[M_j^{12}M_k^{21}\bigr]}+\mathcal E_{21}\,\Re{\bigl[M_j^{21}M_k^{12}\bigr]}\right)\theta^j\otimes\theta^k\\
&=\frac{2z}{q(1-q)}\sum_{j,k=1,2}\mathcal E_{12}\,\Re{\bigl[M_j^{12}M_k^{21}\bigr]}\,\theta^j\otimes\theta^k\\
&=\frac{2z}{q(1-q)}\,\mathcal E_{12}\,(\theta^1\otimes\theta^1+\theta^2\otimes\theta^2)\;.
\end{split}
\ee
Collecting these two expressions, and remembering the explicit form for the coefficients $\mathcal E_{\alpha\beta}$ given in equation \eqref{coeff}, we obtain:
\be
g_{q,z}=\frac{1}{1-w^2}dw\otimes dw + \frac{2wz}{q(1-q)}\frac{(a_{\frac{q}{z}}-b_{\frac{q}{z}})(a_{\frac{1-q}{z}}-b_{\frac{1-q}{z}})}{(a_{\frac{1}{z}}-b_{\frac{1}{z}})}\,(\theta^1\otimes\theta^1+\theta^2\otimes\theta^2)
\ee
 is in accordance with the result obtained in section \ref{2level} for the qubit case (see equation \eqref{qubitmetric}).

\subsection{The special limit $z=1$}

For $z=1$, the two-parameters family of metrics reduces to the 1-parameter family of metrics derived from the Tsallis relative entropy in \cite{vitale}. Indeed, for such value of the parameter $z$, the expression \eqref{gNlevel} for the $q-z$-metric tensor becomes
\be\label{Ts1}
g_{q,1}=\sum_{\alpha=1}^n\frac{1}{p_\alpha}dp_\alpha\otimes dp_\alpha+\frac{1}{q(1-q)}\sideset{}{'}\sum_{\alpha,\beta=1}^{n}(p_\alpha^q-p_\beta^q)(p_\alpha^{1-q}-p_\beta^{1-q})\,\theta^{\alpha\beta}\otimes\theta^{\beta\alpha}\;,
\ee
or equivalently equation \eqref{gNlevel} becomes
\be\label{Ts2}
g_{q,1}=\sum_{\alpha=1}^n\frac{1}{p_\alpha}dp_\alpha\otimes dp_\alpha+\frac{1}{q(1-q)}\sideset{}{'}\sum_{j,k=1}^{n^2-1}\mathcal C_{jk}\,\theta^j\otimes\theta^k\;,
\ee
with
\be
\mathcal C_{jk}=\sideset{}{'}\sum_{\alpha,\beta=1}^{n}(p_\alpha^q-p_\beta^q)(p_\alpha^{1-q}-p_\beta^{1-q})\,\Re{\bigl[M_j^{\alpha\beta}M_k^{\beta\alpha}\bigr]}\;.
\ee
This essentially provides the expression of the quantum metric tensor derived from the Tsallis q-relative entropy for a generic n-level system and, in agreement with \cite{vitale}, for $n=2$ both equation \eqref{Ts1} and equation \eqref{Ts2} actually reproduce the expression \eqref{2levtsallis} for the qubit case.

\subsection{The special limit  $z=1, q=\frac{1}{2}$}

According to the result of Sec. \ref{WY2}, for such values of the parameters, the expression \eqref{gNlevel}  provides us with the general form of (the pullback of) the Wigner-Yanase metric tensor for a generic n-level system, that is
\be
g_{WY}\equiv g_{\frac{1}{2},1}=\sum_{\alpha=1}^n\frac{1}{p_\alpha}dp_\alpha\otimes dp_\alpha+4\sideset{}{'}\sum_{j,k=1}^{n^2-1}\sideset{}{'}\sum_{\alpha,\beta=1}^{n}\left(p_\alpha^{1/2}-p_\beta^{1/2}\right)^2\,\Re{\bigl[M_j^{\alpha\beta}M_k^{\beta\alpha}\bigr]}\,\theta^j\otimes\theta^k\;.
\ee
In particular, using equation \eqref{eigenval}, for $n=2$ we have
\be
\begin{split}
g_{WY}\equiv g_{\frac{1}{2},1}^{(n=2)}&=\frac{1}{1-w^2}dw\otimes dw+8(a_{1/2}-b_{1/2})^2(\theta^1\otimes\theta^1+\theta^2\otimes\theta^2)\\
&=\frac{1}{1-w^2}dw\otimes dw+8(1-\sqrt{1-w^2})\,(\theta^1\otimes\theta^1+\theta^2\otimes\theta^2)\;,
\end{split}
\ee
i.e., we recover the expression \eqref{qubitWY} of Sec. \ref{WY2} for the qubit case.  

\subsection{The special limit  $z=\frac{1}{2},q=\frac{1}{2}$}

When $z=\frac{1}{2}$ and $q=\frac{1}{2}$, according to the result of section \ref{bures}, the expression in equation \eqref{gNlevel} provides us with the general form of (the pullback of) the Bures metric tensor for a generic n-level system, that is:
\be
\begin{split}
g_{B}\equiv g_{\frac{1}{2},\frac{1}{2}}&=\sum_{\alpha=1}^n\frac{1}{p_\alpha}dp_\alpha\otimes dp_\alpha+2\sideset{}{'}\sum_{\alpha,\beta=1}^{n}\frac{(p_\alpha-p_\beta)^3}{(p_\alpha^2-p_\beta^2)}\,\Re{\bigl[M_j^{\alpha\beta}M_k^{\beta\alpha}\bigr]}\,\theta^j\otimes\theta^k\\
&=\sum_{\alpha=1}^n\frac{1}{p_\alpha}dp_\alpha\otimes dp_\alpha+2\sideset{}{'}\sum_{\alpha,\beta=1}^{n}\frac{(p_\alpha-p_\beta)^2}{(p_\alpha+p_\beta)}\,\Re{\bigl[M_j^{\alpha\beta}M_k^{\beta\alpha}\bigr]}\,\theta^j\otimes\theta^k\\
&=\sum_{\alpha=1}^n\frac{1}{p_\alpha}dp_\alpha\otimes dp_\alpha+2\sideset{}{'}\sum_{j,k=1}^{n^2-1}\sideset{}{'}\sum_{\alpha,\beta=1}^{n}\frac{(p_\alpha-p_\beta)^2}{(p_\alpha+p_\beta)}\,\Re{\bigl[M_j^{\alpha\beta}M_k^{\beta\alpha}\bigr]}\,\theta^j\otimes\theta^k\;.
\end{split}
\ee
In particular, using equation \eqref{eigenval}, for $n=2$ we have:
\be
\begin{split}
g_{B}\equiv g_{\frac{1}{2},\frac{1}{2}}^{(n=2)}&=\frac{1}{1-w^2}dw\otimes dw+4\,\frac{(a-b)^2}{a+b}\,(\theta^1\otimes\theta^1+\theta^2\otimes\theta^2)\\
&=\frac{1}{1-w^2}dw\otimes dw+4\,w^2\,(\theta^1\otimes\theta^1+\theta^2\otimes\theta^2)\;,
\end{split}
\ee
i.e., we recover the expression \eqref{2levbures} of section \ref{bures} for the qubit case. 

\subsection{A less trivial example: three level system}
\label{subsec: the qutrit case}

For a three-level system the relevant group of unitary transformations is $SU(3)$. 
The bulk region $\stsp_{3}^{0}$ of the space $\stsp_{3}$ of invertible quantum states, i.e., $\stsp_{3}$ without the totally mixed state, is the union of $SU(3)$ orbits (four- and six-dimensional sub-manifolds in $\mathbb R^8$)   \cite{em3, orbite, ercolessi1}. 
In this case, we have a basis in $\mathcal{B}(\mathcal{H})$ made up of the Gell-Mann matrices denoted by $\lambda_j$, $j=1,\dots,8$, and the identity matrix $\lambda_{0}=\mathds1_{3\times3}$.
Selecting the orthonormal basis $\{|j\rangle\}_{j=0,1,2}$ in $\mathcal{H}$ made up of eigenvectors of $\lambda_{3}$, we have the following matrix realization of the Gell-Mann matrices:
{\normalsize \be\label{gellmann}
\begin{split}
&\lambda_1=\left(\begin{array}{ccc}
0& 1&0\\1& 0&0\\ 0& 0& 0
\end{array} \right),\; \lambda_2=\left(\begin{array}{ccc}
0&-i &0\\i& 0&0\\ 0&0 &0 
\end{array} \right),\;\lambda_3=\left(\begin{array}{ccc}
1&0 &0\\0&-1 &0\\0 &0 & 0
\end{array} \right), \\ \\
& \lambda_4=\left(\begin{array}{ccc}
0&0 &1\\0& 0&0\\ 1& 0&0 
\end{array} \right),\; \lambda_5=\left(\begin{array}{ccc}
0& 0&-i\\0&0 &0\\ i& 0& 0
\end{array} \right),\;  \lambda_6=\left(\begin{array}{ccc}
0&0 &0\\0& 0&1\\0 & 1& 0
\end{array} \right), \\ \\
& \lambda_7=\left(\begin{array}{ccc}
0&0 &0\\0&0 &-i\\ 0&i & 0
\end{array} \right),\;  \lambda_8=\frac{1}{\sqrt 3}\left(\begin{array}{ccc}
1&0 &0\\0&1 &0\\ 0& 0& -2
\end{array} \right),
\end{split}
\ee}
The density matrix $\rho$ representing a quantum state can be written as
\be\label{state}
\rho=U\rho_0U^{-1}=\lambda_\mu x^\mu=\frac{1}{3}\lambda_0+\lambda_jx^j\;,
\ee
where $\mu=0,\dots,8$ and the condition $\text{Tr}\rho=1$ imposes $x_0=\frac{1}{3}$. 
Repeating the analysis of the two-level case, the diagonal density matrix $\rho_0$ in equation \eqref{state} can be written in terms of three real parameters $p_1,p_2,p_3>0$ as
\be\label{diagstate}
\rho_0=\begin{pmatrix} p_1 & 0 & 0\\
0 & p_2 & 0\\
0 & 0 & p_3\\
\end{pmatrix}\qquad\text{with}\qquad\sum_{k=1}^3p_k=1\;,
\ee
or in terms of the diagonal Gell-mann matrices of the $\mathfrak{su}(3)$-Cartan subalgebra ($\lambda_0,\lambda_3$ and $\lambda_8$ in the chosen basis) as
\be\label{lambdadiagstate}
\rho_0=\frac{1}{3}\lambda_0+\beta\lambda_3+\gamma\lambda_8\;,
\ee 
with
\be
\beta=\frac{1}{2}(p_1-p_2)\qquad,\qquad\gamma=\frac{1}{2\sqrt{3}}(p_1+p_2-2p_3)\;.
\ee
From equation \eqref{state}, we have then
\be
U(\beta\lambda_3+ \gamma\lambda_8)U^{-1} =\lambda_jx^j\quad,\quad j=1,\dots,8
\ee
which implies, after taking the square on both sides and the trace
\be
\beta^2+\gamma^2= \sum_{j=1}^8 x_j^2\;,
\ee
or, in terms of the parameters $p_1,p_2,p_3$
\be
\sum_{j=1}^8 x_j^2=\frac{1}{3}-(p_1p_2+p_1p_3+p_2p_3)\;,
\ee
which identifies the manifold of parameters as a submanifold with boundary in $\mathbb R^8$. 
The unitary orbits passing through rank-three density states are diffeomorphic to
\be\label{rank3orbits}
\mathcal O_3\cong\frac{U(3)}{U(1)\times U(1)\times U(1)}\qquad\qquad (\text{dim}\,\mathcal O_3=6)\;,
\ee
if the three eigenvalues are all different (i.e., $p_1\neq p_2\neq p_3$), while for each two eigenvalues being coincident but different from the remaining one, i.e., $p_j=p_k\neq p_\ell, j\neq k\neq\ell$, we have four-dimensional unitary orbits diffeomorphic to
\be\label{rank2orbits}
\mathcal O_2\cong\frac{U(3)}{U(2)\times U(1)}\qquad\qquad (\text{dim}\,\mathcal O_2=4)\;.
\ee

Let us now evaluate the quantum metric \eqref{gNlevel} for this case. In terms of the $3\times3$ matrices $\mathbf{E}$ of the standard basis, the Gell-mann matrices \eqref{gellmann} read as
{\footnotesize \be\label{eqn: sigmatau 3-level}
\begin{split}
&\lambda_1=\mathbf{E}_{12}+\mathbf{E}_{31}\quad,\quad\lambda_2=i(\mathbf{E}_{21}-\mathbf{E}_{12})\quad,\quad\lambda_3=\mathbf{E}_{11}-\mathbf{E}_{22}\quad,\quad\lambda_4=\mathbf{E}_{13}+\mathbf{E}_{31}\;,\\
&\lambda_5=i(\mathbf{E}_{31}-\mathbf{E}_{13})\quad,\quad\lambda_6=\mathbf{E}_{23}+\mathbf{E}_{32}\quad,\quad\lambda_7=i(\mathbf{E}_{32}-\mathbf{E}_{23})\quad,\quad\lambda_8=\frac{1}{\sqrt{3}}(\mathbf{E}_{11}+\mathbf{E}_{22}-2\mathbf{E}_{33})\;,
\end{split}
\ee}
and obviously $\mathbb{I}=\mathbf{E}_{11}+\mathbf{E}_{22}+\mathbf{E}_{33}$. Therefore, imposing that $\lambda_\mu=\sum_{\alpha,\beta=1}^3M_\mu^{\alpha\beta}\mathbf{E}_{\alpha\beta}$ for any $\mu=0,\dots,8$, it follows that the only non-zero coefficients $M_j^{\alpha\beta}$ are
{\footnotesize \be\label{nonzerocoeff}
\begin{split}
&M_0^{11}=M_0^{22}=M_0^{33}=M_3^{11}=-M_3^{22}= M_6^{23}=M_6^{32}=M_1^{12}=M_1^{21}=M_4^{13}=M_4^{31}=1\,, \\
&M_2^{21}=-M_2^{12}=M_7^{32}=-M_7^{23}=M_5^{31}=-M_5^{13}=i\,,\\
&M_8^{11}=M_8^{22}=\frac{1}{\sqrt{3}},M_8^{22}=-\frac{2}{\sqrt{3}}\,.
\end{split}
\ee}
Hence, according to equation \eqref{gNlevel}, in the qutrit case the two contributions to the quantum metric tensor are given by
\be\label{3levtransv}
g_{q,z}^{transv}=\sum_{\alpha=1}^{3}p_\alpha d\ln{p_\alpha}\otimes d\ln{p_\alpha}\;,
\ee
and
{\small\be\label{3levtang}
\begin{split}
g_{q,z}^{tang}&=\frac{2z}{q(1-q)}\sum_{j,k=1}^8\Bigl(\mathcal E_{12}\,\Re{\bigl[M_j^{12}M_k^{21}\bigr]}+\mathcal E_{13}\,\Re{\bigl[M_j^{13}M_k^{31}\bigr]}+\mathcal E_{23}\,\Re{\bigl[M_j^{23}M_k^{32}\bigr]}\Bigr)\theta^{j}\otimes\theta^k =\\
&=\frac{2z}{q(1-q)}\biggl[\frac{(p_1-p_2)(p_1^{\frac{q}{z}}-p_2^{\frac{q}{z}})(p_1^{\frac{1-q}{z}}-p_2^{\frac{1-q}{z}})}{(p_1^{\frac{1}{z}}-p_2^{\frac{1}{z}})}\,(\theta^1\otimes\theta^1+\theta^2\otimes\theta^2)+\\
&\qquad\qquad\;+\frac{(p_1-p_3)(p_1^{\frac{q}{z}}-p_3^{\frac{q}{z}})(p_1^{\frac{1-q}{z}}-p_3^{\frac{1-q}{z}})}{(p_1^{\frac{1}{z}}-p_3^{\frac{1}{z}})}\,(\theta^4\otimes\theta^4+\theta^5\otimes\theta^5)+\\
&\qquad\qquad\;+\frac{(p_2-p_3)(p_2^{\frac{q}{z}}-p_3^{\frac{q}{z}})(p_2^{\frac{1-q}{z}}-p_3^{\frac{1-q}{z}})}{(p_2^{\frac{1}{z}}-p_3^{\frac{1}{z}})}\,(\theta^6\otimes\theta^6+\theta^7\otimes\theta^7)\biggr]\;.
\end{split}
\ee}

The above result shows that in the passage from the first to the second line of equation \eqref{3levtang}  the only non-zero terms are those for $j=k\neq0,3,8$ since the mixed terms and those with $j,k=0,3,8$ are characterized by vanishing $\mathcal E$ coefficients or, being the product $M_j^{\alpha\beta}M_k^{\beta\alpha}$ zero or imaginary, by $\Re[M_j^{\alpha\beta}M_k^{\beta\alpha}]=0$.\\ \\Finally, let us close this section with the following remarks:
\begin{itemize}
\item[\textbf{i)}] From the expression \eqref{3levtang} of $g_{q,z}^{tang}$ it is evident the splitting into three $SU(2)$ copies (cfr. equation \eqref{qubitmetric}). Furthermore, we see that the structure of the metric retrieves the stratification in terms of the unitary orbits discussed in equations \eqref{rank3orbits} and \eqref{rank2orbits}. Indeed, for $p_1\neq p_2\neq p_3$, the tangential part of the metric is the pull-back to the parameters manifold of that on the six-dimensional orbit $\mathcal O_3$, whereas for each two coincident eigenvalues, but different from the remaining one, e.g., for $p_1=p_2\neq p_3$, the coefficient in the first term of equation \eqref{3levtang} vanishes and the other two become equal so that we find the pull-back of four-dimensional unitary orbits $\mathcal O_2$ \cite{orbite, ercolessi1,vitale}.
\item[\textbf{ii)}] For $z\rightarrow1$ the quantum metric $g_{q,z}^{(n=3)}=\eqref{3levtransv}+\eqref{3levtang}$ obtained here for the qutrit case reduces to the one derived in \cite{vitale} using the Tsallis relative entropy (cfr. equation (3.47) in \cite{vitale}). Moreover, our result has the same structure of the symmetric part of the quantum Fisher information tensor obtained in \cite{ercolessi2}, i.e., it splits into three $SU(2)$-related contributions. The explicit form of the coefficients is however different because of the different regularization procedures employed. Indeed, as discussed in \cite{vitale}, the starting point in \cite{ercolessi1,ercolessi2} is a generalization of the Fisher tensor for mixed states defined by means of the so-called symmetric logarithmic derivative, while here we used the $q$-$z$-relative entropy as a potential function for the metric, which cannot give rise to antisymmetric terms and it essentially amounts to define the logarithmic derivative through the introduction of q-logarithms. 
\item[\textbf{iii)}] As discussed in \cite{vitale}, the radial limit procedure illustrated in Sec. \ref{radiallimit} for the qubit case can be performed also in the present case for rank-one pure states and rank-two density states. Indeed, by getting rid of the transversal part of the metric by means of tangent vectors to the orbits and choosing for instance $p_1>p_2>p_3$, we first perform the limit $p_3\rightarrow0, p_1+p_2\rightarrow1$ in order to obtain the metric for the stratum of rank-two density states
{\small \be
\begin{split}
\tilde g_{q,z}&=\frac{2z}{q(1-q)}\biggl[\frac{(p_1-p_2)(p_1^{\frac{q}{z}}-p_2^{\frac{q}{z}})(p_1^{\frac{1-q}{z}}-p_2^{\frac{1-q}{z}})}{(p_1^{\frac{1}{z}}-p_2^{\frac{1}{z}})}\,(\theta^1\otimes\theta^1+\theta^2\otimes\theta^2)+\\
&\qquad\qquad\qquad+p_1\,(\theta^4\otimes\theta^4+\theta^5\otimes\theta^5)+p_2\,(\theta^6\otimes\theta^6+\theta^7\otimes\theta^7)\biggl]_{p_1+p_2=1}\;,
\end{split}
\ee}
and then, if we further perform the limit $p_2\rightarrow0,p_1\rightarrow1$, we obtain the metric for pure states
\be
g^0_{q,z}=\frac{2z}{q(1-q)}\sum_{j\neq3,6,7,8}\theta^j\otimes\theta^j\;,
\ee
which is singular for $q\rightarrow1,0$ as in the two-level case discussed in section \ref{radiallimit} (cfr. equation \eqref{wrlpure}). 
\end{itemize}

\section{Conclusions and outlook}
\label{sec: Conclusions and Outlook}

To summarize, we point out the main results of our work. In the first part of the paper, we developed a coordinate-free formalism for Information Geometry.
In Section \ref{gfromS} we introduced the notion of potential functions as the most general type of two-point function from which it is possible to extract symmetric covariant (0,2) and (0,3) tensors by means of a coordinate-free algorithm.
This algorithm is the coordinate-free counterpart of the standard one used in Information Geometry \cite{amari, Amari-Nagaoka}.
The set of divergence functions used in Information Geometry turns out to be a subset of the set of potential functions introduced here.  
Then, we focused on Quantum Information Geometry.
We reviewed the notion of quantum stochastic map, the so-called monotonicity property for a family of metric tensors on the manifold of invertible quantum states, and the so-called Data Processing Inequality (DPI) for a family of divergence functions on the manifold of invertible quantum states.
By means of the abstract formalism developed in Section \ref{gfromS} we are able to prove that the DPI for a family of quantum divergence functions implies the monotonicity property for the associated family of quantum metric tensors.

Once the formal framework has been settled, we used it in order to compute the family of metric tensors on the space of invertible quantum states (in finite dimensions) associated with the family of $q$-$z$-relative entropies.
This is a two-parameter family of quantum relative entropies from which it is possible to recover almost all known relative entropies by suitably varying the parameters. 
We further investigated possible ranges of the parameters $q$ and $z$ allowing to recover  well-known quantities in Information Geometry. 
In particular, we showed that a proper definition of both the Bures metric and the Wigner-Yanase metric can be derived from this family of divergence functions when $q=z=\frac{1}{2}$ and $q=\frac{1}{2},z=1$ respectively.
The general expression for the $q$-$z$ quantum metric allows us to give explicit expressions both of the Bures and Wigner-Yanase metric for an arbitrary $n$-level system.
 
We performed the calculation starting from the qubit case.
This allows the reader to start with a friendly use of the computational tools used in the paper.
Then, we extended the derivation of the metric tensor to a generic $n$-level system. 
In particular, we worked out explicitly the less trivial case of a three-level system showing how the structure of the metric reflects the structure of the foliation of the stratum of rank-three invertible quantum states into unitary orbits.

In conclusion, a few comments are in order. 
In this work we mainly focused on the computation of the metric tensor but the intrinsic formalism developed here can be also used to extract symmetric covariant (0,3) tensors (skewness tensors). We leave such analysis as well as the extension to higher rank structures to forthcoming publications. Moreover, within the framework of the tomographic reconstruction of metrics on the space of quantum states introduced in \cite{vitale}, in a previous paper \cite{LMMVV12} some of the authors have shown that there exists a one-to-one correspondence between the choice of a tomographic scheme and the quantum metric associated with a particular relative entropy. From this perspective, it would be very interesting to investigate if different $q$-$z$ quantum metrics (corresponding to different values of the parameters) might be related by a change of tomographic scheme. More specifically, since the change of tomographic schemes actually induces a diffeomorphism mapping Petz metrics into Petz metrics and being the operator monotone function \eqref{omf} associated to the $q$-$z$ metric dependent also on the parameters $q$ and $z$, the diffeomorphism induced by changing the tomographic scheme would now involve also $q$ and $z$ thus resulting into a different quantum metric within the $q$-$z$ family itself. According to the generality of such a family of metrics and its relation with well-know quantities in Information Geometry for specific values of the parameters, this may be useful to better understand the connections between the different quantum metrics employed in the literature towards a proper extension of the quantum Fisher metric on the full space of quantum states.\\

\section*{Acknowledgements}
The authors want to thank Prof. Paolo Facchi for driving their attention to the family of  $q$-$z$-\textit{R\'enyi Relative Entropies} and reference \cite{DATTA}.

The authors wish to thank N. Datta for allowing them to take the figure in \cite{DATTA} as source of inspiration for the figure of the present manuscript.
 
The authors want to thank the two anonymous referees for useful comments and remarks, as well as for having pointed out references \cite{carlen_frank_lieb-some_operator_and_trace_function_convexity_theorems, hasegawa_petz-noncommutative_extension_of_information_geometry_II, petz_sudar-geometries_of_quantum_states}.
 
G.M.  would like to acknowledge the grant ``Santander-UC3M Excellence Chairs 2016/2017". P.V.  acknowledges  support by COST (European Cooperation in Science  and  Technology)  in  the  framework  of  COST  Action  MP1405  QSPACE.

\appendix

\section{Explicit computations}
\label{subsec: explicit computations}

Here we will perform the detailed computation of the covariant tensor field:

\be
g_{q,z}(X\,,Y):=-i_{d}^{*}\left(L_{\mathbb{X}_{l}}\,L_{\mathbb{Y}_{r}}\,\mathbb{D}_{q,z}\right)\,,
\ee
where:

\be
D_{q,z}\left(\mathbf{U}\,,\vec{p}_{1}\,;\mathbf{V}\,,\vec{p}_{2}\right)=\frac{1}{q(1-q)}\left(1 - \Tr\left[\left(
(\mathbf{U}\,\bar{\rho}_{0}\,\mathbf{U}^{\dagger})^{\frac{q}{z}} 
(\mathbf{V}\,\bar{\varrho}_{0}\,\mathbf{V}^{\dagger})^{\frac{1-q}{z}}\right)^{z}\right]\right)\,,
\ee
with $\bar{\rho}_{0}=\mathrm{diag}(\vec{p}_{1})$ and $\bar{\varrho}_{0}=\mathrm{diag}(\vec{p}_{2})$.
At this purpose, we start setting:

\begin{equation}\label{A,B}
A=\bar{\rho}^{\frac{q}{z}}=\mathbf{U}\,\bar{\rho}_{0}^{\frac{q}{z}}\,\mathbf{U}^{\dagger}\qquad,\qquad B=\bar{\varrho}^{\frac{1-q}{z}}=\mathbf{V}\,\bar{\varrho}_{0}^{\frac{1-q}{z}}\,\mathbf{V}^{\dagger}\,.
\end{equation}
Since $z\in\mathbb R_+$, it can take both integer and noninteger values. 
Therefore, in order to have a well defined expression, we consider the analytical expansion of the function $(AB)^z$ with respect to the identity, say:

\begin{equation}\label{serie}
(AB)^z=\sum_{n=0}^{\infty}c_n(z)(AB-\mathds1)^n\;.
\end{equation}
Let us notice that, as stressed in \cite{DATTA}, even if $AB=\bar{\rho}^{\frac{q}{z}}\,\bar{\varrho}^{\frac{1-q}{z}}$ is not Hermitian, the spectrum of $AB$ coincides with the spectrum of $BA=(AB)^\dagger$ for $A$ and $B$ Hermitian operators as in equation \eqref{A,B}. 
This ensures that the spectrum of $AB$ is real and hence $(AB)^z$ as a function of $z$ does not have nonanalyticity branches and can be expanded as in \eqref{serie}.

Next, we consider:

\be
L_{\mathbb{X}_{l}}\,L_{\mathbb{Y}_{r}}\,\text{Tr}\left[\left(A\,B\right)^{z}\right]=\sum_{m=0}^{\infty}c_{m}(z)\,L_{\mathbb{X}_{l}}\,L_{\mathbb{Y}_{r}}\,\text{Tr}\left[(AB - \mathds{1})^{m}\right]\,.
\ee
Using the Leibniz rule together with the cyclic property of the trace and with the relation $L_{\mathbf{Y}_{r}}=i_{\mathbf{Y}_{r}}\mathrm{d}$ which is valid on functions, we have:
$$
L_{\mathbb{Y}_{r}}\,\text{Tr}\left[\left(A\,B\right)^{z}\right]=\sum_{m=0}^{\infty}c_{m}(z)\,L_{\mathbb{Y}_{r}}\,\text{Tr}\left(\underset{m}{\underbrace{(AB-\mathds1)\dots(AB-\mathds1)}}\right)=
$$
$$
=\sum_{m=0}^{\infty}c_{m}(z)\,\text{Tr}\left(A\,\left(i_{\mathbb{Y}_{r}}\mathrm{d}B\right)\,(AB-\mathds1)^{m-1}+(AB-\mathds1)\,A\,\left(i_{\mathbb{Y}_{r}}\mathrm{d}B\right)\,(AB-\mathds1)^{m-2}\right. + 
$$
$$
+\dots + \left.(AB-\mathds1)^{m-1}\,A\,\left(i_{\mathbb{Y}_{r}}\mathrm{d}B\right)\right)=
$$
\be\label{dB}
=\sum_{m=0}^{\infty}m\,c_{m}(z)\,\text{Tr}\left((AB-\mathds1)^{m-1}\,A\,\left(i_{\mathbb{Y}_{r}}\mathrm{d}B\right)\right)\,
\ee
where we used the fact that $i_{\mathbf{Y}_{r}}\mathrm{d}A=0$ because $A$ depends only on the elements of the left factor of $\mathcal{M}_{n}\times\mathcal{M}_{n}$.
Then:

{\footnotesize
$$
L_{\mathbb{X}_{l}}\,L_{\mathbb{Y}_{r}}\,\text{Tr}\left[\left(A\,B\right)^{z}\right]=\sum_{m=0}^{\infty}m\,c_{m}(z)\,L_{\mathbb{X}_{l}}\,\text{Tr}\left(\underset{m-1}{\underbrace{(AB-\mathds1)\dots(AB-\mathds1)}}\,A\,\left(i_{\mathbb{Y}_{r}}\mathrm{d}B\right)\right)=
$$
$$
=\sum_{m=0}^{\infty}m\,c_{m}(z)\,\left[\,\text{Tr}\left((AB-\mathds1)^{m-1}\,\left(i_{\mathbb{X}_{l}}\mathrm{d}A\right)\,\left(i_{\mathbb{Y}_{r}}\mathrm{d}B\right)\right) + \text{Tr}\left(\left(i_{\mathbb{X}_{l}}\mathrm{d}A\right)\,B\,(AB-\mathds1)^{m-2}\,\left(i_{\mathbb{Y}_{r}}\mathrm{d}B\right) +\right.\right.
$$
$$
+\left. \left. (AB-\mathds1)\, \left(i_{\mathbb{X}_{l}}\mathrm{d}A\right)\,B\,(AB-\mathds1)^{m-2}\,A\,\left(i_{\mathbb{Y}_{r}}\mathrm{d}B\right)+\cdots + (AB-\mathds1)^{m-2}\, \left(i_{\mathbb{X}_{l}}\mathrm{d}A\right)\,B\,A\,\left(i_{\mathbb{Y}_{r}}\mathrm{d}B\right)\right)\right]=
$$
$$
=z\,\text{Tr}\left((AB)^{z-1}\,\left(i_{\mathbb{X}_{l}}\mathrm{d}A\right)\,\left(i_{\mathbb{Y}_{r}}\mathrm{d}B\right)\right) +
$$
\be\label{dA}
+\sum_{m=0}^{\infty}m\,c_{m}(z)\sum_{a=0}^{m-2}\text{Tr}\left((AB-\mathds1)^{a}\,\left(i_{\mathbb{X}_{l}}\mathrm{d}A\right)\,B\,(AB-\mathds1)^{m-a-2}\,A\,\left(i_{\mathbb{Y}_{r}}\mathrm{d}B\right)\right)\,,
\ee}
where we used the fact that $i_{\mathbf{X}_{l}}\mathrm{d}B=0$ because $B$ depends only on the elements of the right factor of $\mathcal{M}_{n}\times\mathcal{M}_{n}$, and, in the first term of the last equality, we have used the relation

\begin{equation}
z(AB)^{z-1}=\sum_{m=0}^\infty\,m\,c_{m}(z)(AB-\mathds1)^{m-1}
\end{equation}
for the first-order derivative of a analytical function. 
The metric is then:

$$
g_{q,z}(X\,,Y)=\frac{z}{q(1-q)}i_{d}^{*}\left[\text{Tr}\left((AB)^{z-1}\,\left(i_{\mathbb{X}_{l}}\mathrm{d}A\right)\,\left(i_{\mathbb{Y}_{r}}\mathrm{d}B\right)\right)\right] +
$$
{\footnotesize
\be\label{g1}
+\frac{1}{q(1-q)}i_{d}^{*}\left[\sum_{m=0}^{\infty}m\,c_{m}(z)\sum_{a=0}^{m-2}\text{Tr}\left((AB-\mathds1)^{a}\,\left(i_{\mathbb{X}_{l}}\mathrm{d}A\right)\,B\,(AB-\mathds1)^{m-a-2}\,A\,\left(i_{\mathbb{Y}_{r}}\mathrm{d}B\right)\right)\right]\,.
\ee}

In order to perform computations for a generic $N$-level system it is useful to use the canonical basis $\{\mathbf{E}_{jk}\}_{j,k=1,...,n}$ of $\mathcal{B}(\mathcal{H})$ introduced before.
The left-invariant Maurer-Cartan 1-form $U^{-1}dU$ can be then written as:

\be\label{MCleft}
U^{-1}dU=i\sigma_k\theta^k=i\mathbf{E}_{\alpha\beta}\,M^{\alpha\beta}_{k}\,\theta^{k}
\ee
where $\{\sigma_k\}_{k=1,\dots,n^{2-1}}$ denote the basis for the $\mathfrak{su}(n)$, $\{\theta^k\}_{k=1,\dots,n^{2-1}}$ is the dual basis of left-invariant 1-forms, and we have expressed the matrices $\sigma$  as a linear combination of the $\mathbf{E}_{\alpha\beta}$ matrices with complex coefficients\footnote{See equation \eqref{2levsigmatau} in subsection \ref{subsec: the qubit case} and equation \eqref{eqn: sigmatau 3-level} in subsection \ref{subsec: the qutrit case} for the $SU(2)$ and $SU(3)$ case respectively. 
}, say:

\be\label{sigmatau}
\sigma_k=\sum_{\alpha,\beta=1}^N M_k^{\alpha\beta}\mathbf{E}_{\alpha\beta}\qquad,\qquad M_k^{\alpha\beta}\in\mathbb C \,.
\ee
The complex coefficients $M_k^{\alpha\beta}$ have to satisfy the following property:

\be\label{Mcoeff}
M_k^{\beta\alpha}=\overline{M_k^{\alpha\beta}}\qquad\forall\;k=0,\dots,N^2-1
\ee
where $\overline{M_k^{\alpha\beta}}$ denotes the complex conjugate of $M_k^{\alpha\beta}$.
This can be seen by taking the Hermitian conjugate of equation \eqref{sigmatau} which yields:

\be
\sigma_k=\overline{M_k^{\alpha\beta}}\left(\mathbf{E}_{\alpha\beta}\right)^\dagger=\overline{M_k^{\alpha\beta}}\mathbf{E}_{\beta\alpha}=M_k^{\beta\alpha}\mathbf{E}_{\beta\alpha}\;,
\ee
where we have used the fact that the $\sigma$'s are Hermitian and the property of the real matrices $\mathbf{E}$ according to which $\left(\mathbf{E}_{\alpha\beta}\right)^\dagger=\left(\mathbf{E}_{\alpha\beta}\right)^T=\mathbf{E}_{\beta\alpha}$.

The diagonal density matrix $\bar{\rho}_0$ can be written in terms of the $\mathbf{E}$ basis as:

\be\label{rho0}
\bar{\rho}_0=\sum_{\alpha=1}^{n}\,p_{\alpha}\mathbf{E}_{\alpha\alpha}\;,
\ee
where the $p_\alpha$ denote the $n$ eigenvalues of $\bar{\rho}_0$ satisfying the constraint $\text{Tr}(\bar{\rho}_0)=\sum_\alpha p_\alpha=1$, and the sum \eqref{rho0} involves only the diagonal matrices $\mathbf{E}_{\alpha\alpha}$. 
Moreover, by using the decomposition \eqref{rho0} and the commutation relations:

\be
[\mathbf{E}_{\alpha\beta},\mathbf{E}_{\alpha'\beta'}]=\delta_{\beta\alpha'}\mathbf{E}_{\alpha\beta'}-\delta_{\beta'\alpha}\mathbf{E}_{\alpha'\beta}\;,
\ee
we have:

\be
\bigl[U^{-1}dU,\bar{\rho}_{0}^{r}\bigr]=i\sum_{\alpha,\beta}(p_{\beta}^{r}-p_{\alpha}^{r})\mathbf{E}_{\alpha\beta}\,M^{\alpha\beta}_{j}\,\theta^{j}\;,
\ee
for any power $\bar{\rho}_{0}^{r}$ of  $\bar{\rho}_{0}$.
Consequently, we have:

{\footnotesize
\be\label{eqn: lie derivatives wrt left vector field for rho}
\begin{split}
&L_{\mathbb{X}_{l}}\,\bar{\rho}^{\frac{q}{z}}=\mathrm{d}\bar{\rho}^{\frac{q}{z}}(\mathbb{X}_{l})=\mathbf{U}\,\left(\mathrm{d}\bar{\rho}^{\frac{q}{z}}(\mathbb{X}_{l})\right)\,\mathbf{U}^{\dagger} + \mathbf{U}\,\left[\left(\mathbf{U}^{\dagger}\mathrm{d}\mathbf{U}((\mathbb{X}_{l}))\right)\,,\bar{\rho}^{\frac{q}{z}}\right]\,\mathbf{U}^{\dagger}=\\&=\frac{q}{z}\,\sum_{\alpha}\,\mathbf{U}\,\mathbf{E}_{\alpha\alpha}\,\mathbf{U}^{\dagger}\,p_{\alpha}^{\frac{q-z}{z}}\,\mathrm{d}p_{\alpha}(\mathbb{X}_{l}) + \mathbf{U}\,i\sum_{\alpha,\beta}(p_{\beta}^{\frac{q}{z}}-p_{\alpha}^{\frac{q}{z}})\mathbf{E}_{\alpha\beta}\,M^{\alpha\beta}_{j}\,\theta^{j}(\mathbb{X}_{l})\,\mathbf{U}^{\dagger}\,,
\end{split}
\ee

\be\label{eqn: lie derivatives wrt right vector field for rho}
\begin{split}
&L_{\mathbb{Y}_{r}}\,\bar{\varrho}^{\frac{1-q}{z}}=\mathrm{d}\bar{\varrho}^{\frac{1-q}{z}}(\mathbb{Y}_{r})=\mathbf{V}\,\left(\mathrm{d}\bar{\varrho}^{\frac{1-q}{z}}(\mathbb{Y}_{r})\right)\,\mathbf{V}^{\dagger} + \mathbf{V}\,\left[\left(\mathbf{V}^{\dagger}\mathrm{d}\mathbf{V}(\mathbb{Y}_{r})\right)\,,\bar{\varrho}^{\frac{1-q}{z}}\right]\,\mathbf{V}^{\dagger}=\\&=\frac{1-q}{z}\,\sum_{\alpha}\,\mathbf{V}\,\mathbf{E}_{\alpha\alpha}\,\mathbf{V}^{\dagger}\,\widetilde{p}_{\alpha}^{\frac{1-q-z}{z}}\,\mathrm{d}\widetilde{p}_{\alpha}(\mathbb{Y}_{r}) + \mathbf{V}\,i\sum_{\alpha,\beta}(\widetilde{p}_{\beta}^{\frac{1-q}{z}}-\widetilde{p}_{\alpha}^{\frac{1-q}{z}})\mathbf{E}_{\alpha\beta}\,M^{\alpha\beta}_{k}\,\eta^{k}(\mathbb{Y}_{r})\,\mathbf{V}^{\dagger}\,,
\end{split}
\ee}
where $\bar{\varrho}_0=\sum_{\alpha=1}^{n}\,\widetilde{p}_{\alpha}\mathbf{E}_{\alpha\alpha}$.
Now, coming back to the expression \eqref{g1} for the  tensor:
{\footnotesize \be\label{g2}
\begin{split}
&g_{q,z}(X\,,Y)=\frac{z}{q(1-q)}i_{d}^{*}\left[\text{Tr}\left((AB)^{z-1}\,\left(i_{\mathbb{X}_{l}}\mathrm{d}A\right)\,\left(i_{\mathbb{Y}_{r}}\mathrm{d}B\right)\right)\right] +\\
&+\frac{1}{q(1-q)}i_{d}^{*}\left[\sum_{m=0}^{\infty}m\,c_{m}(z)\sum_{a=0}^{m-2}\text{Tr}\left((AB-\mathds1)^{a}\,\left(i_{\mathbb{X}_{l}}\mathrm{d}A\right)\,B\,(AB-\mathds1)^{m-a-2}\,A\,\left(i_{\mathbb{Y}_{r}}\mathrm{d}B\right)\right)\right]\,,
\end{split}
\ee}
we focus on the first term in the RHS, and, using \eqref{eqn: lie derivatives wrt left vector field for rho} and \eqref{eqn: lie derivatives wrt right vector field for rho} and performing the pullback along $i_{d}$, we obtain:
{\footnotesize
\be
\begin{split}
&\frac{z}{q(1-q)}\,\left[\text{Tr}\left(\bar{\rho}_{0}^{\frac{z-1}{z}}\,\left( \frac{q}{z}\,\sum_{\alpha}\,\mathbf{E}_{\alpha\alpha}\,\,p_{\alpha}^{\frac{q-z}{z}}\,\mathrm{d}p_{\alpha}(X) + i\sum_{\alpha,\beta}(p_{\beta}^{\frac{q}{z}}-p_{\alpha}^{\frac{q}{z}})\mathbf{E}_{\alpha\beta}\,M^{\alpha\beta}_{j}\,\theta^{j}(X)\right)\cdot\right.\right.\\
&\left.\left.\cdot\left(\frac{1-q}{z}\,\sum_{\alpha}\,\mathbf{E}_{\gamma\gamma}\,p_{\gamma}^{\frac{1-q-z}{z}}\,\mathrm{d}p_{\gamma}(Y) + i\sum_{\gamma,\mu}(p_{\mu}^{\frac{1-q}{z}}-p_{\gamma}^{\frac{1-q}{z}})\mathbf{E}_{\gamma\mu}\,M^{\gamma\mu}_{k}\,\theta^{k}(Y)\right)\right)\right]\,.
\end{split}
\ee}
It is easy to see that the terms with $\mathrm{d}p_{\alpha}(X)\cdot\theta^{k}(Y)$ and $\theta^{j}(X)\cdot \mathrm{d}p_{\gamma}(Y)$ vanish, indeed:
\be
\frac{i}{1-q}\sum_{\alpha\,\beta\,\gamma\,\mu}\,p_{\beta}^{\frac{z-1}{z}}\,p_{\alpha}^{\frac{q-z}{z}}\,(p_{\mu}^{\frac{1-q}{z}}-p_{\gamma}^{\frac{1-q}{z}})\,\underset{\delta_{\beta\mu}\delta_{\beta\alpha}\delta_{\alpha\gamma}}{\underbrace{\text{Tr}\left(\mathbf{E}_{\beta\beta}\,\mathbf{E}_{\alpha\alpha}\,\mathbf{E}_{\gamma\mu}\right)}}\,\mathrm{d}p_{\alpha}(X)\cdot\,M^{\gamma\mu}_{k}\theta^{k}(Y)=0\,,
\ee
\be
\frac{i}{q}\sum_{\alpha\,\beta\,\gamma\,\mu}\,p_{\mu}^{\frac{z-1}{z}}\,p_{\gamma}^{\frac{1-q-z}{z}}\,(p_{\beta}^{\frac{q}{z}}-p_{\alpha}^{\frac{q}{z}})\,\underset{\delta_{\mu\gamma}\delta_{\mu\alpha}\delta_{\beta\gamma}}{\underbrace{\text{Tr}\left(\mathbf{E}_{\mu\mu}\,\mathbf{E}_{\alpha\beta}\,\mathbf{E}_{\gamma\gamma}\right)}}\,\mathrm{d}p_{\gamma}(Y)\cdot \,M^{\alpha\beta}_{j}\,\theta^{j}(X)=0\,.
\ee
On the other hand, the terms with $\mathrm{d}p_{\alpha}(X)\cdot p_{\gamma}(Y)$ and $\theta^{j}(X)\cdot\theta^{k}(Y)$ are:

\be\label{eqn: first term transversal part of the metric}
\frac{1}{z}\,\sum_{\alpha}\,p_{\alpha}^{-1}\,\mathrm{d}p_{\alpha}(X)\cdot \mathrm{d}p_{\alpha}(Y)\,,
\ee

\be\label{eqn: first term parallel part of the metric}
\frac{z}{q(1-q)}\, \sum_{\alpha\beta}\,p_{\alpha}^{\frac{z-1}{z}}\,(p_{\beta}^{\frac{q}{z}}-p_{\alpha}^{\frac{q}{z}})\,(p_{\beta}^{\frac{1-q}{z}}-p_{\alpha}^{\frac{1-q}{z}})\,\,M^{\alpha\beta}_{j}\,\theta^{j}(X)\cdot\,M^{\beta\alpha}_{k}\,\theta^{k}(Y)\,,
\ee
where it is clear that, in the second expression, all the terms with $\alpha=\beta$ vanish in the sum.
Moreover, note that, even if $\alpha\neq\beta$, the terms in the sum always vanish at every point for which it is $p_{\alpha}=p_{\beta}$.

Concerning the second term in the RHS of \eqref{g1} or, equivalently, of \eqref{g2}, using \eqref{eqn: lie derivatives wrt left vector field for rho} and \eqref{eqn: lie derivatives wrt right vector field for rho} and performing the pullback along $i_{d}$, we obtain:

{\footnotesize
$$
\frac{1}{q(1-q)}\,\sum_{m=0}^{\infty}m\,c_{m}(z)\sum_{a=0}^{m-2}\text{Tr}\left((\bar{\rho}_{0}^{\frac{1}{z}}-\mathds1)^{a}\,\left(\frac{q}{z}\,\sum_{\alpha}\,\mathbf{E}_{\alpha\alpha}\,p_{\alpha}^{\frac{q-z}{z}}\,\mathrm{d}p_{\alpha}(\mathbb{X}_{l}) + i\sum_{\alpha,\beta}(p_{\beta}^{\frac{q}{z}}-p_{\alpha}^{\frac{q}{z}})\mathbf{E}_{\alpha\beta}\,M^{\alpha\beta}_{j}\,\theta^{j}(X)\right)\cdot\right.
$$
$$
\left.\cdot\bar{\rho}_{0}^{\frac{1-q}{z}}\,(\bar{\rho}_{0}^{\frac{1}{z}}-\mathds1)^{m-a-2}\,\bar{\rho}_{0}^{\frac{q}{z}}\,\left(\frac{1-q}{z}\,\sum_{\gamma}\,\mathbf{E}_{\gamma\gamma}\,p_{\gamma}^{\frac{1-q-z}{z}}\,\mathrm{d}p_{\gamma}(Y) + i\sum_{\gamma,\mu}(p_{\mu}^{\frac{1-q}{z}}-p_{\gamma}^{\frac{1-q}{z}})\mathbf{E}_{\gamma\mu}\,M^{\gamma\mu}_{k}\,\theta^{k}(Y)\right)\right)\,.
$$}
Again, it is easy to see that the terms with $\mathrm{d}p_{\alpha}(X)\cdot\,\theta^{k}(Y)$ and $\theta^{j}(X)\cdot \mathrm{d}p_{\gamma}(Y)$ vanish, and we are left with the term in $\mathrm{d}p_{\alpha}(X)\cdot  \mathrm{d}p_{\alpha}(Y)$:

\be\label{eqn: second term transversal part of the metric}
\begin{split}
&\frac{1}{z^{2}}\,\sum_{m=0}^{\infty}\,\sum_{\alpha}\,m(m-1)\,c_{m}(z)\,(p_{\alpha}^{\frac{1}{z}} - 1)^{m-2}\,p_{\alpha}^{2\frac{1-z}{z}}\,\mathrm{d}p_{\alpha}(X)\cdot \mathrm{d}p_{\alpha}(Y)=\\ &=\frac{z-1}{z}\,\sum_{\alpha}\,p_{\alpha}^{-1}\,\mathrm{d}p_{\alpha}(X)\cdot  \mathrm{d}p_{\alpha}(Y)\,,
\end{split}
\ee
where we have used the expression:

\begin{equation}
z(z-1)(p_{\alpha}^{\frac{1}{z}})^{z-2}=\sum_{m=0}^\infty\,m(m-1)\,c_{m}(z)(p_{\alpha}^{\frac{1}{z}}-1)^{m-2}
\end{equation}
for the second-order derivative of an analytical function, and the term in $\theta^{j}(X)\cdot\,\theta^{k}(Y)$:

{\footnotesize
$$
\frac{-1}{q(1-q)}\,\sum_{m=0}^{\infty}\,\sum_{a=0}^{m-2}\,\sum_{\alpha,\beta,\gamma,\mu}\,m\,c_{m}(z)\,\text{Tr}\left((\bar{\rho}_{0}^{\frac{1}{z}}-\mathds1)^{a}\,\mathbf{E}_{\alpha\beta}\,\bar{\rho}_{0}^{\frac{1-q}{z}}\,(\bar{\rho}_{0}^{\frac{1}{z}}-\mathds1)^{m-a-2}\,\bar{\rho}_{0}^{\frac{q}{z}}\, \mathbf{E}_{\gamma\mu}\right)\cdot
$$
$$
 \cdot (p_{\beta}^{\frac{q}{z}}-p_{\alpha}^{\frac{q}{z}})\,(p_{\mu}^{\frac{1-q}{z}}-p_{\gamma}^{\frac{1-q}{z}})\,M^{\alpha\beta}_{j}\theta^{j}(X)\cdot\,M^{\gamma\mu}_{k}\theta^{k}(Y)=
$$

$$
=\frac{1}{q(1-q)}\,\sum_{m=0}^{\infty}\,\sum_{a=0}^{m-2}\,\sum_{b=0}^{a}\,\sum_{c=0}^{m-a-2}\,\sum_{\alpha,\beta}\,m\,c_{m}(z)\,(-1)^{b+c}\,\binom{a}{b}\,\binom{m-a-2}{c}\cdot
$$
$$
\cdot p_{\alpha}^{\frac{a-b}{z}}\,p_{\beta}^{\frac{m-a-1-c}{z}}\,(p_{\beta}^{\frac{q}{z}}-p_{\alpha}^{\frac{q}{z}})\,(p_{\beta}^{\frac{1-q}{z}}-p_{\alpha}^{\frac{1-q}{z}})\,M^{\alpha\beta}_{j}\theta^{j}(X)\cdot\,M^{\beta\alpha}_{k}\theta^{k}(Y)=
$$

$$
=\frac{1}{q(1-q)}\,\sum_{m=0}^{\infty}\,\sum_{a=0}^{m-2}\,\sum_{\alpha,\beta}\,m\,c_{m}(z)\,(p_{\alpha}^{\frac{1}{z}}-1)^{a}\,p_{\beta}^{\frac{1}{z}}\,(p_{\beta}^{\frac{1}{z}}-1)^{m-a-2}\cdot 
$$
$$
\cdot(p_{\beta}^{\frac{q}{z}}-p_{\alpha}^{\frac{q}{z}})\,(p_{\beta}^{\frac{1-q}{z}}-p_{\alpha}^{\frac{1-q}{z}})\,M^{\alpha\beta}_{j}\theta^{j}(X)\cdot\,M^{\beta\alpha}_{k}\theta^{k}(Y)=
$$

$$
=\frac{1}{q(1-q)}\,\sum_{m=0}^{\infty}\,\sum_{\alpha,\beta}\,m\,c_{m}(z)\,\frac{(p_{\beta}^{\frac{1}{z}}-1)^{m-1} - (p_{\alpha}^{\frac{1}{z}}-1)^{m-1}}{p_{\beta}^{\frac{1}{z}} - p_{\alpha}^{\frac{1}{z}}}\, p_{\beta}^{\frac{1}{z}}
$$
$$
(p_{\beta}^{\frac{q}{z}}-p_{\alpha}^{\frac{q}{z}})\,(p_{\beta}^{\frac{1-q}{z}}-p_{\alpha}^{\frac{1-q}{z}})\,M^{\alpha\beta}_{j}\theta^{j}(X)\cdot\,M^{\beta\alpha}_{k}\theta^{k}(Y)=
$$

\be\label{eqn: second term parallel part of the metric}
=\frac{z}{q(1-q)}\,\sideset{}{'}\sum_{\alpha,\beta}\,\left[\frac{p_{\beta}^{\frac{1}{z}}\,(p_{\beta}^{\frac{z-1}{z}} - p_{\alpha}^{\frac{z-1}{z}})\, (p_{\beta}^{\frac{q}{z}}-p_{\alpha}^{\frac{q}{z}})\,(p_{\beta}^{\frac{1-q}{z}}-p_{\alpha}^{\frac{1-q}{z}})}{p_{\beta}^{\frac{1}{z}} - p_{\alpha}^{\frac{1}{z}}}\right]\,M^{\alpha\beta}_{j}\theta^{j}(X)\cdot\,M^{\beta\alpha}_{k}\theta^{k}(Y)\,.
\ee}
where we have used equation \eqref{rho0} and the binomial expansions:

\begin{align}
\label{binom}
(\bar{\rho}_0^{\frac{1}{z}}-\mathbb{I})^a&=\sum_{b=0}^{a}\binom{a}{b}(-1)^b\bar{\rho}_0^{\frac{a-b}{z}}\\
\label{binomial2}
(\bar{\rho}_0^{\frac{1}{z}}-\mathbb{I})^{m-a-2}&=\sum_{c=0}^{m-a-2}\binom{m-a-2}{c}(-1)^c\bar{\rho}_0^{\frac{m-a-2-c}{z}},
\end{align}
in the first equality; the relations:

\begin{align}\label{binom2}
p_{\alpha}^{\frac{1}{z}}(p_{\alpha}^{\frac{1}{z}}-1)^{m-a-2}&=\sum_{c=0}^{m-a-2}\binom{m-a-2}{c}(-1)^{c}\,p_{\alpha}^{\frac{m-a-1-c}{z}}\\
\label{binom2b}
(p_{\alpha}^{\frac{1}{z}}-1)^{a}&=\sum_{b=0}^{a}\binom{a}{b}(-1)^{b}\,p_{\alpha}^{\frac{a-b}{z}}
\end{align}
in the second equality; the expression for the finite sum of a geometric series:

\begin{equation}\label{seriegeom}
\sum_{a=0}^{m-2}x^{a}=\frac{1-x^{m-1}}{1-x}\,,\; \mbox{ with } x=\frac{p_{\alpha}^{\frac{1}{z}} -1}{p_{\beta}^{\frac{1}{z}} -1}
\end{equation}
in the third equality, and the expression:

\be
\sum_{m=0}^{\infty}m\,c_m(z)(p_{\alpha}^{\frac{1}{z}}-1)^{m-1}=z\,p_{\alpha}^{\frac{z-1}{z}}
\ee
for the first-order derivative of an analytical function in the last equality.
Note that we have introduced the primed notation $\sum'$ in equation \eqref{eqn: second term parallel part of the metric} in order to recall that when $\alpha=\beta$, as well as for all those points such that $p_{\alpha}=p_{\beta}$, the terms in the sum vanish.

Collecting the terms in $\mathrm{d}p_{\alpha}(X)\cdot  \mathrm{d}p_{\alpha}(Y)$ (equations \eqref{eqn: first term transversal part of the metric} and \eqref{eqn: second term transversal part of the metric}) we obtain:
\be
\begin{split}
g_{q,z}^{\perp}(X\,,Y)&=\frac{1}{z}\,\sum_{\alpha}\,p_{\alpha}^{-1}\,\mathrm{d}p_{\alpha}(X)\cdot \mathrm{d}p_{\alpha}(Y)+\frac{z-1}{z}\,\sum_{\alpha}\,p_{\alpha}^{-1}\,\mathrm{d}p_{\alpha}(X)\cdot  \mathrm{d}p_{\alpha}(Y)=\\ &=\sum_{\alpha=1}^{n}\,\frac{1}{p_\alpha}\mathrm{d}p_\alpha(X)\cdot \mathrm{d}p_\alpha(Y)\,.
\end{split}
\ee
From this it follows that:
\be\label{gtrans}
g_{q,z}^{\perp}=\sum_{\alpha=1}^{n}\,\frac{1}{p_\alpha}\mathrm{d}p_\alpha\otimes \mathrm{d}p_\alpha=\sum_{\alpha=1}^{n}\,p_\alpha \mathrm{d}\ln p_\alpha\otimes \mathrm{d}\ln p_\alpha
\ee
which is the Fisher-Rao metric related to the component of the ``classical'' probability vector $\vec p=(p_1,\dots,p_n)$ identified with the diagonal elements of the invertible density matrix.

Collecting the terms in $\theta^{j}(X)\cdot\theta^{k}(Y)$ (equations \eqref{eqn: first term parallel part of the metric} and \eqref{eqn: second term parallel part of the metric}) we obtain:

{\small
\be
\begin{split}
&g^{\parallel}_{q,z}(X\,,Y)=\frac{z}{q(1-q)}\,\sideset{}{'}\sum_{\alpha\beta}\,p_{\alpha}^{\frac{z-1}{z}}\,(p_{\beta}^{\frac{q}{z}}-p_{\alpha}^{\frac{q}{z}})\,(p_{\beta}^{\frac{1-q}{z}}-p_{\alpha}^{\frac{1-q}{z}})\,M^{\alpha\beta}_{j}\,\theta^{j}(X)\cdot\,M^{\beta\alpha}_{k}\theta^{k}(Y) +\\
&+\frac{z}{q(1-q)}\,\sideset{}{'}\sum_{\alpha,\beta}\,\left[\frac{p_{\beta}^{\frac{1}{z}}\,(p_{\beta}^{\frac{z-1}{z}} - p_{\alpha}^{\frac{z-1}{z}})\, (p_{\beta}^{\frac{q}{z}}-p_{\alpha}^{\frac{q}{z}})\,(p_{\beta}^{\frac{1-q}{z}}-p_{\alpha}^{\frac{1-q}{z}})}{p_{\beta}^{\frac{1}{z}} - p_{\alpha}^{\frac{1}{z}}}\right]\,M^{\alpha\beta}_{j}\,\theta^{j}(X)\cdot\,M^{\beta\alpha}_{k}\theta^{k}(Y)=\\
=&\frac{z}{q(1-q)}\,\sideset{}{'}\sum_{\alpha,\beta}\,\left[\frac{(p_{\beta}- p_{\alpha})\, (p_{\beta}^{\frac{q}{z}}-p_{\alpha}^{\frac{q}{z}})\,(p_{\beta}^{\frac{1-q}{z}}-p_{\alpha}^{\frac{1-q}{z}})}{p_{\beta}^{\frac{1}{z}} - p_{\alpha}^{\frac{1}{z}}}\right]\,M^{\alpha\beta}_{j}\,\theta^{j}(X)\cdot\,M^{\beta\alpha}_{k}\theta^{k}(Y)\,.
\end{split}
\ee}
Now, let us introduce the shorthand notation for the coefficients:

\be\label{coeff}
\mathcal E_{\alpha\beta}:=\frac{(p_\alpha-p_\beta)(p_\alpha^{\frac{q}{z}}-p_\beta^{\frac{q}{z}})(p_\alpha^{\frac{1-q}{z}}-p_\beta^{\frac{1-q}{z}})}{(p_\alpha^{\frac{1}{z}}-p_\beta^{\frac{1}{z}})}\;\in\;\mathbb R\qquad\text{s.t.}\qquad\mathcal E_{\alpha\beta}=\mathcal E_{\beta\alpha}\,.
\ee 
Then, $g^{\parallel}_{q,z}$ can be written as

\be\label{Ntang}
\begin{split}
g^{\parallel}_{q,z}&=\frac{z}{q(1-q)}\sideset{}{'}\sum_{\alpha,\beta=1}^{n}\,\mathcal E_{\alpha\beta}\,\,M^{\alpha\beta}_{j}\,M^{\beta\alpha}_{k}\,\theta^{j}\otimes\theta^{k}\\
&=\frac{z}{q(1-q)}\sideset{}{'}\sum_{\alpha,\beta=1}^{n}\frac{1}{2}\mathcal E_{\alpha\beta}(M^{\alpha\beta}_{j}\,M^{\beta\alpha}_{k} + M^{\alpha\beta}_{k}\,M^{\beta\alpha}_{j})\theta^{j}\otimes\theta^{k}\;,
\end{split}
\ee
from which, according to the property \eqref{Mcoeff}, it follows that:

\be\label{gtangtheta}
g^{\parallel}_{q,z}=\frac{z}{q(1-q)}\sideset{}{'}\sum_{j,k=1}^{n^2-1}\mathcal C_{jk}\,\theta^j\otimes\theta^k\;,
\ee
with:

\be\label{jkcoeff}
\mathcal C_{jk}=\sideset{}{'}\sum_{\alpha,\beta=1}^{n}\mathcal E_{\alpha\beta}\,\Re{\bigl[M_j^{\alpha\beta}M_k^{\beta\alpha}\bigr]}\;.
\ee
The coefficients $\mathcal C_{jk}$ in equation \eqref{jkcoeff} depend only on the eigenvalues of the density matrix $\bar{\rho}$ and on the transformation matrices relating the two basis. 
Moreover, the $\mathcal C_{jk}$ are symmetric with respect to the exchange of $j$ and $k$ (i.e., $\mathcal C_{jk}=\mathcal C_{kj}$) and, being the $\mathcal{E}_{\alpha\beta}$ defined in equation \eqref{coeff} real, they are also real.

Eventually, summing equations \eqref{gtrans} and \eqref{gtangtheta} we obtain the expression of the tensor $g_{q,z}$:
\be\label{gNlevelapp}
g_{q,z}=g^{\perp}_{q,z} + g^{\parallel}_{q,z}=\sum_{\alpha=1}^{n}\,p_\alpha \mathrm{d}\ln p_\alpha\otimes \mathrm{d}\ln p_\alpha+\frac{z}{q(1-q)}\sideset{}{'}\sum_{j,k=1}^{n^2-1}\mathcal C_{jk}\,\theta^j\otimes\theta^k\;. 
\ee
Note that we kept the notation $\Sigma'$ introduced after equation \eqref{eqn: second term parallel part of the metric} in order to recall that when $\alpha=\beta$, as well as for all those points such that $p_{\alpha}=p_{\beta}$, the terms in the sum vanish.

%


\addcontentsline{toc}{section}{References}
\begin{thebibliography}{}

\bibitem{Marsden} R. Abraham, J. E. Marsden, T. Ratiu, \textit{Manifolds, tensor analysis, and applications}, Springer-Verlag, New York,  3-rd edition (2012).

\bibitem{amari} S. I. Amari, \textit{Information Geometry and Its Applications}, Applied Mathematical Sciences 194, Springer Japan (2016).

\bibitem{Amari-Nagaoka} S. I. Amari and H. Nagaoka, \textit{Methods of Information Geometry}, American Mathematical Society, Providence RI (2000).

\bibitem{andai} A. Andai, \textit{Monotone Riemannian metrics on density matrices with non-monotone scalar curvature}, J. Math. Phys. 44, 3675 (2003) 

\bibitem{araki-relative_entropy_of_states_of_von_neumann_algebras} H. Araki, \textit{Relative Entropy of States of von Neumann Algebras} Publ. RIMS, Kyoto Univ. 11, 809--833 (1976).

\bibitem{DATTA} K. M. R. Audenaert and N. Datta, \textit{alpha-z-relative Renyi entropies}, J. Math.  Phys. 56, 022202 (2015).

\bibitem{Beigi17} S. Beigi, \textit{Quantum R\'enyi Divergence Satisfies Data Processing Inequality}, J. Math. Phys. 54, 122202 (2013).

\bibitem{bengtsson} I. Bengtsson, K. Zyczkowski, \textit{Geometry of Quantum States: An Introduction to Quantum Entanglement}, Cambridge University Press (2007).

\bibitem{carlen_frank_lieb-some_operator_and_trace_function_convexity_theorems} E. A. Carlen and R. L. Frank and E. H. Lieb, \textit{Some operator and trace function convexity
theorems}, Linear ALgebra and its Applications 490, 174--185 (2016).


\bibitem{cencov} N. N. Cencov, \textit{Statistical Decision Rules and Optimal Inference}, American Mathematical Society, Providence, Rhode Island (1982).

\bibitem{cencov_morozowa-markov_invariant_geometry_on_state_manifolds} N. N. Cencov and E. A. Morozowa, \textit{Markov invariant geometry on state manifolds}, Journal of Soviet Mathematics 56(5), 2648--2669 (1991).

\bibitem{CDCFMMP17}  F. M. Ciaglia, F.  Di Cosmo, D. Felice, S. Mancini, G. Marmo, J. M. P\'{e}rez-Pardo,
  \textit{Hamilton-Jacobi approach to potential functions in information geometry}, J.  Math. Phys. 58, 063506 (2017).
  
\bibitem{datta} N. Datta and F. Leditzky, \textit{A limit of the quantum R\'{e}nyi divergence}, J. Phys. A: Math. Gen. 47, 045304 (2014)

\bibitem{ercolessi2} I. Contreras, E. Ercolessi, and M. Schiavina, \textit{On the geometry of mixed states and the Fisher information tensor}, J. Math. Phys. 57 062209 (2016).
  
\bibitem{Epstein31} H. Epstein, \textit{Remarks on two Theorem of Lieb}, Commun. Math. Phys. 31 (1973).  

\bibitem{em3} E. Ercolessi, G. Marmo, G. Morandi, N. Mukunda, \textit{Geometry of mixed states and degeneracy structure of geometric phases for multi-level quantum systems. A unitary group approach}, Int. J. Mod. Phys. A 16 (2001).

\bibitem{ercolessi1} E. Ercolessi and M. Schiavina, \textit{Symmetric logarithmic derivative for general n-level systems and the quantum Fisher information tensor for three-level systems}, Phys. Lett. A 377 (2013).

\bibitem{FL16} R.L. Frank, E.H. Lieb, \textit{Monotonicity of a Relative R\'enyi Entropy}, J. Math. Phys. 54, 122201 (2013).

\bibitem{gibilisco_isola-a_characterization_of_wigner-yanase_skew_information_among_statistically_monotone_metrics} P. Gibilisco and T. Isola, \textit{A characterization of Wigner-Yanase skew information among statistically monotone metrics}, Infin. Dimens. Anal. Quantum. Probab. Relat. Top. 4(4), 553--557 (2001).

\bibitem{gibilisco} P. Gibilisco and T. Isola, \textit{Wigner-Yanase information on quantum state space: the geometric approach}, Journal of Mathematical Physics 44, 3752--3762, (2003). 

\bibitem{grab} J. Grabowski, G. Marmo and M. Kus, \textit{Geometry of quantum systems: Density states and entanglement}, J. Phys. A 38 (2005).

\bibitem{hasegawa_petz-noncommutative_extension_of_information_geometry_II} H. Hasegawa and D. Petz, \textit{Non-Commutative Extension of Information Geometry II}, in \textit{Quantum Communication, Computing, and Measurement}, 109--118, Springer Science + Business Media, New York (1997).

\bibitem{Hiai22} F. Hiai, \textit{Concavity of Certain Matrix Trace and Norm Functions}, Linear Algebra Appl., 439(5), 1568--1589 (2013).

\bibitem{datta2} F. Hiai, D. Petz, \textit{The Proper Formula for Relative Entropy and its Asymptotics in Quantum Probability}, Commun. Math. Phys. 143 (1991).

\bibitem{holevo-probabilistic_and_statistical_aspects_of_quantum_theory} A. S. Holevo \textit{Probabilistic and Statistical Aspects of Quantum Theory}, Edizioni della Normale (2011).

\bibitem{holevo-statistical_structure_of_quantum_theory} A. S. Holevo \textit{Statistical Structure of Quantum Theory}, Springer-Verlag, Berlin Heidelberg (2001).

\bibitem{jaksic_ogata_pautrat_pillet-entropic_fluctuations_in_quantum_mechanics_an_introduction} V. Jaksic and Y. Ogata and Y. Pautrat and Pillet C.-A., \textit{Entropic Fluctuations in Quantum
Statistical Mechanics. An Introduction}, in \textit{Quantum Theory from Small to Large Scales, August 2010, Lecture Notes of the Les Houches Summer School: Volume 95}, 417--553, Oxford University Presss (2012).

\bibitem{kubo} F. Kubo and T. Ando, \textit{Means of positive linear operators}, Math. Ann. 246 (1980).

\bibitem{LMMVV12} M.~Laudato, G.~Marmo, F.~M.~Mele, F.~Ventriglia and P.~Vitale, \textit{Tomographic Reconstruction of Quantum Metrics}, J. Phys. A: Math. Theor. 51(5) 2018.
    
\bibitem{a.a.v.v.-differential_geometry_in_statistical_inference} S. L. Lauritzen, \textit{Differential geometry in statistical inference}, eds. S. I. Amari and O. E. Barndorff-Nielsen and R. E. Kass and S. L. Lauritzen and C. R. Rao, Institute of Mathematical Statistics, Hayward, California (1987).
    
\bibitem{vitale} V.I. Man'ko, G. Marmo, F. Ventriglia, P. Vitale, \textit{Metric on the Space of Quantum States from Relative Entropy. Tomographic Reconstruction}, J. Phys. A: Math. Theor.  50, 335302,   (2017).

\bibitem{Matumoto} T. Matumoto, \textit{Any statistical manifold has a contrast function: on the $C^{3}$-functions taking the minimum at the diagonal of the product manifold}, Hir. Math. Jour. 23(2), 327--332, (1993).

\bibitem{datta6} M. Mosonyi, F. Hiai, \textit{On the Quantum R\'enyi Relative Entropies and Related Capacity Formulas}, IEEE Trans. Inf. Th. 57, 2474-2487 (2011).

\bibitem{ohya_petz-quantum_entropy_and_its_use} M. Ohya and D. Petz, \textit{Quantum Entropy and Its Use}, Springer-Verlag, Berlin Heidelberg (1993).

\bibitem{petz1} D. Petz, \textit{Monotone Metrics on Matrix Spaces}, Linear Algebra Appl. 244, 8196 (1996).

\bibitem{petz-quantum_information_theory_and_quantum_statistics} D. Petz, \textit{Quantum information theory and quantum statistics}, Springer-Verlag, Berlin Heidelberg (2007).

\bibitem{petz-from_f-divergence_to_quantum_quasi-entropies_and_their_use} D. Petz, \textit{From f-Divergence to Quantum Quasi-Entropies and Their Use}, Entropy 12, 304--325 (2010).

\bibitem{petz_sudar-geometries_of_quantum_states}  D. Petz and C. Sudar, \textit{Geometries of Quantum States}, Journal of Mathematical Physics 37, 2662 (1996).


\bibitem{wrl1}D. Petz, C. Sudar, \textit{Extending the Fisher metric to density matrices}, Geometry in Present Days Science, eds. O.E. Barndorff-Nielsen and E.B. Vendel Jensen, World Scientific (1999). 

\bibitem{petz2} D. Petz and C. Sudar, \textit{On the curvature of a certain Riemannian space of matrice}, J. Math. Phys. 37, 2662 (1996).

\bibitem{orbite} S. G. Schirmer, T. Zhang, J. V. Leahy, \textit{Orbits of quantum states and geometry of Bloch vectors for N-level systems}, J. Phys. A 37, 1389 (2004). 

\bibitem{wrl2} C. Sudar, \textit{Radial extension of mononone Riemannian metrics on density matrices}, Publ. Math. Debrecen 49, 243 (1996).

\bibitem{dpiqrd} K. Takahashi and A. Fujiwara, \textit{Information geometry of sandwiched R\'enyi $\alpha$-divergence}, J. Phys. A: Math. Theor. 50, 165301 (2017).

\bibitem{tomamichel-quantum_information_theory_and_quantum_statistics} M. Tomamichel, \textit{Quantum Information Processing with Finite Resources Mathematical Foundations}, Springer International Publishing (2016).

\bibitem{umegaki-conditional_expectation_in_an_operator_algebra_IV} H. Umegaki, \textit{Conditional expectation in an operator algebra IV: Entropy and information}, Kod. Math. Sem. Rep. 14(2): 59--85 (1962).





































\end{thebibliography}
\end{document}